\documentclass[aps,pra,twocolumn,nofootinbib]{revtex4}

\usepackage{bm}
\usepackage{graphicx,graphics,psfrag}
\usepackage{amsbsy}
\usepackage{amsmath}
\usepackage{amsfonts}
\usepackage{amsthm}

\begin{document}

\theoremstyle{plain}
\newtheorem{theorem}{Theorem}
\newtheorem{lemma}[theorem]{Lemma}
\newtheorem{corollary}[theorem]{Corollary}
\newtheorem{conjecture}[theorem]{Conjecture}
\newtheorem{proposition}[theorem]{Proposition}

\theoremstyle{definition}
\newtheorem{definition}{Definition}

\theoremstyle{remark}
\newtheorem*{remark}{Remark}
\newtheorem{example}{Example}

\title{The resource theory of quantum reference frames: manipulations and
monotones}
\author{Gilad Gour}
\email{gour@math.ucalgary.ca}
\affiliation{Institute for Quantum Information Science and
Department of Mathematics and Statistics,
University of Calgary, 2500 University Drive NW,
Calgary, Alberta, Canada T2N 1N4}
\author{Robert W. Spekkens}
\email{r.w.spekkens@damtp.cam.ac.uk}
\affiliation{Department of Applied Mathematics and Theoretical Physics, University of
Cambridge, Cambridge CB3 0WA, United Kingdom}
\date{Oct. 4, 2007}

\begin{abstract}
Every restriction on quantum operations defines a resource theory,
determining how quantum states that cannot be prepared under the
restriction may be manipulated and used to circumvent the
restriction. A superselection rule is a restriction that arises
through the lack of a classical reference frame and the states that
circumvent it (the resource) are quantum reference frames. We
consider the resource theories that arise from three types of
superselection rule, associated respectively with lacking: (i) a
phase reference, (ii) a frame for chirality, and (iii) a frame for
spatial orientation. Focussing on pure unipartite quantum states
(and in some cases restricting our attention even further to subsets
of these), we explore single-copy and asymptotic manipulations. In
particular, we identify the necessary and sufficient conditions for
a deterministic transformation between two resource states to be
possible and, when these conditions are not met, the maximum
probability with which the transformation can be achieved. We also
determine when a particular transformation can be achieved
reversibly in the limit of arbitrarily many copies and find the
maximum rate of conversion.  A comparison of the three resource
theories demonstrates that the extent to which resources can be
interconverted decreases as the strength of the restriction
increases. Along the way, we introduce several measures of frameness
and prove that these are monotonically nonincreasing under various
classes of operations that are permitted by the superselection rule.
\end{abstract}

\pacs{03.67.Mn, 03.67.Hk, 03.65.Ud}
\maketitle
\tableofcontents

\section{Introduction}

For every interesting restriction on operations, there is a
resulting resource theory~\cite{Sch03}. For instance, the
restriction of local operations and classical communication (LOCC)
leads to the theory of entanglement. Against the backdrop of the
LOCC restriction (and an implicit restriction that the parties do
not share any entanglement at the outset), a single entangled pair
or a single use of a noiseless quantum channel are both considered
resources, one static, the other dynamic~\cite{Ple07}. A resource
theory specifies the manner in which one can inter-convert between
various resources, for instance, whether one entangled state can be
transformed by LOCC into another. Indeed, much of quantum
information theory is simply a theory of the inter-conversion
between resources~\cite{Dev04}. We are here interested in the
restriction of a superselection rule (SSR). Specifically, we imagine
a party that is restricted to operations that are invariant under
the action of a group $G$ and refer to this as a
\emph{superselection rule for G} or simply a \emph{G-SSR}. Although
SSRs are often considered to be axiomatic, it is better to consider
them as arising from practical restrictions. Indeed, most SSRs can
be lifted if one has access to a reference frame for the group in
question, so the restriction is ultimately one of access to an
appropriate reference frame~\cite{Aha67,BRS06,BRS07}. In this
context, the analogue of an entangled state -- that which can be
used to temporarily overcome the LOCC restriction -- is a state that
can be used to temporarily overcome the restriction of the SSR. Such
states have been referred to as \emph{bounded-size reference frames}
or simply \emph{quantum reference frames}. In this article, we study
the manner in which one can interconvert between such states under
the SSR. We are therefore exploring the resource theory of quantum
reference frames.

Just as we say that a state is entangled or has entanglement if it cannot be
prepared by LOCC, one can say that a state is \emph{G-asymmetric} or has
nonzero \emph{G-frameness} if it cannot be prepared by G-invariant
operations. One of the goals of this article is to provide operational
measures of frameness. The minimal requirement on such a measure is that
it be monotonically nonincreasing under G-invariant operations, in which
case it will be called a G-frameness monotone, in analogy with the
requirement that entanglement measures be monotonically nonincreasing under
LOCC operations. We distinguish three sorts of monotones: deterministic,
ensemble, and stochastic. These correspond respectively to monotonicity
under deterministic operations between states, under deterministic
operations between states and ensembles of states, and under stochastic
operations between states. An ensemble monotone is the standard notion of
a monotone in entanglement theory (with LOCC standing in for G-invariant
operations), while a deterministic monotone has been recently studied in
entanglement theory under the name of a type 2 monotone~\cite{Nest07}. The
notion of an ensemble frameness monotone is present (though unnamed) in
Vaccaro \emph{et al.}~\cite{Vac05} and Schuch
\emph{et al. }~\cite{SVC04a,SVC04b} while that of a deterministic
frameness monotone is found in
Appendix A of Ref.~\cite{BRST06}. We provide examples of each sort of
monotone through the various SSRs we consider. As in entanglement theory,
a focus on monotonicity properties is motivated by its utility in the study
of frame manipulations.

The structure of any quantum resource theory is dependent on the
extent of the restriction; the more restricted the set of allowed
operations, the fewer possibilities there are of inter-conversions
among different forms of a resource. For instance, in entanglement
theory, the restriction to LOCC operations is more substantive the
more parties one considers. Consider the question of whether two
pure entangled states can be interconverted in the sense of being
transformed one to the other with some nonzero probability using
stochastic LOCC~\cite{Dur00}. For bipartite pure states, one finds
that any two entangled states are interconvertible in this sense.
However, in the tripartite case, one finds that the intrinsically
3-way entangled pure states are divided into two classes, the GHZ
and the W states, with interconvertibility being possible only
within but not between the classes. For intrinsically 4-way
entangled pure states, the number of classes becomes
infinite~\cite{Wal04,Ver02}. Thus $n$-way LOCC constrains
manipulations among $n$-partite pure entangled states more strongly
than $(n-1)$-way LOCC constrains manipulations among $(n-1)$-partite
pure entangled states. The higher the number of parties, the
stronger is the constraint of LOCC.

Similarly, as we demonstrate in this article, an increase in the
strength of the superselection rule (where one SSR is stronger than
another if it allows fewer operations), leads to a decrease in the
number of possibilities for interconversion among quantum reference
frames. In particular, we show that this is the case as one
progresses through the relatively mild restriction of a Z$_2$-SSR,
to the stronger restriction of a U(1)-SSR, to the very strong
restriction of an SU(2)-SSR.  Each resource theory has its own
section in this article. Sections III, IV and V deal respectively
with the resource theories for the U(1)-SSR, the Z$_{2}$-SSR, and
the SU(2)-SSR. (The U(1) case is considered first because it is
likely to be the most familiar and it is the one upon which the most
previous work has been done.)  There are many ways in which each SSR
may arise in practice, and a particular example is provided for
each.  Specifically, the Z$_2$-SSR is shown to correspond to lacking
a reference frame for chirality, the U(1)-SSR to lacking a phase
reference, and the SU(2)-SSR to lacking a Cartesian frame.

We consider various types of manipulations of pure states for each of these
resource theories. In the context of single-copy manipulations, we seek to
determine necessary and sufficient conditions for a transformation from one
pure state to another to be possible by deterministic operations under the
SSR. %The conditions we have found are summarized
%in theorems~\ref{thm:dettransfU1},~\ref{prop:conditionsdeterministicZ2}
%and~\ref{D3}.
These results play the same role in the resource theory of quantum
reference frames as Nielsen's theorem plays in the theory of
entanglement~\cite{Nie99}. If a transformation between two pure
states is not possible deterministically, we attempt to find the
maximum probability with which the conversion can be achieved.
%Our results on this front are summarized in
%theorems~\ref{thm:maxprobU1},~\ref{prop:maxprobZ2} and~\ref{S3}
Our results on this front provide the analogues for reference frames
of Vidal's theorem in entanglement theory~\cite{Vid99}.

A comparison of these results provides one of the senses in which
the stronger SSRs allow fewer frame manipulations. For instance, we
can ask, for every type of SSR, whether the stochastic invariant
operations define multiple different classes in the sense of
stochastic interconvertibility being possible only within but not
between the classes.

The case of the Z$_{2}$-SSR is similar to that of pure bipartite
entanglement: for every pair of resource states, one member of the
pair can be converted to the other with some probability. Actually,
the resource theory of the Z$_{2}$-SSR is even nicer than the theory
of pure bipartite entanglement. In the latter, the reverse of a
stochastically-achievable conversion need not be
stochastically-achievable (for instance if the first state has a
larger Schmidt number than the second), whereas under the
Z$_{2}$-SSR, for every pair of states there is a nonzero probability
of converting both the first to the second and the second to the
first.

The amount of interconvertibility is reduced in the case of the
U(1)-SSR. For a given pair of resource states, it need not be the
case that one member of the pair can be converted to the other. For
instance, a single copy of $\left\vert 0\right\rangle +\left\vert
1\right\rangle $ cannot be converted to a single copy of $\left\vert
0\right\rangle +\left\vert 2\right\rangle ,$ or vice-versa, with any
probability (where $\left\vert n\right\rangle $ denotes an
eigenstate of the number operator). If we introduce an ordering
relation among states wherein one state is judged higher than
another if it can be converted to the other with some probability,
then the states form a partially ordered set under the U(1)-SSR. The
pair of $\left\vert 0\right\rangle +\left\vert 1\right\rangle $ and
$\left\vert 0\right\rangle +\left\vert 2\right\rangle $ provide an
example of two elements that are not ordered. (Note, however, that
for every pair of states, there is a third that is above both in the
partial order. For instance, both $\left\vert 0\right\rangle
+\left\vert 1\right\rangle $ and $\left\vert 0\right\rangle
+\left\vert 2\right\rangle $ can be obtained with some probability
from the state $\left\vert 0\right\rangle +\left\vert 1\right\rangle
+\left\vert 2\right\rangle .$) This is similar to the situation that
exists in the theory of pure \emph{tripartite} entanglement, where
GHZ states and W states cannot be interconverted one to the other
with any probability.

The amount of interconvertibility is reduced even further in the
case of the SU(2)-SSR. Under the U(1)-SSR, the pair of states
$\left\vert 2\right\rangle +\left\vert 3\right\rangle $ and
$\left\vert 0\right\rangle +\left\vert 1\right\rangle +\left\vert
2\right\rangle $ are ordered with respect to one another: although
the latter can't be obtained from the former with any probability,
the opposite conversion is possible. However, an analogous pair of
resource states under the SU(2)-SSR, $\left\vert 2,2\right\rangle
+\left\vert 3,3\right\rangle $ and $\left\vert 0,0\right\rangle
+\left\vert 1,1\right\rangle +\left\vert 2,2\right\rangle $ (where
$\left\vert j,m\right\rangle $ denotes the joint eigenstate of
$J^{2}$ and $J_{z}$ with eigenvalues $j(j+1)$ and $\hbar m$) are not
ordered with respect to one another as neither can be obtained from
the other with any probability.

We also consider \emph{asymptotic} manipulations of pure states. The
question here is: given an arbitrarily large number of copies of one
pure state, with what rate can one deterministically transform these
to (a good approximation of) an arbitrarily large number of copies
of a different pure state under the SSR?  If the asymptotic
interconversion can be achieved \emph{reversibly} between any two
states, then a unique measure of frameness over the states is
sufficient to characterize the rate of interconversion. In the
theory of bipartite entanglement for pure states, the entropy of
entanglement is such a measure. We demonstrate that a unique measure
also exists in the resource theory for the Z$_{2}$-SSR. Under the
U(1)-SSR, we find that certain types of states cannot be
asymptotically interconverted one to the other at any rate (for
instance, one cannot distill $\left\vert 0\right\rangle +\left\vert
1\right\rangle $ from $\left\vert 0\right\rangle +\left\vert
2\right\rangle $). However, we show that for a large class of
states, reversible interconversion \emph{is} possible, and the
unique measure of frameness that determines the rate of
interconversion is simply the variance over number (the connection
of this result to the one of Ref.~\cite{SVC04b} is disussed below).
In the resource theory of the SU(2)-SSR, we again find that there
are pairs of states for which the rate of distillation of one from
the other is strictly zero. In contrast with our U(1) case, however,
we can identify classes of states for which there is a nonzero rate
of interconversion in both directions, but where for certain pairs
the rate in one direction is not the inverse of the rate in the
other. It follows that a single measure of frameness is not in
general sufficient to infer the rate of interconversion of one state
to another in this class. Nonetheless, we show that a \emph{pair} of
measures is sufficient to infer the rates. Furthermore, although
there still exist subclasses of states for which reversible
interconversion is possible, these are much smaller than those
defined by the U(1)-SSR.

Another feature of the resource theory of quantum reference frames that does
not have any analogue in the theory of pure bipartite entanglement is found
in the asymptotic manipulation of resources under the Z$_{2}$-SSR. As we
have said, any two states can be reversibly interconverted asymptotically
under the Z$_{2}$-SSR, in analogy with pure bipartite entanglement theory.
However, unlike the latter, the rate of interconversion fails to be an
ensemble monotone. This result calls into question the widespread tendency
to require ensemble monotonicity of any measure of a resource (such as
entanglement or frameness) and is an example of how the study of quantum
reference frames may yield insights into which features of entanglement
theory are generic to resource theories and which not.

Of course, a prerequisite to answering all the sorts of questions we have
just described is a characterization of the full set of generalized
operations that are permitted under the SSR, that is, the full set of
allowed trace-nonincreasing completely positive maps. Therefore, at the
outset of this article, we demonstrate that a G-invariant operation can be
characterized as one that admits a Kraus decomposition in terms of
\emph{irreducible tensor operators} (see for example p.~232 in~\cite{Sakurai})
for the group G. This connection
allows us to provide convenient expressions for the invariant operations.
In particular, in the context of the SU(2)-SSR, where the characterization
is particularly difficult, the Wigner-Eckart theorem (a well-known result in
nuclear physics~\cite{Sakurai}) specifies the form that irreducible tensor
operators may take.

We end this introduction by placing this article in the context of
previous work in this area. There has been substantial progress on
the theory of quantum reference frames in the last few years. A
problem that has seen a great deal of attention is that of
identifying the optimal state of a quantum reference frame for
transmitting information about some degree of freedom (such as
chirality, phase, or orientation) according to some figure of merit
(such as the fidelity). The pioneering paper in this field is
arguably that of Gisin and Popescu~\cite{Gis99}, who consider the
problem of distributing a single direction in space. This problem,
together with that of distributing a triad of orthogonal directions,
was subsequently studied intensively by various groups (see
Refs.~\cite{Per02b,Bag06,Chi05}). The problem of the distribution of
a phase reference has also received a great deal of attention, with
roots in the field of phase estimation~\cite{Ber00}. In the case of
the problem of distributing chirality, see
Refs.~\cite{Dio00,Gis04,Col05}. A synthesis of much of this work and
more references can be found in Ref.~\cite{BRS07}.

There has also been a great deal of work on the resource theory of
\emph{shared} quantum reference frame, that is, bipartite states
that substitute for a reference frame that is common to many
spatially separated parties. This research has been essentially
confined, however, to the case of phase references. For instance,
van Enk~\cite{Enk05} considers the interconversion of static and
dynamic resources (such as qubits, ebits, cobits, and refbits) in
the presence of a U(1)-SSR, while Bartlett~\emph{et
al.}~\cite{BDSW06} demonstrate some analogies between the theory of
mixed bipartite entangled states and the theory of pure shared phase
references, such as the existence of states that are not locally
preparable but from which free singlets cannot be distilled (a
phenomenon that was also noted in Refs.~\cite{SVC04a,SVC04b,Enk05}).
The work of Schuch, Verstraete, and Cirac \cite{SVC04a,SVC04b},
however, has the most significance for the present article. These
authors considered resource manipulations under the following pair
of restrictions: (i) only LOCC operations can be implemented, (ii)
global and local U(1)-SSRs are in effect. There is a rich interplay
between these two restrictions which is explored in detail in their
article. By contrast, we consider the restriction of a U(1)-SSR
alone in a unipartite context. This allows us to identify which
aspects of the Schuch~\emph{et al.} resource theory are due solely
to the U(1)-SSR and which rely on the further restriction of LOCC.

A statement in Ref. \cite{SVC04b} suggests that the resource theory
for a unipartite scenario with a U(1)-SSR (the one we consider here)
must be trivial: \textquotedblleft the SSR imposes that for any
operator $O,$ $[O,\hat{N}]=0$ must hold [...] As the same
restriction holds for the admissible density operators, all states
can be converted into each other, and no interesting new effects can
be found.\textquotedblright However, this negative assessment is
only defensible under the presumption that SSRs are axiomatic
restrictions. By contrast, a party who lacks a reference frame for
some degree of freedom is restricted to operations that are
invariant -- the same restriction imposed by an axiomatic SSR -- but
he faces no restriction on the states of systems that he might come
to acquire. To be sure, he cannot prepare arbitrary states himself,
but he \emph{can} be provided with systems prepared in such states
by another party who has access to the reference frame. Such a state
would then constitute a resource. It is for this reason that there
is a nontrivial resource theory to be developed even in the
unipartite scenario \footnote{There is a sense in which a local
reference frame is always a shared reference frame with some other
party (although this party may sometimes be quite nebulous, such as
the fixed stars). Therefore, it is possible to extract some results
for the problem in which we are interested from the results of
Schuch et al. Nonetheless, we opt instead to derive all of our
results directly because we believe this to be a more intuitive
approach.}.

Finally, the notion of \emph{generalized entanglement} introduced by
Barnum, Knill, Ortiz, Somma, and Viola~\cite{Bar04} provides a
different approach to the quantification of frameness. In the
present work, resource theories are characterized by what parties
lack, for instance, a quantum channel or a reference frame.
Conversely, the approach of Ref.~\cite{Bar04} characterizes a
resource theory by specifying the set of observables that parties
can access.  For instance, choosing the distinguished set to be the
\emph{local} observables, it is possible to define a criterion for
whether a state is entangled or not. Similarly, by choosing the
distinguished set to be the observables that commute with the action
of the group associated with a given reference frame, one obtains a
criterion for whether a state acts as a quantum reference frame or
not.  Where the approach of Ref.~\cite{Bar04} is lacking, however,
is in characterizing the \emph{operations} that define the resource
theory.  Nonetheless, Ref.~\cite{Bar04} includes a preliminary
exploration of various possibilities for doing so, and the present
article provides further clues.  It is possible, therefore, that
this framework may be developed into something that is sufficiently
general to describe both the resource theories of entanglement and
quantum reference frames and possibly many others besides.

%Just as standard entanglement is relative to a decomposition into
%subsystems, a notion of generalized entanglement can be defined
%relative to a distinguished subspace of observables (standard
%entanglement is a special case because a subsystem decomposition can
%be defined in terms of the subspace of local observables). A
%superselection rule defines a distinguished subspace of observables
%relative to which one can define generalized entangled states. It is
%unclear, however, whether this criterion always yields all and only
%those states that we have judged to be resources, namely, those that
%fail to be G-invariant.

\subsection{Summary of main results}

We here provide a brief synopsis of the main results of this
article. \ For each of the three types of SSRs, we characterize the
operations (i.e. trace-nonincreasing CP maps) that are invariant
under the group G in question. \ We state the necessary and
sufficient conditions for a transformation of one pure resource
state to another to be possible by both deterministic and stochastic
G-invariant operations, and in the latter case, we state what we
have found regarding the maximum probability with which the
transformation can be achieved.
%\ These constitute the analogues for
%the resource theory of unipartite reference frames of Nielsen's
%~\cite{Nie99} and Vidal's \cite{Vid99} theorems in entanglement
%theory. \
Finally, we describe to what extent a pair of resource states can be
interconverted asymptotically.  The results are numbered as in the
text, although the statement of each might differ slightly.
Explanations and proofs can be found in the text.

%\subsubsection{Results for the Z$_{2}$\emph{-SSR}}
\subsubsection{Z$_{2}$\emph{-SSR}}

The Hilbert space decomposition induced by a unitary representation
of Z$_{2} $ is simply $\mathcal{H}=\bigoplus_{b}\mathcal{H}_{b},$
where the bit $b\in
\{0,1\}$ labels the irreducible representations of Z$_{2},$ and the $%
\mathcal{H}_{b}$ are multiplicity spaces. \ Again, transformations
within the multiplicity spaces are clearly Z$_{2}$-invariant and
consequently we
can confine our attention to $\mathcal{H}^{\prime }=\mathrm{span}%
\{\left\vert b\right\rangle \}\subseteq \mathcal{H}$ for some
arbitrary choice of even parity state $\left\vert 0\right\rangle $
and odd parity state $\left\vert 1\right\rangle $ in each
multiplicity space.

\textbf{Lemma \ref{lemma:Z2invariantop}.} A Z$_{2}$-invariant
operation
admits a Kraus decomposition $\{K_{B,\alpha }\},$ where $B\in \{0,1\}$ and $%
\alpha $ is an integer, satisfying
\begin{equation}
K_{B,\alpha }=S_{B}\tilde{K}_{B,\alpha }  \label{eq:KrausZ2}
\end{equation}%
where $\tilde{K}_{B,\alpha }\equiv c_{0}^{(B,\alpha )}\left\vert
0\right\rangle \left\langle 0\right\vert +c_{1}^{(B,\alpha
)}\left\vert 1\right\rangle \left\langle 1\right\vert $ changes the
relative amplitudes of the parity states, and $S_{0}=\left\vert
0\right\rangle \left\langle
0\right\vert +\left\vert 1\right\rangle \left\langle 1\right\vert $ and $%
S_{1}=\left\vert 0\right\rangle \left\langle 1\right\vert
+\left\vert 1\right\rangle \left\langle 0\right\vert $ do nothing or
flip the parity respectively. The coefficients satisfy
$\sum_{B,\alpha }|c_{b}^{(B,\alpha )}|^2 \leq 1$ for all $b$, with
equality if the operation is trace-preserving.

We consider transformations between two states, $\left\vert \psi
\right\rangle $ and $\left\vert \phi \right\rangle ,$ that are Z$_{2}$%
-noninvariant. We define%
\begin{eqnarray*}
p_{b} &\equiv &\left\langle \psi \right\vert \Pi _{b}\left\vert \psi
\right\rangle  \\
q_{b} &\equiv &\left\langle \phi \right\vert \Pi _{b}\left\vert \phi
\right\rangle
\end{eqnarray*}%
where $\Pi _{b}$ is the projector onto $\mathcal{H}_{b}.$ (The
assumed Z$_{2} $-noninvariance of $\left\vert \psi \right\rangle $
and $\left\vert \phi \right\rangle $ imply that all these weights
must be nonzero.) \ We also define
\begin{equation}
\mathcal{C}(|\psi \rangle )\equiv 2\text{min}\{p_{0},\;p_{1}\}.
\end{equation}%
The results we have derived are as follows.

\textbf{Theorem \ref{prop:conditionsdeterministicZ2}.} The
transformation $\left\vert \psi \right\rangle
\rightarrow \left\vert \phi \right\rangle $ is possible by a deterministic Z$%
_{2}$-invariant operation if and only if
\begin{equation}
\mathcal{C}(|\psi \rangle )\geq \mathcal{C}(|\phi \rangle ).
\end{equation}

It turns out that \emph{any} transformation $\left\vert \psi
\right\rangle \rightarrow \left\vert \phi \right\rangle $ can always
be achieved with some probability by stochastic Z$_{2}$-invariant
operations, so we need only specify the maximum achievable
probability.

\textbf{Theorem \ref{prop:maxprobZ2}.} If $\left\vert \psi
\right\rangle \rightarrow \left\vert \phi \right\rangle $ is not
possible by deterministic
Z$_{2}$-invariant operations, then the maximum probability of transforming $%
|\psi \rangle $ into $|\phi \rangle $ using Z$_{2}$-invariant
operations is
\begin{equation*}
P\left( |\psi \rangle \;\rightarrow \;|\phi \rangle \right) =\frac{\mathcal{C%
}(|\psi \rangle )}{\mathcal{C}(|\phi \rangle )}.
\end{equation*}

If a set of resource states is such that for every pair, a
reversible interconversion of arbitrarily many copies of one to
arbitrarily many copies of the other is possible asympototically
with arbitrarily high fidelity, then the maximum rate of any
interconversion is fixed by a unique measure (modulo normalization)
over the set.

\textbf{Theorem \ref{thm:asymptoticz2}.} Under the Z$_2$-SSR,
asymptotic reversible interconversion is possible between any two
pure resource states, and the unique asymptotic measure of
Z$_2$-frameness (modulo normalization) is
\begin{equation}
F^{\infty }(|\psi \rangle )=-\log \left\vert p_{0}-p_{1}\right\vert
. \label{eq:uniqueframenessmonotoneZ2}
\end{equation}

%\subsubsection{Results for the U(1)-SSR}
\subsubsection{U(1)-SSR}

A unitary representation of U(1) induces a decomposition of the
Hilbert
space $\mathcal{H}$ of the form $\mathcal{H}=\oplus _{n}(\mathbb{C\otimes }%
\mathcal{H}_{n})$ where $n\in \mathbb{N}$ labels the irreducible
representations of U(1) and the $\mathcal{H}_{n}$ are multiplicity
spaces. Because any change to the multiplicity index does not
require a phase
reference, we can, without loss of generality, confine our attention to $%
\mathcal{H}^{\prime }=\mathrm{span}\{\left\vert n\right\rangle
\}\subseteq \mathcal{H}$ for some arbitrary choice of number state
$\left\vert n\right\rangle $ in each multiplicity space.

\textbf{Lemma \ref{lemma:U1invariantop}.} An arbitrary
U(1)-invariant operation on $\mathcal{B}(\mathcal{H}^{\prime })$
admits a Kraus decomposition $\{K_{k,\alpha }\},$ where $k$ and
$\alpha $ are integers, such that
\begin{equation*}
K_{k,\alpha }=S_{k}\tilde{K}_{k,a}
\end{equation*}%
where $\tilde{K}_{k,\alpha }=\sum_{n}c_{n}^{(k,\alpha )}|n\rangle
\langle n|$ changes the relative amplitudes of the different number
states, possibly eliminating some, and $S_{k}=\sum_{n=\max
\{0,-k\}}|n+k\rangle \langle n|$
shifts the number of excitations upward by $k,$ that is, upward by $|k|$ if $%
k>0$, and downward by $|k|$ if $k<0$. \ The coefficients satisfy
$\sum_{k,\alpha }|c_{n}^{(k,\alpha )}|^{2}\leq 1$ for all $n$, with
equality if the operation is trace-preserving.

\strut We now consider transformations between a source state
$\left\vert \psi \right\rangle $ and a target state $\left\vert \phi
\right\rangle .$ \ Note that these are assumed to be resources, that
is, states that are
U(1)-noninvariant. \ We denote the weights on $n$ for each of these by%
\begin{eqnarray*}
p_{n} &\equiv &\left\langle \psi \right\vert \Pi _{n}\left\vert \psi
\right\rangle  \\
q_{n} &\equiv &\left\langle \phi \right\vert \Pi _{n}\left\vert \phi
\right\rangle
\end{eqnarray*}%
where $\Pi _{n}$ is the projector onto $\mathcal{H}_{n}.$ \ We also
define
the number spectrum of a state $\left\vert \psi \right\rangle $ by%
\begin{equation*}
\mathrm{Spec}(\psi )\equiv \{n|p_{n}\neq 0\},
\end{equation*}%
the set of $n$ that have nonzero weight in $\left\vert \psi
\right\rangle .$

We derive the following results.

\textbf{Theorem \ref{thm:dettransfU1}.} The transformation
$\left\vert \psi \right\rangle \rightarrow \left\vert \phi
\right\rangle $ is possible by a deterministic U(1)-invariant
operation if and only if $p_{n}$ can be obtained from $q_{n}$ by a
convex sum of shift operations, that is,
\begin{equation*}
p_{n}=\sum_{k=-\infty }^{\infty }w_{k}q_{n-k},
\end{equation*}%
where $0\leq w_{k}\leq 1$ and $\sum_{k}w_{k}=1.$ %\end{theorem}

\textbf{Theorem \ref{prop:criterionstochasticU(1)}.}The transformation $%
|\psi \rangle \rightarrow |\phi \rangle $ is possible by a
stochastic U(1)-invariant operation if and only if the number
spectrum of $\phi $ is a
subset of the shifted number spectrum of $\psi ,$ that is,%
\begin{equation*}
\exists k\in \mathbb{Z}:\mathrm{Spec}(\phi )\subset
\mathrm{Spec}(\psi )+k
\end{equation*}%
where $\mathrm{Spec}(\psi )+k\equiv \{n+k|p_{n}\neq 0\},$

\textbf{Theorem \ref{thm:maxprobU1}.} If there is only a single
value of $k$ such that the condition $\mathrm{Spec}(\phi )\subset
\mathrm{Spec}(\psi )+k$
holds, then the maximum probability of achieving the transformation $%
\left\vert \psi \right\rangle \rightarrow \left\vert \phi
\right\rangle $ using U(1)-invariant operations is
\begin{equation*}
P(\left\vert \psi \right\rangle \rightarrow \left\vert \phi
\right\rangle )=\min_{n}\left( \frac{p_{n}}{q_{n+k}}\right) .
\end{equation*}

\strut Note that in general there will be multiple values of $k$ such that $%
\mathrm{Spec}(\phi )\subset \mathrm{Spec}(\psi )+k.$ \ We have not
identified the maximum probability in these cases.

To state our results for asymptotic transformations, we must first
provide a definition: a number spectrum is said to be \emph{gapless}
if the increment between every successive pair of numbers in the
spectrum is 1.

\textbf{Theorem \ref{thm:asymptoticmeasureU1}.} Under the U(1)-SSR,
asymptotic reversible interconversion is possible among the pure
resource states that have a gapless number spectrum. Within this set
of states, the unique asymptotic measure of U(1)-frameness is the
scaled number variance,
%\begin{equation*}
%F^{\infty }(|\psi \rangle )=V(\left\vert \psi \right\rangle ),
%\end{equation*}%
%where
%\begin{equation*}
%V(\left\vert \psi \right\rangle )\equiv 4\left[ \langle \psi |\hat{N}%
%^{2}|\psi \rangle -\langle \psi |\hat{N}|\psi \rangle ^{2}\right] ,
%\end{equation*}%
\begin{equation*}
F^{\infty }(|\psi \rangle )\equiv 4\left[ \langle \psi |\hat{N}%
^{2}|\psi \rangle -\langle \psi |\hat{N}|\psi \rangle ^{2}\right] ,
\end{equation*}%
and the scaling factor of 4 is chosen so that $(\left\vert
0\right\rangle +\left\vert 1\right\rangle )/\sqrt{2}$ has measure 1.

%\textbf{Theorem} \label{thm:asymptoticmeasureU1} The unique
%asymptotic measure of U(1)-frameness for pure states $|\psi \rangle
%$ that have gapless number spectra is the scaled number variance,
%\begin{equation*}
%F^{\infty }(|\psi \rangle )=V(\left\vert \psi \right\rangle ).
%\end{equation*}%
%\end{theorem}

%\subsubsection{Results for the SU(2)-SSR}
\subsubsection{SU(2)-SSR}

The Hilbert space decomposition induced by a unitary representation
of SU(2) is $\mathcal{H}=\bigoplus_{j}\mathcal{M}_{j}\otimes
\mathcal{N}_{j}$ where $j\in \{0,1/2,1,3/2,...\}$ labels the
irreducible representations of
SU(2), the $\mathcal{M}_{j}$ are the representation spaces and the $\mathcal{%
N}_{j}$ are the multiplicity spaces. \ Again, transformations within
the multiplicity spaces are SU(2)-invariant and consequently we can
confine our
attention to $\mathcal{H}^{\prime }=\bigoplus_{j}\mathcal{M}_{j}=\mathrm{span%
}\{\left\vert j,m\right\rangle \}_{j,m}\subseteq \mathcal{H},$
defined by an arbitrary choice of state in each multiplicity space.

\textbf{Lemma \ref{lemma:U1invariantopbasic}.} An arbitrary
SU(2)-invariant operation on $\mathcal{B}(\mathcal{H}^{\prime })$
admits a Kraus decomposition $\{K_{J,M,\alpha }\},$ where $J\in
\{0,1/2,1,3/2,...\},$ $M\in \{-J,...,J\}$ and $\alpha $ is an
integer, such that
\begin{eqnarray}
K_{J,M,\alpha } &=&\sum_{j^{\prime }=0,1/2,1...}\sum_{m=-j^{\prime
}}^{j^{\prime }}\sum_{\;j=|J-j^{\prime }|}^{J+j^{\prime
}}(-1)^{j^{\prime
}-m}  \notag \\
&&\times
\begin{pmatrix}
j^{\prime } & J & j \\
-m & M & m-M%
\end{pmatrix}
\notag \\
&&\times
%c^{(J,\alpha)}_{j^{\prime },j}\;
f_{J,\alpha}(j^{\prime },j)\; |j^{\prime },m\rangle \langle
j,m-M|\;.
\end{eqnarray}
where the $2\times 3$ matrix is a Wigner $3j$ symbol and the
function $f_{J,\alpha}(j^{\prime },j)$ does not depend on $m$ or
$M$, and satisfies $\sum_{J,j',\alpha}\left|f_{J,\alpha}(j^{\prime
},j)\right|^2\leq 2j+1$ for all $j\;,$ with equality if the
operation is trace-preserving.

Rather than solving the resource theory for the SU(2)-SSR in
complete generality, we have restricted our attention to a subset of
all possible pure states, namely those confined to a subspace
$\mathcal{H}_{\hat{n}}\equiv \mathrm{span}\{\left\vert
j,j\right\rangle _{\hat{n}}|j=0,1/2,1,\dots \}\subset
\mathcal{H}^{\prime }$ (where $\left\vert j,j\right\rangle
_{\hat{n}}$ is the highest weight
eigenstate of $\vec{J}\cdot \hat{n}$) for some $\hat{n}.$ $\mathcal{H}_{\hat{%
n}}$ is the space of linear combinations of SU(2)-coherent states
associated with the quantization axis $\hat{n}$. \ We demonstrate
that SU(2)-invariant maps cannot transform a pure state inside
$\mathcal{H}_{\hat{n}}$ to one outside $\mathcal{H}_{\hat{n}}$ with
any probability, and so the only nontrivial resource theory for such
states corresponds to transformations within a given
$\mathcal{H}_{\hat{n}}.$

\textbf{Lemma \ref{lemma:SU2invariantop}.} An SU(2)-invariant
operation on $\mathcal{H}_{\hat{n}}$ that takes pure states to pure
states admits a Kraus decomposition $\{K_{J,\alpha}\}$, of the form
\begin{equation*}
K_{J,\alpha }=S_{-J}\tilde{K}_{J,\alpha}
\end{equation*}
where $\tilde{K}_{J,\alpha }=\sum_{j}c_{j}^{(J,\alpha)}|j,j\rangle
\langle j,j|$ changes the relative amplitudes of the $|j,j\rangle$
states, possibly eliminating some, and $S_{-J}=\sum_{j\geq
J}|j-J,j-J\rangle \langle j,j|$ shifts the value of $j$ downward by
$J$. The coefficients satisfy $\sum_{J\leq j}\sum_{\alpha}
|c_j^{(J,\alpha)}|^2 \leq 1$ for all $j$, with equality if the
operation is trace-preserving.

Define the weights on $j$ of the source state $\left\vert \psi
\right\rangle
$ and target state $\left\vert \phi \right\rangle $ by%
\begin{eqnarray*}
p_{j} &\equiv &\left\langle \psi \right\vert \Pi _{j}\left\vert \psi
\right\rangle  \\
q_{j} &\equiv &\left\langle \psi \right\vert \Pi _{j}\left\vert \psi
\right\rangle
\end{eqnarray*}%
where $\Pi _{j}$ is the projector onto $\left\vert j,j\right\rangle _{\hat{%
n}},$ and define the $j$-spectrum by%
\begin{equation*}
\mathrm{j}\text{\textrm{-}}\mathrm{Spec}(\psi )\equiv \{j|p_{j}\neq
0\}.
\end{equation*}%
\textbf{Theorem \ref{D3}.} The transformation $\left\vert \psi
\right\rangle \rightarrow \left\vert \phi \right\rangle $ is
possible by a deterministic SU(2)-invariant operation if and only if
\begin{equation}
p_{j}=\sum_{J}w_{J}q_{j+J},
\end{equation}%
where the sum is over $J\in \{0,1/2,1,...\}$ and where  $0\leq
w_{k}\leq 1$
and $\sum_{k}w_{k}=1.$ %\end{theorem}

\textbf{Theorem \ref{prop:criterionstochasticSU(2)}.} The transformation $%
|\psi \rangle \rightarrow |\phi \rangle $ is possible by stochastic
SU(2)-invariant operations if and only if
\begin{equation}
\exists J\in \{0,1/2,1,\dots \}:\mathrm{j}\text{\textrm{-}}\mathrm{Spec}%
(\phi )\subset \mathrm{j}\text{\textrm{-}}\mathrm{Spec}(\psi )-J.
\end{equation}

\textbf{Theorem \ref{S3}.} If there is only a single value of $J$
such that the condition $\mathrm{j}$\textrm{-}$\mathrm{Spec}(\phi
)\subset \mathrm{j}$ \textrm{-}$\mathrm{Spec}(\psi )-J$ holds, then
the maximum probability of achieving the transformation $\left\vert
\psi \right\rangle \rightarrow \left\vert \phi \right\rangle $ using
SU(2)-invariant operations is
\begin{equation*}
P(\left\vert \psi \right\rangle \rightarrow \left\vert \phi
\right\rangle )=\min_{j}\left( \frac{p_{j}}{q_{j-J}}\right) .
\end{equation*}

Finally, defining a gapless j-spectrum to be one wherein the
increment between every successive pair of $j$ values in the
spectrum is 1, we can state our result

\textbf{Theorem \ref{thm:asymptoticsu2}.} In the set of pure
resource states for the SU(2)-SSR that have gapless j-spectra, the
maximum rate at which $n$ copies of $|\psi\rangle$ can be converted
to $m$ copies of $|\phi\rangle$ is determined by a \emph{pair} of
measures: the scaled $j$-mean
\begin{equation}
\mathcal{M}(|\psi \rangle )\equiv 2\langle \psi |\mathcal{J}|\psi
\rangle \;,
\end{equation}
and the the scaled $j$-variance,
\begin{equation}
V(|\psi \rangle )\equiv 4\left[ \langle \psi |\mathcal{J}^{2}|\psi
\rangle -\langle \psi |\mathcal{J}|\psi \rangle ^{2}\right] \;,
\end{equation}
where
\begin{equation}
\mathcal{J}\equiv \sum_{j=0,\frac{1}{2},1,...}j|j,j\rangle \langle
j,j|\;,
\end{equation}
and the scaling factors were chosen such that
$(|0,0\rangle+|1,1\rangle)/\sqrt{2}$ has $j$-mean and $j$-variance of
1.  This rate is given by
\begin{equation}
\lim_{n\rightarrow \infty }\frac{m}{n}=\min
\Big\{\frac{\mathcal{M}(|\psi
\rangle )}{\mathcal{M}(|\varphi \rangle )},\frac{V(|\psi \rangle )}{%
V(|\varphi \rangle )}\Big\}\;.
\end{equation}

\textbf{Corollary \ref{corollary}.} Asymptotic reversible
interconversion is possible within the set of pure resource states
for the SU(2)-SSR that have gapless $j$-spectra and that have the
same ratio of scaled $j$-mean $\mathcal{M}$ to scaled $j$-variance
$V$. Within each such set, the unique asymptotic measure of
SU(2)-frameness (modulo normalization) is
\begin{equation}
F^{\infty}(|\psi\rangle)= V(|\psi\rangle).
\end{equation}

\section{The restriction of lacking a reference frame}

\subsection{\strut Preliminaries}

Reference frames are implicit in the definition of quantum states.
For instance, to assert that the quantum state is an eigenstate of
angular momentum along the $\hat{z}$ direction is to describe the
state relative to some physical system---a reference frame---that
defines the $\hat{z}$ direction. Different degrees of freedom
require different reference frames and are characterized by the
group under which they transform. For instance, if the degree of
freedom is orientation in space, then the group is $SO(3)$ and the
requisite frame is a triad of orthogonal spatial axes -- a Cartesian
frame.

If a party lacks a reference frame for some degree of freedom, then
they are effectively restricted in the sorts of states they can
prepare and the sorts of operations they can implement. Without
access to a Cartesian frame, for example, there is nothing with
respect to which rotations can be defined, and consequently
rotations cannot be implemented. Similarly, the only states that can
be prepared are those that are invariant under rotations.
Consequently, any quantum state that is \emph{not} rotationally
invariant is a resource. More generally, any system that is known to
be aligned with some reference frame is a resource to someone who
does not have access to that reference frame. Such a resource, or
quantum reference frame, is typically useful as a substitute for a
classical reference frame (although perhaps a poor one). For
instance, it may allow one to implement, with some probability, the
sorts of operations and measurements that one could implement if one
had a classical reference frame. Indeed, a non-invariant state
stands in for a classical reference frame in a manner similar to the
way in which an entangled state can stand in for a quantum
channel~\cite{BRS07} (although there are differences). Our task here
will be to characterize the manner in which such resource states can
be transformed under the allowed operations.

We presently sketch the precise nature of this restriction in the general
case. Suppose $G$ denotes the group of transformations associated with the
reference frame. The states that can be prepared without access to the
frame are those that are invariant under these transformations. Assuming
the system is described by a density operator
$\rho \in \mathcal{B}(\mathcal{H)}$,
where $\mathcal{H}$ is a Hilbert space and $\mathcal{B}(\mathcal{H})$
denotes the bounded linear operators on this space, and assuming that
$T:\;G\rightarrow \mathcal{B}(\mathcal{H)}$ denotes the unitary representation
of $G$ that corresponds to the physical transformations in question, the
states that can be prepared satisfy
\begin{equation}
\,T(g)\rho T^{\dag }(g)=\rho ,\quad \forall \ g\in G,
\label{eq:G-invariant_op}
\end{equation}
Equivalently,
\begin{equation}
\lbrack \rho ,T(g)]=0\,,\quad \forall \ g\in G\,.  \label{eq:StatesCommute}
\end{equation}
Such a state is said to be $G$\emph{-invariant}.

If the system is composed of many subsystems,
$\mathcal{H}=\bigotimes_{k}\mathcal{H}_{k},$ where the $k$th
subsystem transforms according to the defining representation
$T_{k},$ then $T$ is the tensor product representation of $G,$ that
is, $T(g)=\bigotimes_{k}T_{k}(g).$

This restriction on states is sometimes referred to as a
\emph{superselection rule} (SSR). Although the latter has often been
considered to be an axiomatic restriction rather than arising from
the lack of a reference frame, the mathematical characterization is
the same. Consequently, we will refer to the restriction of lacking
a reference frame for a group G as a \emph{G-SSR}. (Note that the
most common conception of a SSR, forbidding coherence between
distinguished subspaces, is only appropriate for Abelian groups. For
nonAbelian groups, it is more complicated.)

\strut The operations that can be performed under the $G$-SSR are those
associated with $G$-invariant CP maps. Let $\mathcal{T}$ be the unitary
representation of $G$ on the space of superoperators that corresponds to the
physical transformations in question, so that $\mathcal{T}
(g)[X]=T(g)XT^{\dag }(g).$  A CP map
$\mathcal{E}:\mathcal{B}(\mathcal{H})\rightarrow \mathcal{B}(\mathcal{H})$
is $G$-invariant if it satisfies
\begin{equation}
\mathcal{T}(g)\circ \mathcal{E}\circ \mathcal{T}^{\dag }(g)=\mathcal{E}
\,,\quad \forall \ g\in G\,,  \label{eq:G-invariant_supop}
\end{equation}
where $\mathcal{A}\circ \mathcal{B}[\rho ]=\mathcal{A}[\mathcal{B}[\rho ]]$
denotes a composition of operations, and $\mathcal{F}^{\dag }$, the
Hermitian adjoint for superoperators, is defined by \textrm{Tr}$(X\mathcal{F}
[Y])=$ \textrm{Tr}$(\mathcal{F}^{\dag }[X]Y)$ for all $X,Y\in \mathcal{B}(
\mathcal{H}).$ Equivalently, $\mathcal{E}$ is $G$-invariant if it satisfies
\begin{equation*}
\lbrack \mathcal{E},\mathcal{T}(g)]=0\,,\quad \forall \ g\in G\,,
\end{equation*}
where $[\mathcal{A},\mathcal{B}]=\mathcal{A}\circ \mathcal{B}-\mathcal{B}
\circ \mathcal{A}$ is the superoperator commutator.

\strut It is useful to highlight two ways in which the restriction of
lacking a reference frame may arise. On the one hand, a party may fail to
possess any system that can serve as a reference frame.  Such a restriction
is difficult to imagine in the case of a Cartesian frame, since all that is
required is a system that can define a triad of orthogonal vectors, and
these are ubiquitous (although even in this context, achieving high degrees
of precision and stability is a challenge). Such a restriction is,
however, easy to imagine in the case of more exotic reference frames. For
instance, a Bose-Einstein condensate acts as a reference frame for the phase
conjugate to atom number, and a superconductor acts as a reference frame for
the phase conjugate to number of Cooper pairs~\cite{DBRS06}, and neither is
straightforward to prepare.

The other way in which a superselection rule may arise is if a party
has a local reference frame, but it is uncorrelated with the
reference frame with respect to which the system is ultimately
described. An example serves to illustrate the idea. Suppose that
two parties, Alice and Charlie, each have a reference frame for the
degree of freedom in question, but that these are uncorrelated. If
$g\in G$ is the group element describing the passive transformation
from Alice's to Charlie's frame, the absence of correlation amounts
to assuming that $g$ is completely unknown. It follows that if Alice
prepares a state $\rho $ on $\mathcal{H}$ relative to her frame, the
system is represented relative to Charlie's frame by the
state~\footnote{ The invariant measure is chosen using the maximum
entropy principle: because Charlie has no prior knowledge about
Alice's reference frame, he should assume a uniform measure over all
possibilities.}
\begin{equation}
\mathcal{G}[\rho ]\equiv \int_{G}\text{d}g\,T(g)\rho T^{\dag }(g).
\label{eq:AveragedState}
\end{equation}
where $\text{d}g$ is the group-invariant (Haar) measure. We have
assumed that $G$ is a compact Lie group. If $G$ is instead a finite
group, we simply replace $\int_{G}\mathrm{d}g$ with
$|G|^{-1}\sum_{g\in G}$ where $|G|$ denotes the order of $G.$ We
call the operation $\mathcal{G}$ the \textquotedblleft
$G$-twirling\textquotedblright operation. If we are only interested
in describing Alice's systems relative to Charlie's reference frame
(perhaps because we are only interested in measurements performed
relative to the latter), then we can group the states into
equivalence classes, where the equivalence relation is equality
under G-twirling.  Every equivalence class has a $G$-invariant
member, satisfying $\rho =\mathcal{G}[\rho ]$. Indeed, the totality
of what Alice can predict about the outcomes of Bob's measurements
is always characterized by some G-invariant density operator, so,
relative to Charlie's frame, the density operators satisfying
Eq.~(\ref{eq:G-invariant_op}) are the only ones that Alice can
prepare.

Similarly, if Alice implements an operation $\mathcal{E}$ relative
to her frame, then relative to Charlie's frame she has implemented
the operation
\begin{equation}
\mathfrak{G}(\mathcal{E})\equiv
\int_{G}\text{d}g\,\mathcal{T}(g)\circ \mathcal{E}\circ
\mathcal{T}(g^{-1})\,.  \label{eq:GInvariantOperations}
\end{equation}
Equation~(\ref{eq:GInvariantOperations}) has the form of
Eq.~(\ref{eq:AveragedState}) except with operators replaced by superoperators. It is
therefore appropriate to refer to the map $\mathfrak{G}$ as
\textquotedblleft super-$G$-twirling\textquotedblright . If we again
choose to always represent operations by Alice relative to the reference
frame of Charlie, then all operations are of the form
$\mathcal{E}=\mathfrak{G}(\mathcal{E})$, and any such operation satisfies
Eq.~(\ref{eq:G-invariant_supop}).

\subsection{Kraus representation for G-invariant operations
\label{sec:KrausrepnGinv}}

We now proceed to derive an important result concerning the Kraus
representation of $G$-invariant operations. Suppose the operation
$\mathcal{E}$ has an $N$-term Kraus decomposition $\{K_{\mu}\}.$ It
is then
clear that the operation $\mathcal{T}(g)\circ\mathcal{E}\circ\mathcal{T}%
(g^{-1})$ has an $N$-term Kraus decomposition $\{K_{\mu}^{\prime}\}$
where
$K_{\mu}^{\prime}=T(g)K_{\mu}T^{\dag}(g)=\mathcal{T}(g)[K_{\mu}]$.
But now recall that if two CP maps are equivalent, then the Kraus
operators of one are a unitary remixing of those of the
other~\cite{Nie00}. Equation~(\ref{eq:G-invariant_supop}) then
implies that there exists an $N\times N$ unitary matrix $u(g)$ such
that
\begin{equation}
\mathcal{T}(g)[K_{\mu}]=\sum_{\mu^{\prime}}u_{\mu\mu^{\prime}}(g)K_{\mu
^{\prime}},\quad\forall\ g\in G. \label{Krausfreedom}%
\end{equation}

If the Kraus operators are linearly independent (so that the Kraus
decomposition has the minimal number of elements), then $u$ is a
unitary representation of $G.$ The reason is as follows. \ Suppose
that $\{W_{\mu}\}$
constitutes a dual basis to $\{K_{\mu}\}$ on the operator space $\mathcal{B}%
$($\mathcal{H)}$, so that the elements of one are orthonormal to
those of the
other relative to the Hilbert-Schmidt inner product, that is, \textrm{Tr}%
$(W_{\mu}^{\dag}K_{\mu^{\prime}})=\delta_{\mu\mu^{\prime}}.$ \ It is
always possible to find such a dual basis if the $\{K_{\mu}\}$ are
linearly
independent. \ It follows that%
\[
\mathrm{Tr}\left(
W_{\mu^{\prime\prime}}^{\dag}\mathcal{T}(g)[K_{\mu }]\right)
=u_{\mu\mu^{\prime\prime}}(g),\quad\forall\ g\in G,
\]
and consequently that $u(g)$ is simply a matrix representation of
the superoperator $\mathcal{T}(g).$ \ Because $\mathcal{T}$ is a
representation of $G,$ so is $u.$

By virtue of the unitary freedom in the Kraus decomposition, it is
always possible to choose the Kraus operators such that $u(g)$ is in
block-diagonal form, with the blocks labeled by the irreducible
representations (irreps) of $G$ and possibly a multiplicity index,
and with the dimensionality of each block corresponding to the
dimensionality of the associated irrep. We summarize this result in
the following lemma.

\begin{lemma}
\label{lemma:Ginvariantoperations} A $G$-invariant operation admits
a Kraus decomposition with Kraus operators $K_{jm\alpha},$ where $j$
denotes an irrep, $m$ a basis for the irrep, and $\alpha$ a
multiplicity index, satisfying
\begin{equation}
\mathcal{T}(g)[K_{jm\alpha}]=\sum_{m^{\prime}}u_{mm^{\prime}}^{(j)}%
(g)K_{jm^{\prime}\alpha},\quad\forall\ g\in
G,\label{eq:irreducibletensoroperators}%
\end{equation}
where $u^{(j)}$ is an irreducible unitary representation of $G.$
\end{lemma}

Notice that the action of the group only mixes Kraus operators
associated with the same $j$ and $\alpha.$ The set of operators
$\{K_{jm\alpha}|m\}$ for fixed
$j$ and $\alpha$ is called an \emph{irreducible tensor operator}%
~\cite{Sakurai} of rank $j$ in nuclear and atomic physics.
\footnote{We thank Matthias Christandl for bringing this to our
attention.}

From Eq.~(\ref{eq:irreducibletensoroperators}), it is clear that the
$K_{j,m,\alpha}$ play the same role in the Hilbert-Schmidt operator
space $\mathcal{B}$($\mathcal{H)}$ as the joint eigenstates
$\left\vert j,m,\beta\right\rangle $ of $J^{2}$ and $J_{z}$ play in
the Hilbert space $\mathcal{H}$.

A $G$-invariant operation which is of the form $\mathcal{E}_{j,\alpha}%
(\cdot)=\sum_{m}K_{jm\alpha}(\cdot)K_{jm\alpha}^{\dag}$ where the
$K_{jm\alpha}$ satisfy Eq.~(\ref{eq:irreducibletensoroperators}) for
some irrep $j$ and multiplicity index $\alpha$ will be called an
\emph{irreducible} $G$-invariant operation. Every $G$-invariant
operation is clearly a sum of irreducible $G$-invariant operations.

An obvious question to ask at this stage is whether it is always
possible to physically implement any given $G$-invariant operation.
That it \emph{is} possible is guaranteed by an application of the
Stinespring dilation theorem to $G$-invariant
operations~\cite{KW99}. The theorem ensures that it suffices to
prepare a $G$ -invariant state of an ancilla, couple this to the
system via a $G$ -invariant unitary, and then implement a
$G$-invariant measurement upon the ancilla.

\subsection{Frameness monotones\label{sec:framenessmonotones}}

There are three sorts of frameness monotones that we will consider in this
work. We term these deterministic, ensemble and stochastic frameness
monotones. We consider each in turn.

We define a \emph{deterministic} $G$-\emph{frameness monotone} as a function
$F:\;\mathcal{B}(\mathcal{H})\;\rightarrow \;\mathbb{R}^{+}$ that does not
increase under deterministic $G$-invariant operations. Specifically, $F$
is a $G$-frameness monotone if for all $\rho \in \mathcal{B}(\mathcal{H})$
and for all trace-preserving CP maps $\mathcal{E}$ satisfying
Eq.~(\ref{eq:G-invariant_supop}),
\begin{equation}
F(\mathcal{E}(\rho ))\leq F(\rho ).  \label{eq:dmonotonecond}
\end{equation}
This definition is in analogy with that of a type 2 entanglement
monotone, introduced in Ref.~\cite{Nest07}, which is a function
$E:\;\mathcal{B}(\mathcal{H}^{A}\otimes \mathcal{H}^{B}\otimes
\cdots ) \;\rightarrow \;\mathbb{R}^{+}$ that is non-increasing
under deterministic LOCC operations. The notion of a deterministic
frameness monotone was first made explicit in Appendix A of
Ref.~\cite{BRST06}.

We define an \emph{ensemble }$G$-\emph{frameness monotone }as a
function $F:\;\mathcal{B}(\mathcal{H})\;\rightarrow
\;\mathbb{R}^{+}$ that does not increase \emph{on average} under
$G$-invariant operations. This definition is in analogy with that of
standard entanglement monotones, functions
$E:\;\mathcal{B}(\mathcal{H}^{A}\otimes \mathcal{H}^{B}\otimes
\cdots )\;\rightarrow \;\mathbb{R}^{+}$ that are non-increasing on
average under LOCC operations~\cite{Ben96,Ben96b}. To make the
definition explicit, we note that the most general sorts of
$G$-invariant operations include: (1) $G$-invariant measurements,
and (2) forgetting information. $G$-invariant measurements generate
a transformation from a state $\rho $ to an ensemble $\{(w_{\mu
},\sigma _{\mu })\}$ (i.e. $\rho $ collapses to $\sigma _{\mu }$
with probability $w_{\mu })$ where $w_{\mu }\sigma _{\mu
}=\mathcal{E}_{\mu }(\rho )$ for some trace-nonincreasing
$G$-invariant operation $\mathcal{E}_{\mu }.$ For a frameness
monotone $F$ to be nonincreasing on average, we require
\begin{equation}
\sum_{\mu }w_{\mu }F(\sigma _{\mu })\leq F(\rho ).  \label{eq:monotonecond1}
\end{equation}
The second requirement is that if one knows the state to be $\sigma
_{\mu }$ with probability $w_{\mu }$ and then discards the
information about $\mu ,$ resulting in the state $\sigma =\sum_{\mu
}w_{\mu }\sigma _{\mu },$ then $F$ is nonincreasing,
\begin{equation}
F(\sigma )\leq \sum_{\mu }w_{\mu }F(\sigma _{\mu }).
\label{eq:monotonecond2}
\end{equation}

Note that any non-decreasing concave function of an ensemble frameness
monotone is also an ensemble frameness monotone. To see that this is the
case, consider the $G$-invariant transformation $\rho \rightarrow \{(w_{\mu
},\sigma _{\mu })\}$ and let $f:[0,1]\rightarrow \mathbb{R}$ be a
non-decreasing concave function (i.e. $f(tx+(1-t)y)\geq tf(x)+(1-t)f(y)$ for
all $t,x,y\in \lbrack 0,1]$). It is then straightforward to see that if $F$
is an ensemble frameness monotone, so that $F(\rho )\geq \sum_{\mu }w_{\mu
}F(\sigma _{\mu }),$ then $f(F(\rho ))\geq \sum_{\mu }w_{\mu }f(F(\sigma
_{\mu })).$

The idea of requiring that a measure of frameness be nonincreasing on
average under invariant operations is present in
Vaccaro~\emph{et al.}~\cite{Vac05} and Schuch~\emph{et al.}~\cite{SVC04b}.

Finally, we define a \emph{stochastic }$G$-\emph{frameness monotone}
as a $G$ -frameness monotone that is nonincreasing even under
stochastic (that is, nondeterministic) $G$-invariant operations.
Specifically, $F$ is a stochastic $G$-frameness monotone if for all
$\rho \in \mathcal{B}(\mathcal{H})$ and for all $\sigma \in
\mathcal{B}(\mathcal{H})$ such that the transformation $\rho
\rightarrow \sigma $ can be achieved either deterministically or
indeterministically by a $G$-invariant operation, $F(\sigma )\leq
F(\rho ).$ Equivalently, for all $\rho \in \mathcal{B}(\mathcal{H})$
and for all trace-nonincreasing completely positive maps
$\mathcal{S}$ that are $G$ -invariant, we require that
\begin{equation}
F(\mathcal{S}(\rho )/\mathrm{Tr}\left( \mathcal{S}(\rho )\right)
)\leq F(\rho ).  \label{eq:smonotonecond}
\end{equation}
An example from entanglement theory of a stochastic monotone is the Schmidt
number, which can not be increased even with probability less than 1 using
LOCC.

It is sometimes useful to consider a measure of frameness that is
only defined on a subset of all states. For instance, it may be defined
only on pure states (or only on a subset of pure states). Indeed, this
situation will be the norm in the present work. In this case, the
condition of Eq.~(\ref{eq:dmonotonecond})
(respectively Eq.~(\ref{eq:smonotonecond})) is only required to hold if the map
$\mathcal{E}$ (respectively $\mathcal{S}$) takes states in the subset of interest to
others in that subset -- otherwise the left-hand side of the condition is
not well-defined. Similarly, the condition of Eq.~(\ref{eq:monotonecond1})
is only required to hold when every outcome of the measurement yields a
state in the subset of interest, and that of (\ref{eq:monotonecond2}) is
only required to hold if the subset is closed under convex combination.
These weaker conditions are all that are required to hold for a measure with
a restricted domain of definition to be deemed a monotone of each type.

\begin{figure}[tp]
\includegraphics[scale=.6]{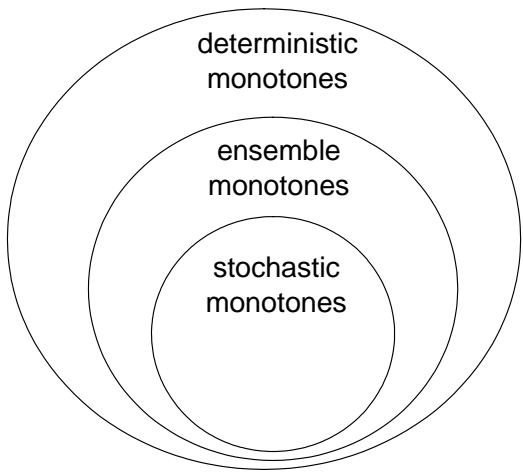}
\caption{A Venn diagram of frameness monotones for pure states.}
\label{mono-fig}
\end{figure}

For measures that are only defined on pure states, ensemble
monotones are only required to satisfy Eq.~(\ref{eq:monotonecond1})
(because the pure states are not closed under convex combination).
In this case, if a measure is a stochastic frameness monotone then
it is an ensemble frameness monotone because
Eq.~(\ref{eq:smonotonecond}) implies $F(\sigma _{\mu })\leq F(\rho
)$, which implies Eq.~(\ref{eq:monotonecond1}).  Furthermore, if a
measure is an ensemble frameness monotone then it is a deterministic
frameness monotone because deterministic $G$-invariant
transformations are a special case of $G$-invariant measurements
wherein there is only a single outcome. These inclusions are denoted
schematically in Fig.~\ref{mono-fig}.

There are a couple of other features that are nice for a measure of
frameness to have although these are, strictly speaking, only a choice of
convention: (1) Positivity, $F(\rho )\geq 0$ for all $\rho $ in the domain
of definition, and (2) Zero on $G$-invariant states $F(\rho )=0$ if $[\rho
,T(g)]=0$ for all $g\in G.$ Where there is a freedom in the definition of a
frameness measure, we will choose conventions ensuring that these features
hold.

\subsection{The motivation for requiring monotonicity}

\strut The motivation for demanding that a measure of the resource be
monotonically nonincreasing under the allowed operations is that it is a
necessary condition if the measure is to have operational significance.
This is an important point that is worth making precise.

We shall say that a measure of a resource is operational if and only if it
quantifies the \emph{optimal} figure of merit for some task that requires
the resource for its implementation. Specifically, we imagine a task that
is described entirely operationally (that is, in terms of empirically
observable consequences) and a figure of merit that quantifies the degree of
success achieved by every possible protocol for implementing the task (under
the restriction that defines the resource theory). Success might be
measured in terms of the probability of achieving some outcome, or the yield
of some other resource, etcetera. The key point is that
\emph{any processing of the resource} (consistent with the restriction that defines
the resource theory) cannot increase an operational measure of that resource
because the definition of an operational measure \emph{already incorporates}
an optimization over protocols and thus an optimization over all such
processings.

Because the sorts of operations that can appear in a protocol for the task
may be restricted, an operational measure might only be monotonically
nonincreasing for a restricted set of operations. The various monotones
described above -- deterministic, ensemble and stochastic -- are appropriate
for different sorts of tasks.

Some tasks may be achieved by protocols that at their end yield an ensemble
of states $\{(w_{\mu },\sigma _{\mu })\}.$ If the figure of merit for the
task is an average $\sum_{\mu }w_{\mu }f(\sigma _{\mu })$ of some figure of
merit $f$ for the final state, then the \emph{optimal} figure of merit for
the task (optimized over all protocols achieving the task) is an ensemble
monotone by definition.

If the figure of merit $f$ is a linear function of the density
operator, $f(\sum_{\mu }w_{\mu }\sigma _{\mu })=\sum_{\mu }w_{\mu
}f(\sigma _{\mu })$, then $f$ is unchanged by forgetting
information. Furthermore, the condition of being nonincreasing on
average under measurements becomes the condition of being
nonincreasing under deterministic operations. Consequently, the
notion of a deterministic frameness monotone is only distinct from
that of an ensemble frameness monotone for nonlinear figures of
merit. As an example, if one has a figure of merit over $\rho $ that
quantifies what can be achieved with $N>1$ copies of $\rho$, then
even if the achievement is itself some linear function of $\rho
^{\otimes N}$, the figure of merit need not be a linear function of
$\rho$. As noted above, the resource theory for the Z$_{2}$-SSR
provides an example of an operationally well-motivated measure of
frameness, the asymptotic rate of reversible interconversion of
resources, which is a deterministic monotone but not an ensemble
monotone.

Other tasks might incorporate post-selection in their definition.
Consequently, if the protocol yields an ensemble of outcomes, the figure of
merit for the protocol may be the maximum of some figure of merit for each
possible outcome rather than the average. This is the case, for instance,
when one is interested in the best-case or worst-case scenarios. The
measures of the resource for such tasks satisfy the strongest possible
constraint of monotonicity: they must be stochastic monotones.

\subsection{Single-copy frame manipulations}

For each sort of SSR considered in the paper, we seek to find necessary and
sufficient conditions for the existence of a deterministic $G$-invariant
operation that converts a pure state $\left\vert \psi \right\rangle$ into
another $\left\vert \phi \right\rangle$. In the context of entanglement
theory, these are provided by Nielsen's theorems~\cite{Nie99}. If a
particular conversion cannot be achieved deterministically, then we wish to
know the maximum probability with which it can be achieved. This is the
analogue of Vidal's formula in the theory of entanglement~\cite{Vid99}.

\subsection{Asymptotic frame manipulations and the unique asymptotic measure
of frameness\label{sec:asympframemanip}}

Even though a single copy of $\rho$ may not be converted to a single
copy of $\sigma$ deterministically under the $G$-SSR, a
transformation of $N$ copies of $\rho $ to $M$ copies of $\sigma$
might still be achievable.  Of particular interest is the question
of whether the transformation
\begin{equation*}
\rho ^{\otimes N}\rightarrow \sigma ^{\otimes M}
\end{equation*}
can be achieved in the limit $N\rightarrow \infty ,$ in the sense that there
exists a deterministic $G$-invariant operation $\mathcal{E}$ such that
\begin{equation*}
\textrm{Fid}(\mathcal{E}(\rho ^{\otimes N}),\;\sigma ^{\otimes
M})\simeq 1
\end{equation*}
where $\textrm{Fid}(\rho ,\sigma )\equiv Tr|\sqrt{\rho }\sqrt{\sigma
}|$ is the fidelity. The maximum ratio of $M$ to $N$ in the
asymptotic limit, $R_{\sigma }(\rho )\equiv \lim_{N\rightarrow
\infty }M/N,$ is called the asymptotic rate of conversion of $\rho $
to $\sigma $.

Clearly, the asymptotic rate of conversion to $\sigma$, $R_{\sigma
}$, is a deterministic frameness monotone. The proof is by
contradiction. If it were not, then there would exist a
deterministic $G$-invariant operation $\mathcal{E}$ that could be
performed on each of the $N$ copies of $\rho$ such that one could
then generate copies of $\sigma$ at an asymptotic rate of $R_{\sigma
}(\mathcal{E}(\rho ))\geq $ $R_{\sigma }(\rho )$, contradicting the
assumption that $R_{\sigma }$ quantified the optimal rate. An
analogue of this result holds in any resource theory.

Note also that $R_{\sigma }$ is weakly additive,
\begin{equation*}
R_{\sigma }(\rho ^{\otimes 2})=2R_{\sigma }(\rho ).
\end{equation*}
The proof is simply that $N$ copies of $\rho ^{\otimes 2}$ are
equivalent to $2N$ copies of $\rho$ and consequently can yield twice
as many copies of $\sigma$.

The resource theory that arises from a restriction on operations is
particularly simple if any form of the resource can be
\emph{reversibly} transformed (in an asymptotic sense) to any other
form under the restricted operations. In this case, for any pair of
states, $\rho$ and $\sigma$, one can reversibly transform $N$ copies
of $\rho $ into $M$ copies of $\sigma $ (or a good approximation
thereof) in the limit of large $N$. That is,
\begin{equation}
\rho ^{\otimes N}\Leftrightarrow \sigma ^{\otimes M},
\end{equation}
in the sense that there exist $G$-invariant operations $\mathcal{E}$ and
$\mathcal{E}^{\prime }$ such that
\begin{eqnarray*}
\textrm{Fid}(\mathcal{E}(\rho ^{\otimes N}),\;\sigma ^{\otimes M}) &\simeq &1 \\
\textrm{Fid}(\mathcal{E}^{\prime }(\sigma ^{\otimes M}),\;\rho
^{\otimes N}) &\simeq &1
\end{eqnarray*}
in the limit $N\rightarrow \infty $.

If there exist such asymptotic reversible transformations between
any two states then a single measure of $G$-frameness over the
states is sufficient to characterize the rate of interconversion
between any two. Specifically, if $\rho ^{\otimes N}\Leftrightarrow
\sigma ^{\otimes M}$, then we can define a measure of $G$-frameness
over all states, $F^{\infty },$ by
\begin{equation}
\lim_{N\rightarrow \infty }\frac{M}{N}=\frac{F^{\infty }(\rho
)}{F^{\infty }(\sigma )}.  \label{eq:uniquemeasure}
\end{equation}
This clearly does not fix the normalization of $F^{\infty }$, however, a
useful convention for doing so is to choose a particular state $\sigma$ to
be the \textquotedblleft standard\textquotedblright against which all
others are compared and to set $F^{\infty }(\sigma )=1$ for this state.

One of the most celebrated results in the theory of entanglement is
that there is a unique measure of entanglement for bipartite pure
states, the entropy of entanglement, which quantifies the number of
e-bits (i.e. maximally entangled states of two qubits) that can be
distilled from a given pure state $\left\vert \psi \right\rangle$ in
the asymptotic limit of many copies~\cite{Ben96,Ben96b}. We show
that whether one can obtain a unique measure of frameness for pure
states depends on the nature of the group associated with the frame.
In particular, a unique measure arises for the pure states under the
$Z_{2}$-SSR, but asymptotically reversible transformations exist
only for certain subsets of pure states for the U(1)-SSR and the
SU(2)-SSR.

If there is a unique measure of frameness $F^{\infty}$, then it is a
deterministic monotone and is weakly additive.  This follows from
the fact that such a measure is an instance of an asymptotic
conversion rate and the fact that such rates are deterministic
monotones and are weakly additive (as shown above). As it turns out,
however, $F^{\infty}$ need not be an ensemble frameness monotone. A
counterexample is provided by the resource theory for the
Z$_{2}$-SSR. This result is particularly interesting because it has
no analogue in pure state bipartite entanglement theory: the entropy
of entanglement, which quantifies the asymptotic rate of reversible
interconversion between entangled states, \emph{is} an ensemble
monotone. To the authors' knowledge, it is an open question whether
there exist subsets of the mixed bipartite entangled states or
multipartite entangled states that exhibit similar behaviour,
namely, that any two states in the subset can be reversibly
interconverted asymptotically but the rate of interconversion is not
an ensemble monotone.

\section{Resource theory of the U(1)-SSR\label{sec:RFsforphase}}

\subsection{\strut Phase references}

The first example we consider is that of a \emph{phase reference}, for which
the relevant group of transformations is U(1), the group of real numbers
modulo $2\pi$ under addition. \strut One requires a phase reference, for
instance, to prepare a coherent state of the electromagnetic field. The
phase reference typically takes the form of a strong classical field (a
local oscillator) with respect to which the phase of the coherent state is
defined. For two parties to share a phase reference, their local
oscillators must have a well-known relative phase, which is to say that they
must be phase-locked.

A phase reference is also required to prepare coherent superpositions of
eigenstates of any additively conserved charge. A charge operator differs
from a number operator because there is no lower bound on its spectrum. In
what follows, we shall presume a phase conjugate to number rather than
charge, although the results could easily be adapted to the case of charge.

Finally, note that if one possesses a reference frame consisting of a single
direction in space --the frame relative to which a system can be described
as pointing up or down -- then what one lacks to achieve a full Cartesian
frame (a triad of orthogonal directions) is a phase reference. In this
sense, the lack of a phase reference is a milder restriction than the lack
of a full Cartesian frame.

A phase shift of $\phi\in(0,2\pi)$ is represented by the unitary
\begin{equation*}
T(\phi )=e^{i\phi \hat{N}},
\end{equation*}
where $\hat{N}$ is the number operator.  $T$ is a unitary
representation of U(1). The states that are U(1)-invariant (or
\textquotedblleft phase-shift-invariant\textquotedblright ) are
those satisfying
\begin{equation*}
T(\phi )\rho T^{\dag }(\phi )=\rho \text{ \ \ \ \ }\forall \phi \in \mathrm{U(1)}.
\end{equation*}
This is equivalent to the condition
\begin{equation*}
\lbrack \rho ,\hat{N}]=0,
\end{equation*}
so that the invariant $\rho$ are block-diagonal relative to the
eigenspaces of $\hat{N}$.

It will be useful for us to decompose the Hilbert space $\mathcal{H}$ into a
direct sum of the carrier spaces $\mathcal{H}_{n}$ for the irreducible
representations of U(1), that is, the eigenspaces of the total number
operator $\hat{N},$
\begin{equation}
\mathcal{H}=\bigoplus_{n=0}^{\infty }\mathcal{H}_{n}\,.
\end{equation}
The dimensionality of each $\mathcal{H}_{n}$\strut is simply the
multiplicity of the $n$th irreducible representation of U(1) on the system.
As an example, if our system is $K$ optical modes, then $\mathcal{H}_{n}$ is
the eigenspace of states containing $n$ photons and has dimension equal to
the number of ways of distributing these $n$ photons among $K$ modes.

Let $\beta $ be a multiplicity index, so that we may denote a basis for
$\mathcal{H}_{n}$ by $\left\vert n,\beta \right\rangle$. An arbitrary
state can then be written as
\begin{equation*}
\left\vert \psi \right\rangle =\sum_{n,\beta }c_{n,\beta }\left\vert n,\beta\right\rangle ,
\end{equation*}
and transforms under phase shifts as
\begin{equation}
T(\phi )|\psi \rangle =\sum_{n}e^{in\phi }\sum_{\beta }c_{n,\beta
}\left\vert n,\beta \right\rangle \,.  \label{eq:U(1)actiononpurestate}
\end{equation}

\strut In this article, we will be considering the resource theory for pure
states only. From Eq.~(\ref{eq:U(1)actiononpurestate}), it is clear that
all and only those pure states that are confined to a single
$\mathcal{H}_{n} $ ($c_{n,\beta }\neq 0$ for only a single value of $n)$ are
U(1)-invariant and thus preparable under the U(1)-SSR. In the theory of
entanglement any state that cannot be prepared by LOCC can be considered a
resource. In our case, any state that cannot be prepared by $U(1)$-invariant
operations is considered a resource. The one-mode state $|0\rangle $ or
$|1\rangle $ or the two-mode state $(a|01\rangle +b|10\rangle )/\sqrt{2}$ are
not resources because they can be prepared under the U(1)-SSR (i.e. they are
considered cheap). On the other hand, the one-mode state $a|0\rangle
+b|1\rangle $ or the two mode state $a|01\rangle +b|12\rangle$ \emph{cannot}
be prepared under the U(1)-SSR and therefore \emph{do} constitute resources.

Because the multiplicity space carries a trivial representation of U(1)
(phase shifts act as identity upon it), it is clear that any change to the
multiplicity index does not require a phase reference. In other words, any
operation \emph{within} one of the $\mathcal{H}_{n}$ is possible under the
U(1)-SSR. Consequently, any pure state
$\left\vert \psi \right\rangle=\sum_{n,\beta }a_{n,\beta }|n,\beta \rangle$
can be taken, by a U(1)-invariant unitary operation, to the form
\begin{equation}
\left\vert \psi \right\rangle =\sum_{n}a_{n}|n\rangle ,
\label{eq:standardform1}
\end{equation}
where $|n\rangle$ is some particular element of $\mathcal{H}_n$.  We
will presume this form for states in what follows. We are thereby
restricting ourselves to the subspace $\mathcal{H}^{\prime
}=\mathrm{span}\{\left\vert n\right\rangle \}\subseteq \mathcal{H}$.
In the optical context, for example, this corresponds to
transforming all multi-mode states into single-mode states. The
analogue of this convention in the context of entanglement theory
for pure bipartite states would be to work with a particular choice
of Schmidt basis. Because the transformation from one Schmidt basis
to another can always be achieved by local unitaries, the Schmidt
basis is irrelevant to the question of entanglement manipulation and
it is therefore convenient to factor it out of the problem.

Relative to this standard form, the resource states are simply those
for which $a_{n}\neq 0$ for more than one value of $n$. To see how
such resource states can be manipulated, we must determine what can
be achieved using $U(1)$-invariant operations.

\subsection{U(1)-invariant operations\label{sec:U1invariantops}}

We now apply Lemma \ref{lemma:Ginvariantoperations} to the characterization
of U(1)-invariant operations. Note first that the irreducible
representations of U(1) are labeled by an integer $k,$ and are all
1-dimensional. The $k$th irreducible representation
$u_{k}:\mathrm{U(1)}\rightarrow \mathbb{C}$ has the form
\begin{equation*}
u_{k}(\phi )=e^{-ik\phi }.
\end{equation*}
It follows that the Kraus operators $K_{k,\alpha }$ of a
U(1)-invariant operation are labeled by an irrep $k$ and a
multiplicity index $\alpha$ and satisfy
\begin{equation}
e^{i\phi \hat{N}}K_{k,\alpha }e^{-i\phi \hat{N}}=e^{ik\phi }K_{k,\alpha
},\quad \forall \ \phi \in \mathrm{U(1).}  \label{eq:U(1)Krausconstraint}
\end{equation}
Note that by virtue of the fact that the irreps are 1d, the Kraus operators
do not get mixed with one another under the action of U(1). This provides
a significant simplification relative to the non-Abelian case.

As we are confining ourselves to the subspace $\mathcal{H}^{\prime
}=\mathrm{span}\{\left\vert n\right\rangle \}$, the most general
expression for $K_{k,\alpha }$ is
\begin{equation}
K_{k,\alpha }=\sum_{n,n^{\prime }}c_{nn^{\prime }}^{(k,\alpha )}|n\rangle
\langle n^{\prime }|\;,
\end{equation}
where the $c_{nn^{\prime }}^{(k,\alpha )}$ are complex coefficients.
Plugging this into Eq.~(\ref{eq:U(1)Krausconstraint}) yields the constraint
\begin{equation*}
\sum_{n,n^{\prime }}c_{nn^{\prime }}^{(k,\alpha )}e^{i(n-n^{\prime })\phi
}|n\rangle \langle n^{\prime }|=\sum_{n,n^{\prime }}c_{nn^{\prime
}}^{(k,\alpha )}e^{ik\phi }|n\rangle \langle n^{\prime }|,
\end{equation*}
from which it follows that $n^{\prime }=n-k$ and consequently
\begin{equation}
K_{k,\alpha }=\sum_{n=\max \{0,k\}}^{\infty }c_{n}^{(k,\alpha
)}|n\rangle \langle n-k|,  \label{eq:U(1)Kraus}
\end{equation}
for some amplitudes $c_n^{(k,\alpha)}$.(Note that if we were
considering a phase degree of freedom conjugate to charge rather
than number, the sum would have no lower bound.)

In order for the operation to be trace-nonincreasing, we require
$\sum_{k,\alpha }K_{k,\alpha }^{\dag }K_{k,\alpha }\leq I,$ which implies
that $\sum_{k,\alpha }|c_{n}^{(k,\alpha )}|^{2}\leq 1$ for all $n$, where
the inequalities are saturated if the operation is trace-preserving.

We summarize this result in the following lemma, where we also introduce a
useful factorization for the Kraus operators.

\begin{lemma} \label{lemma:U1invariantop} An arbitrary U(1)-invariant
operation admits a Kraus decomposition $\{K_{k,\alpha } \},$ where
$k$ and $\alpha $ are integers, such that
\begin{equation}
K_{k,\alpha }=S_{k}\tilde{K}_{k,a}  \label{eq:krausforU1}
\end{equation}
where $\tilde{K}_{k,\alpha }=\sum_{n}c_{n}^{(k,\alpha )}|n\rangle
\langle n|$ changes the relative amplitudes of the different number
states, possibly eliminating some, and $S_{k}=\sum_{n=\max
\{0,-k\}}|n+k\rangle \langle n|$ shifts the number of excitations
upward by $k,$ that is, upward by $|k|$ if $k>0$, and downward by
$|k|$ if $k<0$.  The coefficients satisfy $\sum_{k,\alpha
}|c_{n}^{(k,\alpha )}|^{2}\leq 1$ for all $n$, with equality if the
operation is trace-preserving.
\end{lemma}

As was mentioned in Sec.~\ref{sec:KrausrepnGinv}, the Stinespring
dilation theorem implies that there is always a way of physically
implementing any U(1)-invariant operation. Nonetheless, it is worth
saying a few words about how this is achieved. Just as the
restriction of LOCC still permits one to add and discard local
ancillae for free, in the resource theory for a U(1)-SSR, one can
add and discard ancillae prepared in U(1)-invariant states for free.
In order to shift the number of the system up by $k$ (i.e. to
implement the operation $S_{k}(\cdot )S_{k}^{\dag }$), one simply
adds an ancilla in an eigenstate $\left\vert k\right\rangle $ of the
number operator and implements the $\mathrm{U(1)}$-invariant unitary
operation that transforms the two-mode state $|n\rangle |k\rangle$
into the one mode state $|n+k\rangle $. To shift the number down by
$k$, one simply implements the $\mathrm{U(1)}$-invariant unitary
operation that takes the one-mode state $\left\vert
n+k\right\rangle$ to the two-mode state $|n\rangle |k\rangle$, and
then discards the second mode. This sort of argument was used in
Schuch \emph{et al.}~\cite{SVC04b} to justify
Eq.~(\ref{eq:U(1)Kraus}).

\subsubsection{U(1)-invariant unitaries}

As discussed above, all unitary operations within a given subspace
$\mathcal{H}_{n}$ are U(1)-invariant. However, there are more
U(1)-invariant unitaries besides these, specifically, nontrivial
unitaries on the subspace $\mathcal{H}^{\prime }=$
$\mathrm{span}\{\left\vert n\right\rangle \}$. Because unitary
operations have a single Kraus operator, they are irreducible
U(1)-invariant operations. However, the only way in which a single
Kraus operator $K$ can be unitary is if $k=0$ in
Eq.~(\ref{eq:krausforU1}), i.e. the operation does not allow shifts
in the number, and $|c_{n}|=1,$ so that $K$ must have the form
$\sum_{n}e^{i\chi_{n}}|n\rangle \langle n|.$ All told, the unitary
operations that are U(1)-invariant have the effect of merely
changing the relative phases of the $\left\vert n\right\rangle$.

At first glance, it might seem surprising that the phase of a state
can be changed without requiring a phase reference. Perhaps the
easiest way to develop an intuition for why this is true is to
consider two parties who don't share a phase reference. If Alice and
Bob share only a notion of what is up, that is, the $\hat{z}$ axis
of a Cartesian frame, then what they are lacking, relative to a full
Cartesian frame, is the angle between their local $\hat{x}$ axes.
This scenario is an example of lacking a phase reference. Alice
certainly cannot prepare a state of definite phase relative to Bob's
frame, nor gain any information about this phase, because this
requires sharing a common $\hat{x}$ axis with Bob. However, she
\emph{can} change the phase of a state relative to Bob's frame by a
fixed amount because this only requires performing a rotation about
the common $\hat{z}$ axis.

By a U(1)-invariant unitary, any state $\left\vert \psi
\right\rangle =\sum_{n}a_{n}|n\rangle $ can be taken to the form
\begin{equation}
\left\vert \psi \right\rangle =\sum_{n}\sqrt{p_{n}}|n\rangle ,
\label{eq:standardform2}
\end{equation}
where $\sum_{n}p_{n}=1,$ that is, a form with real-amplitude coefficients.
Consequently, to understand the possible resource manipulations, it suffices
to consider states of this standard form -- the real-amplitude states on
$\mathcal{H}^{\prime }=\mathrm{span}\{\left\vert n\right\rangle \}\subseteq \mathcal{H}.$
%Within this restricted set, the U(1)-invariant operations have Kraus operators of the form
%of Eq. (<ref>eq:krausforU1</ref>) but where the c_n^(k,a) are real coefficients.
%Such Kraus operators merely change the relative weights of the |n> but not the
%relative phases. The only unitary operation of this form is identity.
This convention is analogous, in the entanglement theory of pure bipartite
states, to restricting attention not just to states with a fixed Schmidt
basis but with real-amplitude Schmidt coefficients (because the phases of
the Schmidt coefficients can be changed by local unitaries).

\subsection{Deterministic single-copy transformations}

Consider first the question of which transformations
$|\psi \rangle\rightarrow |\phi \rangle$ can be achieved deterministically using
only $U(1)$-invariant operations. We assume the states to be in the standard
form, $\left\vert \psi \right\rangle =\sum_{n}\sqrt{p_{n}}\left\vert
n\right\rangle $ and $\left\vert \phi \right\rangle =\sum_{n}\sqrt{q_{n}}
\left\vert n\right\rangle ,$ and we denote the vector with components $q_{n}$
by $\vec{q}$. \ We also define a shift operator $\Upsilon _{k}$ on this
vector space by $\Upsilon _{k}\vec{q}=\vec{q}^{\,\prime }$ where
$q_{n+k}^{\prime }=q_{n}$.

\begin{theorem}\label{thm:dettransfU1} The necessary and sufficient
conditions for the transformation $\left\vert \psi \right\rangle \rightarrow
\left\vert \phi \right\rangle $ to be possible by a deterministic
U(1)-invariant operation is if $\vec{p}$ can be obtained from $\vec{q}$ by a
convex sum of shift operations, that is,
\begin{equation}
\vec{p}=\sum_{k=-\infty }^{\infty }w_{k}\Upsilon _{k}\vec{q},
\label{eq:NSforDU(1)}
\end{equation}
where $0\leq w_{k}\leq 1$ and $\sum_{k}w_{k}=1.$
\end{theorem}

\textbf{Proof.}
\footnote{The proofs are not required for the intelligibility of the text and we
recommend that they be ignored on a first reading.}
To be U(1)-invariant,
the operation must have Kraus operators $\{K_{k,\alpha }\}$ of the form
specified in lemma~\ref{lemma:U1invariantop}. Given that the operation
implements a pure-to-pure transformation, each Kraus operator must take
$\left\vert \psi \right\rangle $ to the same state, that is, for all
$k,\alpha ,$
\begin{equation}
K_{k,\alpha }\left\vert \psi \right\rangle =\sqrt{w_{k,\alpha }}\left\vert
\phi \right\rangle .
\end{equation}
where $0\leq w_{k,\alpha }\leq 1$. However,
\begin{eqnarray}
K_{k,\alpha }\left\vert \psi \right\rangle  &=&\sum_{n}c_{n}^{(k,\alpha )}
\sqrt{p_{n}}\left\vert n+k\right\rangle  \\
&=&\sqrt{w_{k,\alpha }}\sum_{n^{\prime }}\sqrt{q_{n^{\prime }}}\left\vert
n^{\prime }\right\rangle ,
\end{eqnarray}
and therefore
\begin{equation}
(c_{n}^{(k,\alpha )})^{2}p_{n}=w_{k,\alpha }q_{n+k}.  \label{eq:rrr}
\end{equation}

For the transformation to be deterministic, we require that
$\sum_{k,\alpha}\left\langle \psi \right\vert K_{k,\alpha }^{\dag }K_{k,\alpha }\left\vert
\psi \right\rangle =1,$ which implies that $\sum_{n}\left[ \sum_{k,\alpha
}(c_{n}^{(k,\alpha )})^{2}\right] p_{n}=1,$ and consequently that
$\sum_{k,\alpha }(c_{n}^{(k,\alpha )})^{2}=1$ for all $n$ such that
$p_{n}\neq 0.$

Summing Eq.~(\ref{eq:rrr}) over $k$ and $\alpha $\ and defining $w_{k}\equiv
\sum_{\alpha }w_{k,\alpha }$ yields $p_{n}=\sum_{k}w_{k}q_{n+k},$ which,
modulo a change in the sign of the dummy variable, is equivalent to
Eq.~(\ref{eq:NSforDU(1)}).

Conversely, if Eq.~(\ref{eq:NSforDU(1)}) holds, then we have $
p_{n}=\sum_{k}w_{k}q_{n+k}$ and we can define a set of amplitudes
$c_{n}^{(k)}\equiv \sqrt{w_{k}q_{n+k}/p_{n}}$ (with $c_{n}^{(k)}\equiv 0$ for
$n$ such that $p_{n}=0$). It follows that we can define operators
$K_{k}=S_{k}\tilde{K}_{k}$ where $\tilde{K}_{k}=\sum_{n}c_{n}^{(k)}\left\vert
n\right\rangle \left\langle n\right\vert $ are positive operators and where
\emph{\ }$\sum_{k,\alpha }\left\langle \psi \right\vert K_{k,\alpha }^{\dag
}K_{k,\alpha }\left\vert \psi \right\rangle
=\sum_{n}\sum_{k}(c_{n}^{(k)})^{2}p_{n}=1$. Consequently, the operators
$K_{k}=S_{k}\tilde{K}_{k}$ can constitute the Kraus operators for a
U(1)-invariant operation that is deterministic in its action on $\left\vert
\psi \right\rangle .$ \ Finally, it is straightforward to verify that this
operation achieves the transformation $\left\vert \psi \right\rangle
\rightarrow \left\vert \phi \right\rangle$ QED.

\begin{figure}[tp]
\includegraphics[scale=.8]{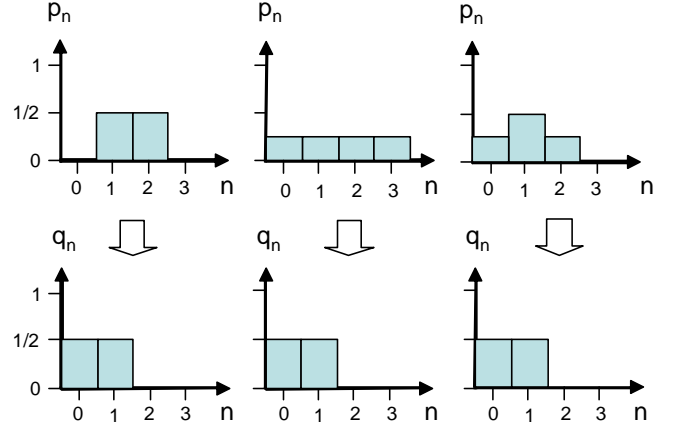}
\caption{Three examples of transformations that can be achieved by
deterministic U(1)-invariant operations.} \label{dete-fig}
\end{figure}

Some examples of transformations that can be achieved
deterministically are illustrated in Fig.~\ref{dete-fig}. The first
example, $\left( \left\vert 1\right\rangle +\left\vert
2\right\rangle \right) /\sqrt{2}\rightarrow \left( \left\vert
0\right\rangle +\left\vert 1\right\rangle \right) /\sqrt{2}$,
satisfies the condition because $\vec{p}=\Upsilon _{1}\vec{q}$.  The
U(1)-invariant operation that achieves the transformation has a
single Kraus operators $S_{-1}$ corresponding to a shift of the
number downward by 1.  (The operation is deterministic because
$S_{-1}^{\dag}S_{-1}$ acts as identity on $\left( \left\vert
1\right\rangle +\left\vert 2\right\rangle \right) /\sqrt{2}$.) The
second example, $\left( \left\vert 0\right\rangle +\left\vert
1\right\rangle +\left\vert 2\right\rangle +\left\vert 3\right\rangle
\right) /2\rightarrow \left( \left\vert 0\right\rangle +\left\vert
1\right\rangle \right) /\sqrt{2}$, satisfies the condition because
$\vec{p}=\frac{1}{2}\vec{q}+\frac{1}{2}\Upsilon _{2}\vec{q}$, and
the U(1)-invariant operation that achieves the transformation has
Kraus operators $K_{0}=\left\vert 0\right\rangle \left\langle
0\right\vert +\left\vert 1\right\rangle \left\langle 1\right\vert $
and $K_{-2}=S_{-2}\left( \left\vert 2\right\rangle \left\langle
2\right\vert +\left\vert 3\right\rangle \left\langle 3\right\vert
\right) ,$ corresponding to implementing a projective-valued measure
$\{\left\vert 0\right\rangle \left\langle 0\right\vert +\left\vert
1\right\rangle \left\langle 1\right\vert ,\left\vert 2\right\rangle
\left\langle 2\right\vert +\left\vert 3\right\rangle \left\langle
3\right\vert \}$ and shifting the number downward by $2$ upon
obtaining the second outcome. Finally, the third transformation,
$\left( \left\vert 0\right\rangle +\sqrt{2 }\left\vert
1\right\rangle +\left\vert 2\right\rangle \right) /2\rightarrow
\left( \left\vert 0\right\rangle +\left\vert 1\right\rangle \right)
/\sqrt{2} $, satisfies the condition because
$\vec{p}=\frac{1}{2}\vec{q}+\frac{1}{2} \Upsilon _{1}\vec{q}$ and
the operation has Kraus operators $ K_{0}=\left\vert 0\right\rangle
\left\langle 0\right\vert +\frac{1}{\sqrt{2}} \left\vert
1\right\rangle \left\langle 1\right\vert $ and $ K_{-1}=S_{-1}\left(
\frac{1}{\sqrt{2}}\left\vert 1\right\rangle \left\langle
1\right\vert +\left\vert 2\right\rangle \left\langle 2\right\vert
\right) $ corresponding to a measurement of the POVM $\{\left\vert
0\right\rangle \left\langle 0\right\vert +\frac{1}{2}\left\vert
1\right\rangle \left\langle 1\right\vert ,\frac{1}{2}\left\vert
1\right\rangle \left\langle 1\right\vert +\left\vert 2\right\rangle
\left\langle 2\right\vert \}$ followed by a shift downward by 1 upon
obtaining the second outcome.

The problem of determining whether Eq.~(\ref{eq:NSforDU(1)}) is satisfied
reduces to determining whether $\vec{p}$ falls in the convex hull of the
$\Upsilon _{k}\vec{q}.$ If $\vec{p}$ has a finite number of nonzero
elements, the number of $k$ values over which one must vary is also finite.
%\ \texttt{I seem to recall that this sort of problem is a semi-definite
%programming problem that can be solved efficiently. }

It is worth noting that a necessary condition for the transformation
$ \left\vert \psi \right\rangle \rightarrow \left\vert \phi
\right\rangle $ to be achieved by a U(1)-invariant operation is that
$\vec{p}$ is majorized by $ \vec{q}$ (an introduction to the notion
of majorization can be found in Ref.~\cite{Batia}). The proof is
simply that the shift operation $\Upsilon _{k}$ is a type of
permutation, and consequently, $\sum_{k}w_{k}\Upsilon _{k}$ is a
doubly-stochastic matrix. Thus if Eq.~(\ref{eq:NSforDU(1)}) holds,
then $\vec{p}$ can be obtained from $\vec{q}$ by a doubly-stochastic
matrix, and it then follows from the Polya-Littlewood-Richardson
theorem~\cite{Batia} that $\vec{p}$ is majorized by $\vec{q}$.

Majorization is well-known in quantum information theory because one
entangled state can be transformed deterministically to another by LOCC if
and only if the spectrum of the reduced density operator of the one is
majorized by that of the other~\cite{Nie99}. In the present context,
majorization is a necessary but not a sufficient condition, so the
conditions that $\vec{p}$ and $\vec{q}$ must satisfy are \emph{stronger }
than the conditions that the spectra of the entangled states must satisfy.
Only if the doubly stochastic matrix connecting $\vec{q}$ to $\vec{p}$ is a
convex sum of permutations of a particular type, namely permutations that
merely shift each nonzero element of $\vec{q}$ by the same fixed amount,
will the transformation $\left\vert \psi \right\rangle \rightarrow
\left\vert \phi \right\rangle $ be possible.

\subsection{Stochastic single-copy transformations}

\subsubsection{Necessary and sufficient conditions}

\strut We now consider the problem of achieving the transformation $|\psi
\rangle \rightarrow |\phi \rangle$ with some non-zero probability, i.e.
stochastically rather than deterministically, using only $U(1)$-invariant
operations. In this case we are able not only to shift the distribution
over number rigidly, but also to change the relative probabilities assigned
to different number eigenstates. Therefore, the only feature of $\psi$
and $\phi$ that is relevant to the question of whether $|\psi \rangle
\rightarrow |\phi \rangle$ under stochastic $G$-invariant operations is the
set of number eigenvalues to which they assign non-zero probability. If $
\left\vert \psi \right\rangle =\sum_{n}\sqrt{p_{n}}\left\vert n\right\rangle
,$ then this set for $\left\vert \psi \right\rangle $ can be specified as $
\{n|p_{n}\neq 0\}$. The cardinality of this set will be denoted by $
\mathcal{S}(\psi )$. It is also useful to list the elements of the set in
ascending order, and to denote the ordered set and its elements by
\begin{equation*}
\mathrm{Spec}(\psi )\equiv \{n_{1}(\psi ),n_{2}(\psi ),...,n_{\mathcal{S}
(\psi )}(\psi )\}.
\end{equation*}
We refer to this set as the \emph{number spectrum }of $\left\vert \psi
\right\rangle .$ \ As an example, if $\left\vert \psi \right\rangle =\sqrt{
1/2}\left\vert 0\right\rangle +\sqrt{3/10}\left\vert 2\right\rangle +\sqrt{
1/5}\left\vert 6\right\rangle$, then $\mathcal{S}(\psi )=3$ and $\mathrm{
Spec}(\psi )=\{0,2,6\}$.

Clearly, if $\mathrm{Spec}(\phi )$ is a rigid translation of $\mathrm{Spec}
(\psi )$ then the transformation is possible. We write this sufficient
condition as
\begin{equation}
\exists k\in \mathbb{Z}:\mathrm{Spec}(\phi )=\mathrm{Spec}(\psi )+k,
\label{eq:suffcondstochtransfU(1)}
\end{equation}
where $\mathrm{Spec}(\psi )+k\equiv \{n_{0}(\psi )+k,n_{1}(\psi
)+k,...,n_{ \mathcal{S}(\psi )}(\psi )+k\}.$ (One could also write
the condition as $ \exists k\in \mathbb{Z}:\forall n\in
\mathrm{Spec}(\psi )$, $n-k\in \mathrm{ Spec}(\phi ).)$ Note that
$k$ can be negative and consequently $\mathrm{ Spec}(\psi )+k$ may
have negative elements. However, if this occurs then
$\mathrm{Spec}(\psi )+k$ cannot equal $\mathrm{Spec}(\phi )$ since
the latter has only positive elements, and the $k$ value in question
is not one for which the transformation is possible.

Although the condition of Eq.~(\ref{eq:suffcondstochtransfU(1)}) is
sufficient, it is not necessary. Because a stochastic transformation can
send a non-zero probability to zero, $\mathrm{Spec}(\phi )$ need only be a
subset of a rigid translation of $\mathrm{Spec}(\psi )$. Consequently, we
have
\begin{theorem} \label{prop:criterionstochasticU(1)}The transformation
$|\psi \rangle \rightarrow |\phi \rangle $ is possible using stochastic
U(1)-invariant operations if and only if
\begin{equation}
\exists k\in \mathbb{Z}:\mathrm{Spec}(\phi )\subset \mathrm{Spec}(\psi )+k.
\label{condition0}
\end{equation}
\end{theorem}
(One could also write the condition as $\exists k\in \mathbb{Z}:\forall n\in
\mathrm{Spec}(\phi )$, $n-k\in \mathrm{Spec}(\psi ).)$ \ Here, we must
include for consideration those $k$ that yield negative elements for
$\mathrm{Spec}(\psi )+k$ because these elements might be given zero amplitude
by the operation.

\begin{figure}[tp]
\includegraphics[scale=.8]{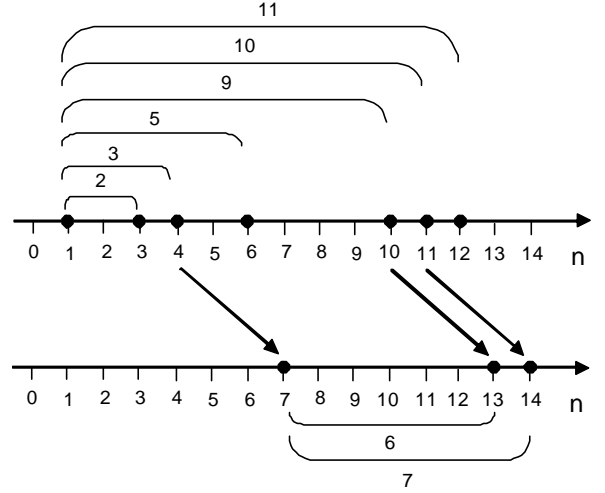}
\caption{An example of a transformation that can be achieved by a stochastic
U(1)-invariant operation.}
\label{fig:stochasticU1}
\end{figure}

An example is illustrated in Fig.~\ref{fig:stochasticU1}. If
$\mathrm{Spec} (\psi )=\{1,3,4,6,10,11,12\}$ and $\mathrm{Spec}(\phi
)=\{7,13,14\},$ the transformation is possible by sending to zero
the weights of the number eigenstates $\left\vert 1\right\rangle
,\left\vert 3\right\rangle ,\left\vert 6\right\rangle $ and
$\left\vert 12\right\rangle ,$ and translating the number upward by
$k=3,$ thereby transforming $\left\vert 4\right\rangle ,\left\vert
10\right\rangle $ and $\left\vert 11\right\rangle $ to $\left\vert
7\right\rangle ,\left\vert 13\right\rangle $ and $\left\vert
14\right\rangle $ respectively, and finally rescaling the weights to
correspond to those of $\left\vert \phi \right\rangle$.

It is not difficult to see that the theorem must be true.
Nevertheless, we provide an explicit proof.

\textbf{Proof. }Let $\left\vert \psi \right\rangle =\sum_{n}\sqrt{
p_{n}}\left\vert n\right\rangle $ and $\left\vert \phi \right\rangle
=\sum_{n}\sqrt{q_{n}}\left\vert n\right\rangle$. Suppose
Eq.~(\ref{condition0}) holds for some $k$, then we can achieve $|\psi \rangle
\rightarrow |\phi \rangle $ using the U(1)-invariant operation defined by
the Kraus operator $K_{k}=S_{k}\tilde{K}_{k}$ where $\tilde{K}
_{k}=\sum_{n}c_{n}^{(k)}\left\vert n\right\rangle \left\langle n\right\vert $
and the $c_{n}^{(k)}$ are defined as follows: If $n+k\in \mathrm{Spec}(\phi
),$ then $c_{n}^{(k)}\propto \sqrt{q_{n+k}}/\sqrt{p_{n}}$ (with norm chosen
such that $c_{n}^{(k)}\leq 1$), otherwise $c_{n}^{(k)}=0$. Note that for
$n+k\in \mathrm{Spec}(\phi )$, we have $q_{n+k}\neq 0$. Also, given
Eq.~(\ref{condition0}), if $n+k\in \mathrm{Spec}(\phi )$ then $n\in \mathrm{Spec}
(\psi )$ and $p_{n}\neq 0$. Thus $c_{n}^{(k)}$ is always well-defined.
It is easily verified that by these definitions, $K_{k}\left\vert \psi
\right\rangle \propto \left\vert \phi \right\rangle$.

Conversely, if $K_{k}\left\vert \psi \right\rangle \propto
\left\vert \phi \right\rangle ,$ then supposing that $\tilde{K}
_{k}=\sum_{n}c_{n}^{(k)}\left\vert n\right\rangle \left\langle n\right\vert ,
$ we require $(c_{n}^{(k)})^{2}p_{n}\propto q_{n+k}$.
Eq.~(\ref{condition0}) follows. QED.

Note that there exist pairs of states, $|\psi \rangle$ and $|\phi
\rangle ,$ for which neither direction of transformation (neither
$|\psi \rangle \rightarrow |\phi \rangle $ nor $|\phi \rangle
\rightarrow |\psi \rangle $) is possible using stochastic
U(1)-invariant operations. A simple example is the pair $\left\vert
0\right\rangle +\left\vert 1\right\rangle $ and $\left\vert
0\right\rangle +\left\vert 2\right\rangle $. \footnote{The
distinction between $\left\vert 0\right\rangle +\left\vert
1\right\rangle $ and $\left\vert 0\right\rangle +\left\vert
2\right\rangle $ as quantum phase references is analogous to the
distinction between $\left( \left\vert 0\right\rangle +\left\vert
1\right\rangle \right) ^{\otimes 2}$ and $\left( \left\vert
0\right\rangle +\left\vert 2\right\rangle \right) ^{\otimes 2}$ as
\emph{shared} quantum phase references which, as van Enk has
noted~\cite{Enk05}, play distinct roles in the theory of static and
dynamic quantum communication resources under a local U(1)-SSR.}

\subsubsection{Maximum probability}

Given two pure states $\left\vert \psi \right\rangle =\sum_{n}\sqrt{p_{n}}
\left\vert n\right\rangle $ and $\left\vert \phi \right\rangle =\sum_{n}
\sqrt{q_{n}}\left\vert n\right\rangle $ such that the transformation $
\left\vert \psi \right\rangle \rightarrow \left\vert \phi \right\rangle $ is
possible by stochastic U(1)-invariant operations, what is the maximum
probability to convert $|\psi \rangle $ into $|\phi \rangle $? We have
only been able to find the complete solution in a special case.

\begin{theorem}\label{thm:maxprobU1} If there is only a single value of $k$
such that the condition $\mathrm{Spec}(\phi )\subset \mathrm{Spec}(\psi )+k$
holds, then the maximum probability of achieving the transformation
$\left\vert \psi \right\rangle \rightarrow \left\vert \phi \right\rangle $
using U(1)-invariant operations is
\begin{equation*}
P(\left\vert \psi \right\rangle \rightarrow \left\vert \phi \right\rangle
)=\min_{n}\left( \frac{p_{n}}{q_{n+k}}\right) .
\end{equation*}
\end{theorem}

\textbf{Proof.} Recall the proof of Thm.~\ref{thm:dettransfU1} where
it was noted that for a U(1)-invariant operation with Kraus
operators $ \{K_{k,\alpha }\}$ to achieve $\left\vert \psi
\right\rangle \rightarrow \left\vert \phi \right\rangle $
deterministically, it must satisfy $K_{k,\alpha }\left\vert \psi
\right\rangle =\sqrt{w_{k,\alpha }}\left\vert \phi \right\rangle $
for all $k$ and $\alpha$. To achieve the transformation
stochastically, this condition need only hold for one or more pairs
of values of $k$ and $\alpha$. We can still deduce
Eq.~(\ref{eq:rrr}) for these pairs of values, which we denote by
$(k,\alpha )\in S$, and it follows that we have $w_{k,\alpha
}=(c_{n}^{(k,\alpha )})^{2}p_{n}/q_{n+k}$ for every $n$. The total
probability of this transformation is therefore
\begin{equation*}
w=\sum_{k,\alpha \in S}\frac{(c_{n}^{(k,\alpha )})^{2}p_{n}}{q_{n+k}}.
\end{equation*}
The task is to maximize this quantity under variations of the
$c_{n}^{(k,\alpha )}$ subject to the constraint that $\sum_{k,\alpha \in
S}(c_{n}^{(k,\alpha )})^{2}\leq 1$ for every $n$.

The assumption that $\left\vert \psi \right\rangle \rightarrow \left\vert
\phi \right\rangle $ can only be achieved for a single value of $k$ implies
that our sum may be restricted to this value,
\begin{equation*}
w=\frac{\left( \sum_{\alpha \in S}(c_{n}^{(k,\alpha )})^{2}\right) p_{n}}{
q_{n+k}},
\end{equation*}
and given that $\sum_{\alpha \in S}(c_{n}^{(k,\alpha )})^{2} \leq 1$
for every $n,$ we infer that $w\leq \frac{p_{n}}{q_{n+k}}$ for every
$n.$ This set of inequalities is captured by the single inequality
$w\leq \min_{n}\left\{ \frac{p_{n}}{q_{n+k}}\right\}$.  By choosing
$\sum_{\alpha \in S}(c_{n}^{(k,\alpha )})^{2}=1$ for the $n$ that
achieves the minimum, we can saturate the inequality. QED.

If $\mathrm{Spec}(\phi )\subset \mathrm{Spec}(\psi )+k$ for several
different values of $k$, then the optimization is much more
difficult. It may be that there is an optimal $k$ to use.
Alternatively, it may be that the probabilities associated with
different $k$ values can be added because one can implement a
measurement upon $\left\vert \psi \right\rangle $ such that more
than one outcome collapses the state to $\left\vert \phi
\right\rangle$. This is what occurs deterministically with
$(\left\vert 0\right\rangle +\left\vert 1\right\rangle +\left\vert
2\right\rangle +\left\vert 3\right\rangle )/2\rightarrow (\left\vert
0\right\rangle +\left\vert 1\right\rangle )/\sqrt{2}$. As another
example, if there are two values, $k_{1}$ and $k_{2},$ that satisfy
$\mathrm{Spec} (\phi )\subset \mathrm{Spec}(\psi )+k$ and for which
$|k_{1}-k_{2}|>n_{ \mathcal{S}(\phi )}(\phi )-n_{1}(\phi )$ (so that
$\mathrm{Spec}(\phi )-k_{1} $ and $\mathrm{Spec}(\phi )-k_{2}$ do
not overlap) then the probability of the transformation is at least
\begin{equation*}
w=\min_{n}\left( \frac{p_{n}}{q_{n+k_{1}}}\right) +\min_{n}\left( \frac{p_{n}
}{q_{n+k_{2}}}\right) .
\end{equation*}
The problem of finding the maximum probability in the general case remains
open, although techniques analogous to those in Ref.~\cite{Vid99} are likely
to yield the solution.

\subsection{\strut Stochastic U(1)-frameness monotones
\label{sec:stochasticU1monotones}}

From Thm.~(\ref{prop:criterionstochasticU(1)}), we see that the cardinality
of the number spectrum is non-increasing under stochastic $U(1)$-invariant
operations, that is, $\mathcal{S}(\phi )\leq \mathcal{S}(\psi )$. This
cardinality therefore satisfies the definition of a stochastic frameness
monotone. (Note that the amplitudes $\sqrt{p_{n}}\equiv \left\langle n|\psi
\right\rangle $ play an analogous role here to that of the Schmidt
coefficients in entanglement theory, and the number spectrum cardinality
$\mathcal{S}(\psi )$ is analogous to the Schmidt number.)

In fact, it is straightforward to see that there are other features of the
number spectrum that define stochastic frameness monotones. Perhaps the most
obvious such feature is the difference between the largest and the smallest
element of the spectrum. But the nonincreasing property also holds true
for the difference between the second-largest and the smallest element of
the spectrum, the third-largest and the smallest, and so forth.
Fig.~\ref{fig:stochasticU1} makes this feature evident.

We can thereby define stochastic frameness monotones in terms of these
difference as
\begin{eqnarray*}
\mathcal{F}_{1}(\psi ) &\equiv &n_{\mathcal{S}(\psi )}(\psi )-n_{1}(\psi ) \\
\mathcal{F}_{2}(\psi ) &\equiv &n_{\mathcal{S}(\psi )-1}(\psi )-n_{1}(\psi )
\\
&&\dots \\
\mathcal{F}_{\mathcal{S}(\psi )-1}(\psi ) &\equiv &n_{2}(\psi )-n_{1}(\psi )
\end{eqnarray*}
and
\begin{equation*}
\mathcal{F}_{j}(\psi )\equiv 0\text{ for }j\geq \mathcal{S}(\psi ).
\end{equation*}
We have presented these in decreasing order, $\mathcal{F}_{j+1}(\psi )<
\mathcal{F}_{j}(\psi )$ for $j<\mathcal{S}(\psi )$.

\strut That all stochastic frameness monotones should be nonincreasing is
clearly a necessary condition for the possibility of a particular
transformation. \ For instance, $\left\vert \psi \right\rangle =\left(
|0\rangle +|1\rangle \right) /\sqrt{2}$ can not be converted to $\left\vert
\phi \right\rangle =(|0\rangle +|2\rangle )/\sqrt{2}$ even by stochastic $
U(1)$-invariant operations because $\mathcal{F}_{1}(\psi )<\mathcal{F}
_{1}(\phi )$~\footnote{However, as noted by van Enk~\cite{Enk05}, two copies of
$\left\vert0\right\rangle +\left\vert 1\right\rangle$ can be converted with some
probability to a single copy of $\left\vert 0\right\rangle +\left\vert
2\right\rangle$.}.

In order to characterize the necessary and sufficient conditions for
stochastic interconversion in terms of these monotones, it is useful to
define the set of nonzero monotones,
\begin{equation*}
\mathrm{Mons}(\psi )\equiv \{\mathcal{F}_{1}(\psi
),\mathcal{F}_{2}(\psi ),\dots , \mathcal{F}_{\mathcal{S}(\psi
)-1}(\psi )\}.
\end{equation*}
We can easily infer from Eq. (\ref{condition0}) the following alternative
form of theorem~\ref{prop:criterionstochasticU(1)}.

\begin{proposition} The transformation $|\psi \rangle \rightarrow |\phi
\rangle$ is possible using stochastic U(1)-invariant operations if and only
if
\begin{equation}
\exists l\in \mathbb{N}:\text{\textrm{Mons}}(\phi )\subset \mathrm{Mons}
(\psi )-l  \label{condition}
\end{equation}
where $\mathrm{Mons}(\psi )-l\equiv \{\mathcal{F}_{1}(\psi )-l,\mathcal{F}
_{2}(\psi )-l,\dots ,\mathcal{F}_{\mathcal{S}(\psi )-1}(\psi )-l\}.$
\end{proposition}
(One can also write this as $\exists l\in \mathbb{N}:\forall k\in $\textrm{
Mons}$(\phi )$, $k+l\in \mathrm{Mons}(\psi )$).

Returning to our previous example of $\mathrm{Spec}(\psi
)=\{1,3,4,6,10,11,12\}$ and $\mathrm{Spec}(\phi )=\{7,13,14\},$ we
have $ \mathrm{Mons}(\psi )=$\strut $\{11,10,9,5,3,2\}$ and
$\mathrm{Mons}(\phi )=\{7,6\}.$ \ Clearly, $\mathrm{Mons}(\psi
)-3=\{8,7,6,2,0,-1\},$ which includes $\{7,6\},$ so the condition is
satisfied for $l=3$. Again, the figure makes this clear.

\subsection{Asymptotic transformations}

\label{sec:asymptransfU1}

In this section, we demonstrate the existence of reversible asymptotic
transformations -- and therefore the existence of a unique measure of
U(1)-frameness -- for pure states that have a \emph{gapless} number
spectrum. \ A gap occurs when there are values of $n$ receiving zero
probability between a pair of values of $n$ receiving nonzero probability.
For instance, the spectra $\{2,3,4,5\}$ and $\{0,1\}$ are gapless, while $
\{1,2,4,5\}$ and $\{1,7,9\}$ have gaps. \ (A gapless number spectrum can
also be characterized as one that is uniform over its support, that is, for
which $n_{i+1}(\psi )=n_{i}(\psi )+1$ for $i=1,...,\mathcal{S}(\psi )-1.$)

The unique measure is the scaled number variance
\begin{equation}
V(\left\vert \psi \right\rangle )\equiv 4\left[ \langle \psi |\hat{N}
^{2}|\psi \rangle -\langle \psi |\hat{N}|\psi \rangle ^{2}\right] ,
\label{eq:variance}
\end{equation}
where the normalization is chosen in such a way that the state $(\left\vert
0\right\rangle +\left\vert 1\right\rangle )/\sqrt{2}$ has unit variance.

\begin{theorem} \label{thm:asymptoticmeasureU1} The unique
asymptotic measure of U(1)-frameness for pure states $|\psi \rangle $ that
have gapless number spectra is the variance,
\begin{equation*}
F^{\infty }(|\psi \rangle )=V(\left\vert \psi \right\rangle ).
\end{equation*}
\end{theorem}

Given the choice of normalization, it follows that $V(\left\vert
\psi \right\rangle )$ quantifies the rate at which one can distill
copies of $ (\left\vert 0\right\rangle +\left\vert 1\right\rangle
)/\sqrt{2},$ which may be considered to be one \textquotedblleft
bit\textquotedblright\ of phase reference. van Enk~\cite{Enk05} has
introduced the term \emph{refbit }for the bipartite state
$(\left\vert 0\right\rangle \left\vert 1\right\rangle +\left\vert
1\right\rangle \left\vert 0\right\rangle )/\sqrt{2},$ which can be
considered to be one \textquotedblleft bit\textquotedblright\ of a
\emph{shared} phase reference. We suggest that it may be judicious
to call the latter a \emph{shared refbit}, while the state
$(\left\vert 0\right\rangle +\left\vert 1\right\rangle )/\sqrt{2}$
is called a \emph{local refbit}.

This theorem is the adaptation to the unipartite context of the main
result from Ref.~\cite{SVC04a} (where the measure was called the
\emph{superselection-induced variance}). Although the proof can be
easily inferred from its bipartite counterpart in
Ref.~\cite{SVC04a}, for the sake of completeness and pedagogy, at
the end of this section we provide a proof that is native to the
unipartite context.

Note that the variance is not only weakly additive (as it must be given the
discussion in Sec.~\ref{sec:asympframemanip}), but strongly additive as
well; that is, given two finite dimensional pure states $|\psi \rangle $ and
$|\phi \rangle $ we have
\begin{equation*}
V(|\psi \rangle \otimes |\phi \rangle )=V(|\psi \rangle )+V(|\phi \rangle ).
\end{equation*}

Finally, we note that not only is the variance a deterministic monotone (as
it must be given the discussion in Sec.~\ref{sec:asympframemanip}), it is an
ensemble monotone as well.

\begin{lemma}\label{lemma:varianceismonotone} $V(|\psi \rangle )$ is an
ensemble frameness monotone.
\end{lemma}

\textbf{Proof.} Under the U(1)-SSR, a transition from a state to an
ensemble is induced by a U(1)-invariant measurement, that is, a
measurement for which each outcome is associated with a
U(1)-invariant operation. Suppose the outcome $\mu $ occurs with
probability $w_{\mu }$ and is associated with a U(1)-invariant
operation with Kraus decomposition $\{K_{k,\alpha }^{(\mu
)}|k,\alpha \}$ of the form specified in
Lemma~\ref{lemma:U1invariantop}.

Given that each outcome leaves the system in a fixed state $\left\vert \phi
_{\mu }\right\rangle$, we have that
\begin{equation*}
K_{k,\alpha }^{(\mu )}\left\vert \psi \right\rangle =\sqrt{w_{\mu ,k,\alpha }
}\left\vert \phi _{\mu }\right\rangle
\end{equation*}
for all $k$ and $\alpha ,$ where
\begin{equation}
w_{\mu }=\sum_{k,\alpha }w_{\mu ,k,\alpha }.
\label{eq:weights}
\end{equation}
The average value of $V$ in the final ensemble is $\sum_{\mu }w_{\mu
}V\left( |\phi _{\mu }\rangle \right)$.

Now note that there is a fine-graining of this measurement where
each outcome is associated with the U(1)-invariant operation that
has the single Kraus operator $K_{k,\alpha }^{(\mu )},$ so that the
outcomes are labeled not only by $\mu ,$ but by $k$ and $\alpha $ as
well and each has probability $w_{\mu ,k,\alpha }$ of occurring. The
average value of $V$ for the ensemble generated by this measurement
is $\sum_{\mu ,k,\alpha }w_{\mu ,k,\alpha }V\left( |\phi _{\mu
}\rangle \right) ,$ but because $|\phi _{\mu }\rangle $ does not
depend on $k$ and $\alpha$, Eq.~(\ref{eq:weights}) implies that the
average value of $V$ is the same as for the original measurement. It
suffices therefore to show that $V$ is an ensemble monotone for the
fine-grained measurement.

We redefine $\mu $ to run over the outcomes of this fine-grained
measurement. Each outcome is associated with a Kraus operation $K_{\mu }$
which, by lemma~\ref{lemma:U1invariantop}, has the form
\begin{equation*}
K_{\mu }=\sum_{n}c_{n}^{(\mu )}|n+k_{\mu }\rangle \langle n|\;,
\end{equation*}
where the $c_{n}^{(\mu )}$ are complex coefficients and the $k_{\mu }$ are
integers. We therefore have
\begin{equation*}
\lbrack \hat{N},K_{\mu }]=k_{\mu }K_{\mu
}\;\;\text{and}\;\;[\hat{N}^{2},K_{\mu }]=2k_{\mu }K_{\mu }\hat{N} +
k_{\mu }^{2}K_{\mu }
\end{equation*}
Now, after an outcome $\mu $ has occurred, the state of the system is $|\phi
_{\mu }\rangle =\frac{1}{\sqrt{w_{\mu }}}K_{\mu }|\psi \rangle ,$ where $
w_{\mu }=\langle \psi |K_{\mu }^{\dag }K_{\mu }|\psi \rangle $. Thus, on
average
\begin{align*}
&\sum_{\mu }w_{\mu }V(|\phi _{\mu }\rangle )\\
&=4\sum_{\mu }\left( \langle \psi |K_{\mu }^{\dag }\hat{N}^{2}K_{\mu
}|\psi \rangle -\frac{\langle \psi |K_{\mu }^{\dag }\hat{N} K_{\mu
}|\psi \rangle ^{2}}{w_{\mu }}\right)
\end{align*}
From the commutation relations above we conclude that:
\begin{align*}
\sum_{\mu }\langle \psi |K_{\mu }^{\dag }\hat{N}^{2}K_{\mu }|\psi
\rangle &
=\langle \psi |\hat{N}^{2}|\psi \rangle  \\
& +2\sum_{\mu }k_{\mu }\langle \psi |K_{\mu }^{\dag }K_{\mu
}\hat{N}|\psi \rangle +\sum_{\mu }w_{\mu }k_{\mu }^{2}\;,
\end{align*}
and
\begin{align*}
\sum_{\mu }\frac{\langle \psi |K_{\mu }^{\dag }\hat{N} K_{\mu }|\psi
\rangle ^{2}}{ w_{\mu }}& =\sum_{\mu }\frac{1}{w_{\mu }}\langle \psi
|K_{\mu }^{\dag
}K_{\mu }\hat{N} |\psi \rangle ^{2} \\
& +2\sum_{\mu }k_{\mu }\langle \psi |K_{\mu }^{\dag }K_{\mu
}\hat{N}|\psi \rangle +\sum_{\mu }w_{\mu }k_{\mu }^{2}
\end{align*}
where for the upper equation we have used the fact that $\sum_{\mu }K_{\mu
}^{\dag }K_{\mu }=I.$ We therefore obtain
\begin{equation*}
\sum_{\mu }w_{\mu }V(|\phi _{\mu }\rangle )=4 \left( \langle \psi
|\hat{N}^{2}|\psi \rangle -\sum_{\mu }\frac{\langle \psi |K_{\mu
}^{\dag }K_{\mu }\hat{N}|\psi \rangle ^{2}}{ w_{\mu }}\right)\;.
\end{equation*}
Now, let $x_{\mu }\equiv \langle \psi |K_{\mu }^{\dag }K_{\mu
}\hat{N}|\psi \rangle $. From the Cauchy-Schwarz inequality we
obtain
\begin{align*}
\sum_{\mu }\frac{x_{\mu }^{2}}{w_{\mu }}& = \sum_{\mu }\frac{x_{\mu
}^{2}}{ w_{\mu }}\sum_{\mu ^{\prime }}w_{\mu ^{\prime }}\geq \left(
\sum_{\mu }\frac{
x_{\mu }}{\sqrt{w_{\mu }}}\sqrt{w_{\mu }}\right) ^{2} \\
& =\left( \sum_{\mu }x_{\mu }\right) ^{2}=\langle \psi |\hat{N}|\psi
\rangle ^{2}\;,
\end{align*}
where for the last equality we have again used the fact that $\sum_{\mu
}K_{\mu }^{\dag }K_{\mu }=I$. We therefore have
\begin{equation*}
\sum_{\mu }w_{\mu }V(|\phi _{\mu }\rangle )\leq 4 \left( \langle
\psi |\hat{N}^{2}|\psi \rangle -\langle \psi |\hat{N}|\psi \rangle
^{2}=V(|\psi \rangle ) \right).
\end{equation*}
Thus, the variance of $\hat{N}$ is non-increasing on average under
U(1)-invariant operations.

Note that this proof follows that of Schuch~\emph{et al.}~\cite{SVC04b} for
the ensemble monotonicity of the superselection-induced variance but
generalizes the latter insofar as it incorporates the possibility of shifts
in the value of $n$. QED.

\textbf{Proof of theorem}~\ref{thm:asymptoticmeasureU1}. Suppose the
state is given in the standard form $|\psi \rangle =\sum_{n}\sqrt{p_{n}}
|n\rangle .$The assumption that $\left\vert \psi \right\rangle $ has a
gapless number spectrum implies that $p_{n}\neq 0$ for all $n$ in the range $
n_{1}(\psi )$ to $n_{\mathcal{S}(\psi )}(\psi )$. However, we can shift
the number downward by $n_{1}(\psi )$ using a U(1)-invariant operation to
obtain
\begin{equation*}
\left\vert \psi \right\rangle =\sum_{n=0}^{d}\sqrt{\tilde{p}_{n}}|n\rangle ,
\end{equation*}
$\mathcal{\newline
}$where $d\equiv \mathcal{S}(\psi )-1$ and $\tilde{p}_{n}=p_{n+n_{1}(\psi )}$.
We therefore assume a state of this form in what follows.

We would like to write an expression for $|\psi \rangle ^{\otimes
N}$ in the standard form. Recall that all the terms in the resulting
expression with the same total number eigenvalue can be transformed,
via a U(1)-invariant operation to a single term which we denote by
$\left\vert n\right\rangle $ (for instance, $\left\vert
0\right\rangle \left\vert 2\right\rangle \left\vert 2\right\rangle
$\emph{, }$\left\vert 1\right\rangle \left\vert 1\right\rangle
\left\vert 2\right\rangle $\emph{, }and $\left\vert 1\right\rangle
\left\vert 0\right\rangle \left\vert 3\right\rangle $\ can all be
transformed to $|4\rangle $). Keeping this in mind and using the
multinomial formula, we have
\begin{equation}
|\psi \rangle ^{\otimes N}=\sum_{n=0}^{dN}\sqrt{r_{n}}|n\rangle \;,
\end{equation}
with
\begin{equation*}
r_{n}\equiv \sum \frac{n!}{N_{1}!N_{2}!\cdots N_{d-1}!}\tilde{p}_{0}^{N_{0}}
\tilde{p}_{1}^{N_{1}}\cdots \tilde{p}_{d-1}^{N_{d-1}}\;,
\end{equation*}
where the sum is taken over all nonnegative integers $N_{0},
N_{1},...,N_{d} $ for which $\sum_{n^{\prime }=0}^{d}N_{n^{\prime
}}=N$ and $\sum_{n^{\prime }=0}^{d}n^{\prime }N_{n^{\prime }}=n$. In
the limit $ N\rightarrow \infty $, the distribution $r_{n}$
approaches a Gaussian as long as for all $n\in \{0,...,d\},$ we have
$\tilde{p}_{n}>0$~\cite{SVC04b}. The proof is blocked if
$\tilde{p}_{n}=0$ for some $n$ in this range and it is for this
reason that our theorem is restricted to pure states with gapless
number spectra.

A Gaussian distribution depends only on two parameters: the mean and
the variance. However, in the limit of large $N,$ the mean can be
shifted freely by $U(1)$-invariant operations. This follows from
lemma \ref{lemma:U1invariantop} and the fact that the amplitude of
the Gaussian at $n=0$ is negligible in the limit of large $N$ .
Hence, in the limit $N\rightarrow \infty ,$ there exists an allowed
$\mathcal{E}$ such that $\textrm{Fid}(\mathcal{E}(|\psi \rangle
^{\otimes N}),\;|\phi \rangle ^{\otimes M})\rightarrow 1$ as long as
$|\psi \rangle ^{\otimes N}$ and $|\phi \rangle ^{\otimes M}$ have
the same variance, $ V(|\psi \rangle ^{\otimes N})=V(|\phi \rangle
^{\otimes M})$. Given the additivity of the variance, we conclude
that in the limit $N\rightarrow \infty ,$ $MV(\left\vert \phi
\right\rangle )=NV(\left\vert \psi \right\rangle ),$ which is the
result we sought to demonstrate.

Note that although the only maps that are reversible on \emph{all }
states are unitary maps, in the present context reversibility of the
maps is only required on states of the form $|\psi\rangle^{\otimes
N}$, a constraint that can be met by nonunitary operations such as
those induced by the shift operator $S_{k}$. This fact is critical
in the present context because the mean can only be changed by such
operations. Of course, the fact that the change must be accomplished
only imperfectly is critical here. QED.

Finding the asymptotic rate of interconversion between a pair of
states $\left\vert \psi \right\rangle $ and $\left\vert \phi
\right\rangle ,$ one or both of which have gapped spectra, remains
an open problem. Nonetheless, we present a few observations on the
general problem here.

First, it is clear that the asymptotic rate of interconversion can
be strictly zero. The case $|\phi\rangle \rightarrow |\psi\rangle$
where $\left\vert \psi \right\rangle =(|0\rangle +|1\rangle
)/\sqrt{2}$ (with a gapless spectrum) and $\left\vert \phi
\right\rangle =(|0\rangle +|2\rangle )/\sqrt{2}$ (with a gapped
spectrum) provide the simplest example. The rate is zero because
$\left\vert \phi \right\rangle ^{\otimes M}$ has no weight on odd
numbers for any $M,$ while $\left\vert \psi \right\rangle ^{\otimes
N}$ does for all $N$. (Schuch~\emph{et al.}~\cite{SVC04b} point out
that small amounts of additional resources can catalyze the
asymptotic interconversion, but strictly speaking the rate is zero.)

Second, the rate of interconversion can be zero in one direction but
nonzero in the other.  Indeed, the states $\left\vert \psi
\right\rangle =(\left\vert 0\right\rangle +\left\vert 1\right\rangle
)/\sqrt{ 2}$ and $\left\vert \phi \right\rangle =(\left\vert
0\right\rangle +\left\vert 2\right\rangle )/\sqrt{2}$ are such a
case.  In the limit $N\rightarrow \infty ,$ the weights
$r_{n}=|\left\langle n\right\vert (|\psi \rangle ^{\otimes N})|^{2}$
form a Gaussian with variance $NV(\left\vert \psi \right\rangle ),$
while the weights $s_{n}=|\left\langle n\right\vert (|\phi \rangle
^{\otimes M})|^{2}$ are zero for all odd values of $n$ but lie under
a Gaussian envelope with variance $MV(\left\vert \phi \right\rangle
).$ Recalling Thm.~\ref{thm:dettransfU1}, $|\psi \rangle ^{\otimes
N}$ can be transformed deterministically to $|\phi \rangle ^{\otimes
M}$ if and only if $\vec{r}=\sum_{k}w_{k}\Upsilon _{k}\vec{s}$. This
condition is indeed satisfied (at least approximately) when the
variances of $\vec{r}$ and $\vec{ s}$ are equal, i.e. when
$\lim_{N\rightarrow \infty }M/N=V(\left\vert \psi \right\rangle
)/V(\left\vert \phi \right\rangle ),$ because in this case $
\vec{r}\simeq (1/2)\vec{s}$ $+(1/2)\Upsilon _{1}\vec{s}$. The
deterministic transformation is achieved by measuring whether $n$ is
even or odd, and upon finding it odd, shifting its value upward by
$k=1$. More precisely, the U(1)-invariant operation that achieves
$|\psi \rangle ^{\otimes N}\rightarrow |\phi \rangle ^{\otimes M}$
is the one that has Kraus operators $K_{0}=\sum_{n\text{
even}}\left\vert n\right\rangle \left\langle n\right\vert $ and
$K_{1}=S_{1}\sum_{n\text{ odd}}\left\vert n\right\rangle
\left\langle n\right\vert .$ As noted above, the opposite
transformation, $|\phi \rangle ^{\otimes N}\rightarrow |\psi \rangle
^{\otimes M},$ cannot be achieved with any rate. It is useful to
justify this in terms of the condition for deterministic
transformations. The condition requires that $\vec{s}
=\sum_{k}w_{k}\Upsilon _{k}\vec{r}$ but this cannot be satisfied
(even approximately) as no convex combination of shifted versions of
a gapless spectrum can yield a gapped spectrum. In brief, under a
deterministic U(1)-invariant operation, gaps can be created but they
cannot be filled. We see once again that there are distinct
inequivalent sorts of resources under the U(1)-SSR.
%, such as the states $(|0\rangle
%+|1\rangle )/\sqrt{2}$ and $(|0\rangle +|2\rangle )/\sqrt{2}$.
%(Although a classification of all the different sorts of resources
%awaits a general solution of the asymptotic interconvertability
%problem.)

Third, the rate of interconversion is not a continuous function.
Consider the unnormalized and gapless states
$|0\rangle+\epsilon|1\rangle+|2\rangle$ and
$|0\rangle+|1\rangle+\epsilon|2\rangle$, where $\epsilon$ is a
positive real small number. Because the variance of the first state
is greater than that of the second, the rate of converting the first
to the second is greater than 1 for sufficiently small epsilon.
However, the rate must jump discontinuously to zero if we take
$\epsilon$ to zero.

Finally, note that the class of states with gapless number spectra
is not the only class for which reversible asymptotic
transformations exist. Many other examples can be found, such as the
class of states with gaps of width $x$ for some fixed $x>0.$

Clearly, there remains much work to be done to completely characterize
asymptotic transformations under a U(1)-SSR for states with arbitrary number
spectra.

\section{Resource theory of the Z$_{2}$-SSR}

\subsection{Chiral frames}

The second example we consider is that of a reference frame for
chirality. Such a frame is the component of a Cartesian frame with respect
to which the handedness of a quantum system is defined. The space
inversion $\vec{x}\;\rightarrow \;-\vec{x}$ is the coordinate transformation
that changes a right-handed system into a left-handed one and
vice-versa. Performing space inversion twice is equivalent to
performing the identity transformation $\vec{x}\;\rightarrow \;\vec{x}$.
These two transformations are a representation of the group $Z_{2}$. We
label the two elements of $Z_{2}$ by $e$ and $f$ (identity and flip).
Their representation on Hilbert space is
\begin{eqnarray*}
T(e) &=&I, \\
T(f) &=&\pi ,
\end{eqnarray*}
where $\pi $ is the parity operator. Because the parity operator is
Hermitian and satisfies $\pi ^{2}=I,$ its eigenvalues are $\pm 1$.

For a single quantum particle, the action of the parity operator is
\begin{equation}
\pi \left\vert l,m\right\rangle =(-1)^{l}\left\vert l,m\right\rangle ,
\label{eq:paritysingleparticle}
\end{equation}
where $l$ and $m$ are the orbital angular momentum quantum numbers
-- $l(l+1)$ is the eigenvalue of $\mathbf{L}^{2}$ and $\hbar m$ is
the eigenvalue of $ \mathbf{L}_{z}$. The parity is even (eigenvalue
+1) for $l$ even and it is odd (eigenvalue -1) for $l$ odd. Equation
(\ref{eq:paritysingleparticle}) is easily verified by noting that
$\left\langle \vec{x}|(\pi |l,m\right\rangle )=\left\langle
-\vec{x}|l,m\right\rangle =Y_{l}^{m}(-\vec{x}
)=(-1)^{l}Y_{l}^{m}(\vec{x})=(-1)^{l}\left\langle
\vec{x}|l,m\right\rangle$, where the $Y_{l}^{m}(\vec{x})$ denote the
spherical harmonics, and where we have made use of the fact that
these are eigenfunctions of the space inversion operation with
eigenvalue $(-1)^{l}$ (see e.g. p.~255 in Sakurai~\cite{Sakurai}).

For $N$ quantum particles, the representation of space inversion is simply
the tensor product representation, $\pi \equiv
T(f)=\bigotimes_{i=1}^{N}T_{i}(f)=\bigotimes_{i=1}^{N}\pi _{i},$ where $\pi
_{i}$ is the parity operator for the $i$th particle. This acts as
\begin{equation*}
\pi \bigotimes_{i}\left\vert l_{i},m_{i}\right\rangle =(-1)^{\sum_{i^{\prime
}}l_{i^{\prime }}}\bigotimes_{i}\left\vert l_{i},m_{i}\right\rangle .
\end{equation*}
Consequently the collective parity is even (odd) if the sum $\sum_{i^{\prime
}}l_{i^{\prime }}$ of the orbital angular momentum quantum numbers of the
components is even (odd).

Note that a spin system is invariant under space inversion. It follows
that no state of a spin system can act as a quantum reference frame for
chirality. Only quantum particles can constitute such a resource
(see e.g. p.~254 in~\cite{Sakurai}).

The states of the quantum particle that are invariant under space inversion
(i.e. the $Z_{2}$-invariant states) are those satisfying
\begin{equation*}
\pi \rho \pi =\rho ,
\end{equation*}
or equivalently,
\begin{equation*}
\lbrack \rho ,\pi ]=0.
\end{equation*}

Again, it is useful to decompose the Hilbert space into the eigenspaces of
even and odd parity (the carrier spaces for the irreducible representations
of $Z_{2}$),
\begin{equation*}
\mathcal{H}=\mathcal{H}_{\text{even}}\oplus \mathcal{H}_{\text{odd}}.
\end{equation*}
Let $\{\left\vert b,\beta \right\rangle \}$ be a basis for $\mathcal{H},$
where $b$ is a bit specifying the parity and $\beta $ is a multiplicity
index,
\begin{equation*}
\pi \left\vert b,\beta \right\rangle =(-1)^{b}\left\vert b,\beta
\right\rangle .
\end{equation*}
Clearly, $\{\left\vert 0,\beta \right\rangle $\} is a basis for $\mathcal{H
}_{\text{even}}$ and $\{\left\vert 1,\beta \right\rangle \}$ is a basis for $
\mathcal{H}_{\text{odd}}$.

\strut For instance, for a single quantum particle, we can take
\begin{equation*}
\left\vert b,\beta \right\rangle =\left\vert l,m\right\rangle ,
\end{equation*}
where
\begin{eqnarray*}
b=l\;{\rm mod}\;2 =\left\{
\begin{array}{c}
0\text{ if }l\text{ is even} \\
1\text{ if }l\text{ is odd}
\end{array}
\right.
\end{eqnarray*}
and $\beta $ is an index that specifies the remaining information in ($l,m),$
specifically,
\begin{equation*}
\beta =(\frac{l-b}{2},m).
\end{equation*}
Similarly, for $N$ quantum particles, we can take
\begin{equation*}
\left\vert b,\beta \right\rangle =\bigotimes_{i=1}^{N}\left\vert
l_{i},m_{i}\right\rangle ,
\end{equation*}
where
\begin{eqnarray*}
b =\left( \sum_{i}l_{i}\right)\; {\rm mod}\; 2
=\left\{
\begin{array}{c}
0\text{ if }\sum_{i}l_{i}\text{ is even} \\
1\text{ if }\sum_{i}l_{i}\text{ is odd}
\end{array}
\right.
\end{eqnarray*}
and $\beta $ is an index that specifies the remaining information in $
l_{1},m_{1},l_{2},m_{2},....$

It is clear that any change to the multiplicity index does not require a
reference frame for chirality. It follows that any operation within $
\mathcal{H}_{\text{even}}$ or within $\mathcal{H}_{\text{odd}}$ is possible
under the Z$_{2}$-SSR and that any pure state $\left\vert \psi \right\rangle
=\sum_{b,\beta }\lambda _{b,\beta }|b,\beta \rangle $ can be taken, by a Z$
_{2}$-invariant unitary operator, to the form
$$
\left\vert \psi \right\rangle =\sum_{b}\lambda _{b}|b\rangle
=\lambda _{0}\left\vert 0\right\rangle +\lambda _{1}\left\vert
1\right\rangle\;,
$$
where $\left\vert 0\right\rangle $ and $\left\vert 1\right\rangle $ are a
pair of standard states of even and odd parity respectively, $\pi \left\vert
b\right\rangle =(-1)^{b}\left\vert b\right\rangle $. In what follows, we
will assume that all pure states have been transformed into this standard
form. We are therefore working in the two-dimensional subspace $\mathcal{H}
^{\prime }=$ $\mathrm{span}(\left\vert 0\right\rangle ,\left\vert
1\right\rangle )$.

Note that only the states $\left\vert 0\right\rangle $ and
$\left\vert 1\right\rangle $ are invariant under space inversion.
Any coherent superposition of these is therefore a resource under
the Z$_{2}$-SSR. Such quantum states, which act as bounded-size
reference frames for chirality, have been dubbed ``quantum gloves''
in recent work~\cite{Gis04,Col05}.

The physical significance of the $\mathrm{Z}_{2}$-SSR is clarified
by considering a scenario wherein two parties, Alice and Charlie,
fail to share a reference frame for chirality. If the state of a
system is $\rho $ relative to Alice's frame, then relative to
Charlie's frame the state is described by the $Z_{2}$-twirling of
$\rho$,
\begin{equation*}
\mathcal{Z}[\rho ]\equiv \frac{1}{2}\rho +\frac{1}{2}\pi \rho \pi .
\end{equation*}
An eigenstate of parity relative to Alice's frame appears as the same state
relative to Charlie's frame. On the other hand, a superposition of such
states is a resource in the sense that it provides for Charlie a token of
Alice's chiral frame, one which Charlie cannot prepare himself.

It should be noted that there are many other restrictions on
operations, besides the lack of a chiral reference frame, that are
described by a Z$_2$-SSR. For instance, having a reference frame for
phase modulo $\pi$ is a restriction relative to possessing a full
phase reference and is described by a Z$_2$-SSR. It may arise, for
instance, if Alice and Bob have local phase references and are
uncertain of whether they are in phase or $\pi$ out of phase.  This
is clearly a milder restriction than knowing nothing about the
relative phase. We highlight this example because it provides a
physical explanation of why the U(1)-SSR is a stronger restriction
than the Z$_2$-SSR. Nonetheless, when we attempt to characterize our
results in physical terms, we shall use the lack of a chiral
reference frame as our example.

\subsection{Z$_{2}$-invariant operations}

We now turn to the $\mathrm{Z}_{2}$-invariant operations.
Lemma~\ref{lemma:Ginvariantoperations} provides a characterization.
First note that $\mathrm{Z}_{2}$ has only two irreducible representations, both
1-dimensional, which we label by $B\in \{0,1\}$ and denote by $u_{B}:\mathrm{
Z}_{2}\rightarrow \mathbb{C}$. Denoting the elements of $Z_{2}$ by $e$ and
$f$, the irreps are
\begin{eqnarray*}
u_{0}(e) &=&1,\text{ }u_{0}(f)=1,\text{ and} \\
u_{1}(e) &=&1,\text{ }u_{1}(f)=-1.
\end{eqnarray*}

It follows from Lemma~\ref{lemma:Ginvariantoperations} that a Z$_{2}$
-invariant operation has Kraus operators $K_{B,\alpha },$ labelled by an
irrep $B$ and a multiplicity index $\alpha $, satisfying
\begin{eqnarray*}
\pi K_{0,\alpha }\pi  &=&K_{0,\alpha }, \\
\pi K_{1,\alpha }\pi  &=&-K_{1,\alpha }.
\end{eqnarray*}
Just as in the U(1) case, the fact that the irreps are 1-dimensional implies
that the action of Z$_{2}$ does not mix these Kraus operators, which
simplifies the analysis.

Confining ourselves to the two-dimensional subspace $\mathcal{H}^{\prime }=$
$\mathrm{span}(\left\vert 0\right\rangle ,\left\vert 1\right\rangle ),$ we
infer that $K_{0,\alpha }$ is a $Z_{2}$-invariant operator of the form
\begin{eqnarray}
K_{0,\alpha } &=&a_{\alpha}\left\vert 0\right\rangle \left\langle
0\right\vert
+b_{\alpha}\left\vert 1\right\rangle \left\langle 1\right\vert ,  \notag \\
&=&\left(
\begin{array}{cc}
a_{\alpha} & 0 \\
0 & b_{\alpha}
\end{array}
\right) ,  \label{eq:ab}
\end{eqnarray}
while $K_{1,\alpha }$ has the form
\begin{eqnarray}
K_{1,\alpha } &=&c_{\alpha}\left\vert 1\right\rangle \left\langle
0\right\vert
+d_{\alpha}\left\vert 0\right\rangle \left\langle 1\right\vert ,  \notag \\
&=&\left(
\begin{array}{cc}
0 & d_{\alpha} \\
c_{\alpha} & 0
\end{array}
\right) .
\end{eqnarray}
%where $|a|,|b|,|c|,|d|$ $\leq 1.$
In order for the operation to be trace-nonincreasing, we require
$\sum_{B,\alpha }K_{B,\alpha }^{\dag }K_{B,\alpha }\leq I,$ which
implies that $\sum_{\alpha }(|a_{\alpha}|^2+|c_{\alpha}|^2) \leq 1$
and $\sum_{\alpha }(|b_{\alpha}|^2+|d_{\alpha}|^2) \leq 1$, where
the inequalities are saturated if the operation is trace-preserving.

We summarize this result by the following lemma:
\begin{lemma}\label{lemma:Z2invariantop} A Z$_{2}$-invariant operation
admits a Kraus decomposition $\{K_{B,\alpha }\},$ where $B\in \{0,1\}$ and $
\alpha $ is an integer, satisfying
\begin{equation}
K_{B,\alpha }=S_{B}\tilde{K}_{B,\alpha }  \label{eq:KrausZ2}
\end{equation}
where
\begin{eqnarray*}
\tilde{K}_{B,\alpha } &\equiv &c_{0}^{(B,\alpha )}\left\vert 0\right\rangle
\left\langle 0\right\vert +c_{1}^{(B,\alpha )}\left\vert 1\right\rangle
\left\langle 1\right\vert  \\
&=&\left(
\begin{array}{cc}
c_{0}^{(B,\alpha )} & 0 \\
0 & c_{1}^{(B,\alpha )}
\end{array}
\right)
\end{eqnarray*}
changes the relative amplitudes of the parity eigenstates, and
\begin{eqnarray*}
S_{0} &\equiv &I=\left(
\begin{array}{cc}
1 & 0 \\
0 & 1
\end{array}
\right)  \\
S_{1} &\equiv &X=\left(
\begin{array}{cc}
0 & 1 \\
1 & 0
\end{array}
\right)
\end{eqnarray*}
are, respectively, the identity operator, which leaves the parity
unchanged, and the Pauli X operator, which flips the parity. The
coefficients satisfy $ \sum_{B,\alpha } |c_{b}^{(B,\alpha )}|^2 \leq
1$ for all $b$, with equality if the operation is trace-preserving.
\end{lemma}
This Kraus decomposition is analogous to the one specified in
Lemma~\ref{lemma:U1invariantop} for U(1)-invariant operations.

\subsubsection{Z$_{2}$-invariant unitaries}

We have already mentioned how all unitaries that act within the
spaces $ \mathcal{H}_{\text{even}}$ and $\mathcal{H}_{\text{odd}}$
are Z$_{2}$ -invariant. \ Indeed, it is because of this fact that
every state can be transformed to one of the form $\left\vert \psi
\right\rangle =\lambda _{0}\left\vert 0\right\rangle +\lambda
_{1}\left\vert 1\right\rangle$. There are, however, additional
Z$_{2}$-invariant unitaries. Consider the two sorts of irreducible
operations described in the previous section. For an operation to be
unitary, the associated Kraus operator must be a unitary operator.

For $K_{0,\alpha }$ to be unitary, we require that
$|a|^{2}=|b|^{2}=1$ in Eq.~(\ref{eq:ab}). Such an operation can
still change the relative phase of $\left\vert 0\right\rangle $ and
$\left\vert 1\right\rangle$. It follows that any state $\left\vert
\psi \right\rangle =\lambda _{0}\left\vert 0\right\rangle +\lambda
_{1}\left\vert 1\right\rangle $ can be transformed by a Z$_{2}$
-invariant unitary into one of the form
\begin{equation*}
\left\vert \psi \right\rangle =\sqrt{p_{0}}\left\vert 0\right\rangle +\sqrt{
p_{1}}\left\vert 1\right\rangle ,
\end{equation*}
where $p_{0}+p_{1}=1,$ which is to say, a form with real-amplitude
coefficients.

In addition, the bit flip operation $X$
is unitary, which implies that any state can be transformed to one of the
form
\begin{equation*}
\left\vert \psi \right\rangle =\sqrt{p_{0}}\left\vert 0\right\rangle +\sqrt{
p_{1}}\left\vert 1\right\rangle ,\text{ where }p_{0}\geq p_{1}.
\end{equation*}
We will use both of these forms in what follows. If the form with ordered
weights is being used, then this assumption will be made explicit.

\subsection{Deterministic single-copy transformations
\label{sec:determtransfsZ2}}

We wish to determine when the transformations $|\psi \rangle
\rightarrow |\phi \rangle $ can be achieved \emph{deterministically}
using only Z$_{2}$-invariant operations.

We begin by defining a measure of Z$_{2}$-frameness that will be
significant in what follows.

\textbf{Definition: }For a state of the form $\left\vert \psi \right\rangle =
\sqrt{p_{0}}\left\vert 0\right\rangle +\sqrt{p_{1}}\left\vert 1\right\rangle
,$ we define the measure $\mathcal{C}$ by
\begin{equation}
\mathcal{C}(|\psi \rangle )\equiv 2\text{min}\{p_{0},\;p_{1}\}
\label{eq:chirality}
\end{equation}
Note that if the
state is written in the standard form where $p_{0}\geq p_{1}\;$then the
measure is simply expressed as
\begin{equation*}
\mathcal{C}(|\psi \rangle )\equiv 2p_{1}.
\end{equation*}
We choose a normalization factor of 2 so that $0\leq \mathcal{C}(|\psi
\rangle )\leq 1$.

As we will see in Sec.~\ref{sec:stochasticZ2}, this measure has an
operational interpretation: $\mathcal{C}(\psi )/\mathcal{C}(\phi )$
determines the maximum probability to convert $\psi $ into $\phi $ using
only Z$_{2}$-invariant operations. Also, in Sec.~\ref{sec:ensembleZ2monotone}
we show that it is an ensemble monotone and
satisfies several interesting properties. This measure also helps us to
express the criterion for deterministic single-copy transformations.

\begin{theorem}\label{prop:conditionsdeterministicZ2}
The necessary and
sufficient condition for the transformation
$\left\vert \psi \right\rangle\rightarrow \left\vert \phi \right\rangle$ to be possible by a
deterministic Z$_{2}$-invariant operation is
\begin{equation}
\mathcal{C}(|\psi \rangle )\geq \mathcal{C}(|\phi \rangle ).
\label{eq:NSforDZ2}
\end{equation}
\end{theorem}

Note that if we take $\left\vert \psi \right\rangle$ and
$\left\vert \phi\right\rangle $ to be in the standard forms,
\begin{eqnarray*}
|\psi \rangle &=&\sqrt{p_{0}}|0\rangle +\sqrt{p_{1}}|1\rangle \;\text{where }
p_{0}\geq p_{1}\;\; \\
|\phi \rangle &=&\sqrt{q_{0}}|0\rangle +\sqrt{q_{1}}|1\rangle \text{ where }
q_{0}\geq q_{1}.
\end{eqnarray*}
then the condition can be expressed as
\begin{equation}
p_{0}\leq q_{0},  \label{eq:NSforDZ2ii}
\end{equation}
which is equivalent to
\begin{equation}
\vec{p}\text{ is majorized by }\vec{q}  \label{eq:NSforDZ2_iii}
\end{equation}
where the notion of majorization is defined in Ref.~\cite{Batia}.

\textbf{Proof.} Recall from lemma \ref{lemma:Z2invariantop} that a general Z$
_{2}$-invariant operation has a Kraus decomposition $\{K_{B,\alpha }\}$ where
\begin{equation*}
K_{B,\alpha }=S_{B}\left(
\begin{array}{cc}
c_{0}^{(B,\alpha )} & 0 \\
0 & c_{1}^{(B,\alpha )}
\end{array}
\right) .
\end{equation*}
For a deterministic transformation $\left\vert \psi \right\rangle
\rightarrow \left\vert \phi \right\rangle ,$ we require
\begin{equation}
K_{B,\alpha }\left\vert \psi \right\rangle =\sqrt{w_{B,\alpha }}\left\vert
\phi \right\rangle ,  \label{eq:zzz}
\end{equation}
for all $B,\alpha $ where $0\leq w_{B,\alpha }\leq 1$ and $
\sum_{B,a}w_{B,\alpha }=1$. \ The case $B=0$ yields
\begin{equation*}
c_{0}^{(B,\alpha )}\sqrt{p_{0}}=\sqrt{w_{B,\alpha }}\sqrt{q_{0}},
\end{equation*}
whereas the case $B=1$ yields
\begin{equation*}
c_{0}^{(B,\alpha )}\sqrt{p_{0}}=\sqrt{w_{B,\alpha }}\sqrt{q_{1}}
\end{equation*}
Therefore,
\begin{equation}
K_{0,\alpha }=\sqrt{w_{0,\alpha }}\left(
\begin{array}{cc}
\sqrt{q_{0}/p_{0}} & 0 \\
0 & \sqrt{q_{1}/p_{1}}
\end{array}
\right)   \label{eq:Z2K1}
\end{equation}
\begin{equation}
K_{1,\alpha }=X\sqrt{w_{1,\alpha }}\left(
\begin{array}{cc}
\sqrt{q_{1}/p_{0}} & 0 \\
0 & \sqrt{q_{0}/p_{1}}
\end{array}
\right) .  \label{eq:Z2K2}
\end{equation}

For the transformation to be deterministic, we require that
$\sum_{B,\alpha }\left\langle \psi \right\vert K_{B,\alpha }^{\dag
}K_{B,\alpha }\left\vert \psi \right\rangle =1,$ which implies that
$\sum_{b} \left[ \sum_{B,\alpha }(c_{b}^{(B,\alpha )})^{2}\right]
p_{b}=1,$ and consequently that $\sum_{B,\alpha }(c_{b}^{(B,\alpha
)})^{2}=1$ for $b=0$ and $b=1.$ It follows that

\begin{eqnarray*}
w_{0}q_{0}+w_{1}q_{1} &=&p_{0}, \\
w_{0}q_{1}+w_{1}q_{0} &=&p_{1}.
\end{eqnarray*}
where
\begin{equation*}
w_{B}\equiv \sum_{\alpha }w_{B,\alpha }\text{,}
\end{equation*}
so that $0\leq w_{B}\leq 1$ and $\sum_{B}w_{B}=1$. Solving these for $w_{0}$
and $w_{1},$ we have
\begin{eqnarray}
w_{0} &=&\frac{p_{0}q_{0}-p_{1}q_{1}}{q_{0}^{2}-q_{1}^{2}}  \label{eq:C} \\
w_{1} &=&\frac{p_{0}q_{1}-p_{1}q_{0}}{q_{1}^{2}-q_{0}^{2}}.  \label{eq:C'}
\end{eqnarray}
Recalling that $q_{0}>q_{1}$ and $p_{0}>p_{1},$ these two conditions imply
that $q_{0}\geq p_{0}.$ \

Conversely, if $q_{0}\geq p_{0},$ then the operation defined by the
pair of Kraus operators of Eqs. (\ref{eq:Z2K1}) and (\ref{eq:Z2K2})
(with no degeneracy index $\alpha$) and with $w_{0}$ and $w_{1}$
given by Eqs. (\ref{eq:C}) and (\ref{eq:C'}) is a Z$_{2}$-invariant
trace-preserving operation that achieves the transformation
$\left\vert \psi \right\rangle \rightarrow \left\vert \phi
\right\rangle .$ QED.

Consider now the state $|+\rangle \equiv (|0\rangle +|1\rangle )/\sqrt{2}$
with degenerate weights, $p_{0}=p_{1}=\frac{1}{2}.$ It is a maximal resource
in the sense that it can be deterministically transformed into any other
state with nondegenerate weights. The reason is that for all states of the
standard form $\left\vert \phi \right\rangle =\sqrt{q_{0}}|0\rangle +\sqrt{
q_{1}}|1\rangle $ where $q_{0}\geq q_{1},$ we have $q_{0}\geq 1/2,$ which
implies that the condition of Eq.~(\ref{eq:NSforDZ2ii}) is satisfied.
Indeed, the Kraus operators of the operation that implements the
transformation $\left\vert +\right\rangle \rightarrow \left\vert \phi
\right\rangle $ are simply $K_{0}=$diag$(\sqrt{2q_{0}},\sqrt{2q_{1}})$ and $
K_{1}=X$diag$(\sqrt{2q_{1}},\sqrt{2q_{0}}).$\ Conversely, any state of the
standard form $\left\vert \psi \right\rangle =\sqrt{p_{0}}|0\rangle +\sqrt{
p_{1}}|1\rangle $ where $p_{0}>p_{1},$ i.e. with nondegenerate weights,
cannot be deterministically transformed into $\left\vert +\right\rangle $
because $p_{0}>1/2$ and Eq.~(\ref{eq:NSforDZ2ii}) fails to be satisfied.

\subsection{Ensemble Z$_{2}$-frameness monotones\label
{sec:ensembleZ2monotone}}

It is shown here that not only is $\mathcal{C}$ an ensemble
Z$_{2}$-frameness monotone, but every such monotone is a
non-decreasing concave function of $\mathcal{C}.$

\begin{lemma} $\mathcal{C}(|\psi \rangle )$ is an ensemble
Z$_{2}$-frameness monotone.
\end{lemma}

\textbf{Proof. }A transition from a state to an ensemble occurs as the
result of a Z$_{2}$-invariant measurement, that is, a measurement for which
each outcome is associated with a Z$_{2}$-invariant operation. For the same
reasons provided in the proof of lemma~\ref{lemma:varianceismonotone}, it
suffices to consider measurements for which each outcome is associated with
an operation with a single Kraus operator (all other measurements can be
obtained by coarse-graining of these and this process does not change the
value of the monotone).

Suppose the outcome $\mu $ occurs with probability $w_{\mu }$ and is
associated with a Kraus operator $K_{\mu }$ which, by
lemma~\ref{lemma:Z2invariantop}, has the form
\begin{equation*}
K_{\mu }=S_{B_{\mu }}\left(
\begin{array}{cc}
c_{0}^{(\mu )} & 0 \\
0 & c_{1}^{(\mu )}
\end{array}
\right) \;,
\end{equation*}
where $\sum_{\mu }|c_{b}^{(\mu )}|^{2}\leq 1$ for $b=0$ and $1$, and where $
B_{\mu }$ is a bit, with $S_{0}=I$ while $S_{1}=X.$ We then have
\begin{eqnarray*}
\sum_{\mu }w_{\mu }\mathcal{C}\left( |\phi _{\mu }\rangle \right)
&=&\sum_{\mu }\mathcal{C}\left( K_{\mu }|\psi \rangle \right)  \\
&=&\sum_{\mu }\min \{|c_{0}^{(\mu )}|^{2}p_{0},\;|c_{1}^{(\mu )}|^{2}p_{1}\}
\\
&\leq &\min \{p_{0},\;p_{1}\}
=\mathcal{C}(|\psi \rangle )\;.
\end{eqnarray*}
QED.

Every non-decreasing concave function of $\mathcal{C}$ is also an ensemble
monotone, as noted in Sec.~\ref{sec:framenessmonotones}. (Recall that $f$ is
concave if $f(wx+(1-w)y)\geq wf(x)+(1-w)f(y)$ for all $w,x,y\in \lbrack
0,1]. $) What is particularly interesting about $\mathcal{C}$ however is
that the opposite implication also holds true.

\begin{theorem}\label{lemma:allZ2monotones}
Every ensemble Z$_{2}$-frameness monotone is a non-decreasing concave function of $\mathcal{C}$.
\end{theorem}

The proof of this theorem makes use of the following theorem concerning the
transformation $\left\vert \psi \right\rangle \rightarrow \{(w_{\mu
},\left\vert \varphi _{\mu }\right\rangle \}$ of a pure state $\left\vert
\psi \right\rangle $ to an ensemble $\{(w_{\mu },\left\vert \varphi _{\mu
}\right\rangle \}.$ \ Such a transformation is achieved if there is a
measurement that collapses $\left\vert \psi \right\rangle $ to $\left\vert
\varphi _{\mu }\right\rangle $ with probability $w_{\mu }.$

\begin{theorem} \label{thm:puretoensembleZ2} Every transformation $\mathcal{T}:\;
\left\vert \psi \right\rangle \rightarrow \{(w_{\mu },\left\vert \varphi
_{\mu }\right\rangle \}$ that does not increase $\mathcal{C}$ on average,
i.e. for which
\begin{equation}
\sum_{\mu }w_{\mu }\mathcal{C}(|\varphi _{\mu }\rangle )\leq \mathcal{C}
(|\psi \rangle ),  \label{1a1}
\end{equation}
can be achieved by some Z$_{2}$-invariant operation.
\end{theorem}
(This theorem has an analog in entanglement theory; see theorem 2 in~\cite{Dan99}.)

\textbf{Proof.} Without loss of generality, we take the states to be in
the standard form
\begin{eqnarray*}
|\psi \rangle &=&\sqrt{p_{0}}|0\rangle +\sqrt{p_{1}}|1\rangle \\
|\varphi _{\mu }\rangle &=&\sqrt{q_{0}^{(\mu )}}|0\rangle +\sqrt{q_{1}^{(\mu
)}}|1\rangle \;,
\end{eqnarray*}
where $p_{0}\geq p_{1}$ and $q_{0}^{(\mu )}\geq q_{1}^{(\mu )}.$

We now define the state $|\bar{\varphi}\rangle $ as
\begin{equation}
|\bar{\varphi}\rangle \equiv \sqrt{t_{0}}|0\rangle +\sqrt{t_{1}}|1\rangle ,
\label{eq:varphibar}
\end{equation}
where
\begin{equation}
t_{1}\equiv \sum_{\mu }w_{\mu }q_{1}^{(\mu )},  \label{eq:t1}
\end{equation}
so that $t_{0}\geq t_{1}.$ Noting that $\mathcal{C}(|\varphi _{\mu
}\rangle )=2q_{1}^{(\mu )}$ and $\mathcal{C}(|\bar{\varphi}\rangle )=2t_{1},$
we infer from Eq.~(\ref{eq:t1}) that
\begin{equation*}
\mathcal{C}(|\bar{\varphi}\rangle )=\sum_{\mu }w_{\mu }\mathcal{C}(|\varphi
_{\mu }\rangle ).
\end{equation*}

It then follows from Eq.~(\ref{1a1}) that
$\mathcal{C}(|\bar{\varphi} \rangle )\leq \mathcal{C}(|\psi \rangle
),$ which implies, by Thm.~\ref{prop:conditionsdeterministicZ2},
that the transformation $|\psi \rangle \rightarrow
|\bar{\varphi}\rangle $ is achievable deterministically by $ Z_{2}
$-invariant operations. Therefore, we need only to show that we can
generate the ensemble $\{(w_{\mu },|\varphi _{\mu }\rangle )\}$
starting from $|\bar{\varphi}\rangle $.

We now define the following set of positive Z$_{2}$-invariant Kraus
operators:
\begin{equation*}
K_{\mu }=\sqrt{\frac{w_{\mu }q_{0}^{(\mu )}}{t_{0}}}|0\rangle \langle 0|+
\sqrt{\frac{w_{\mu }q_{1}^{(\mu )}}{t_{1}}}|1\rangle \langle 1|\;.
\end{equation*}
One can easily see that
\begin{equation*}
\sum_{\mu }K_{\mu }^{\dag }K_{\mu }=I\;,
\end{equation*}
and
\begin{equation*}
K_{\mu }|\bar{\varphi}\rangle =\sqrt{w_{\mu }}|\varphi _{\mu }\rangle \;.
\end{equation*}
Hence, the combination of this measurement with the deterministic
protocol $ |\psi \rangle \rightarrow |\bar{\varphi}\rangle $
realizes the required transformation $\mathcal{T}$. QED.

We are now in a position to prove Thm.~\ref{lemma:allZ2monotones}.

\textbf{Proof of theorem~\ref{lemma:allZ2monotones}.} Let $F$ be an
arbitrary frameness monotone. \ States in the standard form
$\left\vert \psi \right\rangle =\sqrt{p_{0}}\left\vert
0\right\rangle +\sqrt{p_{1}}\left\vert 1\right\rangle $ where
$p_{0}\geq p_{1}$ are completely specified by $p_{1}$, or
equivalently $\mathcal{C}(|\psi \rangle )=2p_{1},$ and so
$F(\left\vert \psi \right\rangle )$ can be written as a function of
$\mathcal{C}(|\psi \rangle ),$ namely, $F(\left\vert \psi
\right\rangle )=$ $f(\mathcal{C} (|\psi \rangle )).$ It remains only
to show that $f$ is a nondecreasing concave function.

If $f$ was a decreasing function, then $F$ would decrease under some
transformation $\left\vert \psi \right\rangle \rightarrow \left\vert
\phi \right\rangle $ for which $\mathcal{C}(|\psi \rangle )\geq
\mathcal{C}(|\phi \rangle ).$ But by
Thm.~\ref{prop:conditionsdeterministicZ2}, every such transformation
can be achieved deterministically and consequently $F$ cannot
decrease under this transformation if it is a monotone.

To prove that $f$ is a concave function, it suffices to show that for all
sets $\{x_{\mu }\}$ such that $x_{\mu }\in \lbrack 0,1]$ and all probability
distributions $w_{\mu },$
\begin{equation}
\sum_{\mu }w_{\mu }f\left( x_{\mu }\right) \leq f\left( \sum_{\mu }w_{\mu
}x_{\mu }\right) .  \label{eq:concavity}
\end{equation}
Define a set of states $\{\left\vert \varphi _{\mu }\right\rangle
\}$ such that $\mathcal{C}(\left\vert \varphi _{\mu }\right\rangle
)=x_{\mu }$, and another state $|\psi\rangle$ such that
$\mathcal{C}(\left\vert \psi\right\rangle )=\sum_{\mu}w_\mu x_{\mu
}$, so that
\begin{equation}
\sum_{\mu }w_{\mu }\mathcal{C}(|\varphi _{\mu }\rangle )=\mathcal{C}(|
\psi\rangle ).  \label{eq:saturateC}
\end{equation}
From Thm.~\ref{thm:puretoensembleZ2} it follows that the
transformation $\left\vert \psi\right\rangle \rightarrow \{(w_{\mu
},\left\vert \varphi _{\mu }\right\rangle \}$ can be achieved by
Z$_{2}$-invariant operations. Given the presumed monotonicity of $F$
under this transformation, we have
\begin{equation}
\sum_{\mu }w_{\mu }f\left( \mathcal{C}(|\varphi _{\mu }\rangle )\right) \leq
f\left( \mathcal{C}(|\psi\rangle )\right) \;.  \label{eq:fC}
\end{equation}
Substituting Eq.~(\ref{eq:saturateC}) into Eq.~(\ref{eq:fC}), we
obtain Eq.~(\ref{eq:concavity}). QED.

\subsection{Stochastic single-copy transformations\label{sec:stochasticZ2}}

We would like now to find the maximum possible probability to inter-convert
one resource into another using only Z$_{2}$-invariant operations.

\begin{theorem}\label{prop:maxprobZ2} If the condition of
Thm.~\ref{prop:conditionsdeterministicZ2} fails to be satisfied, so
that $\mathcal{C}(|\phi \rangle )>\mathcal{C}(|\psi \rangle )$, then
the maximum probability of transforming $|\psi \rangle $ into $|\phi
\rangle $ using Z$_{2}$-invariant operations is
\begin{equation*}
P\left( |\psi \rangle \;\rightarrow \;|\phi \rangle \right)
=\frac{\mathcal{C}(|\psi \rangle )}{\mathcal{C}(|\phi \rangle )}.
\end{equation*}
\end{theorem}

If we express the states in the form where $p_{0}\geq p_{1}$ and $q_{0}\geq
q_{1},$ then the result may be expressed simply as $P\left( |\psi \rangle
\;\rightarrow \;|\phi \rangle \right) =p_{1}/q_{1}$.

\textbf{Proof.} We begin by showing that $P\left( |\psi \rangle
\;\rightarrow \;|\phi \rangle \right) \leq \mathcal{C}(|\psi \rangle
)/\mathcal{C}(|\phi \rangle ).$ The proof is by contradiction. \ If
the transformation could be achieved with a probability $P^{\prime }
> \mathcal{C}(|\psi \rangle )/ \mathcal{C}(|\phi \rangle ),$ then
the average value of $\mathcal{C}$ after the measurement would be at
least $P^{\prime }\mathcal{C}(|\phi \rangle ) > \mathcal{C}(|\psi
\rangle ),$ contradicting the fact that $\mathcal{C}$ is an ensemble
Z$_{2}$-frameness monotone and therefore nonincreasing under
deterministic Z$_{2}$-invariant operations.

All that remains is to show that there is a protocol that saturates the
inequality. The protocol is defined by a measurement with two possible
outcomes corresponding to Kraus operators
\begin{eqnarray*}
K_{0} &=&\left(
\begin{array}{cc}
x & 0 \\
0 & 1
\end{array}
\right) \;\;\text{and}\;\;K_{0}^{\perp }=\left(
\begin{array}{cc}
\sqrt{1-x^{2}} & 0 \\
0 & 0
\end{array}
\right) ,\;\; \\
\text{where}\;\;x &=&\sqrt{\frac{q_{0}p_{1}}{q_{1}p_{0}}}\;.
\end{eqnarray*}
One verifies that this is a possible measurement operation by noting that $
K_{0}^{\dag }K_{0},$ $K_{0}^{\perp \dag }K_{0}^{\perp }\geq 0$ and $
K_{0}^{\dag }K_{0}+K_{0}^{\perp \dag }K_{0}^{\perp }=I.$ The operation is Z$
_{2}$-invariant, because both $K_{0}$ and $K_{0}^{\perp }$ are of the form
specified in lemma \ref{lemma:Z2invariantop}. Finally, by noting that
\begin{equation*}
K_{0}=\sqrt{\frac{p_{1}}{q_{1}}}\left(
\begin{array}{cc}
\sqrt{q_{0}/p_{0}} & 0 \\
0 & \sqrt{q_{1}/p_{1}}
\end{array}
\right)
\end{equation*}
it is straightforward to see that if the $K_{0}$ outcome occurs, the state
collapses to $|\phi \rangle $ with probability $w_{0}=|K_{0}|\psi \rangle
|=p_{1}/q_{1}=\mathcal{C}(|\psi \rangle )/\mathcal{C}(|\phi \rangle )$.
QED.

\subsection{Stochastic Z$_{2}$-frameness monotones}

By Thm.~\ref{prop:maxprobZ2}, the only instance of the probability
of a transformation $\left\vert \psi \right\rangle \rightarrow
\left\vert \varphi \right\rangle $ being zero is if
$\mathcal{C}(|\psi \rangle )=0$ while $\mathcal{C}(|\varphi \rangle
)\neq 0,$ but this is just the obviously impossible case of a
transformation from a state $\left\vert \psi \right\rangle
=\left\vert 0\right\rangle $ or $\left\vert 1\right\rangle $ that is
Z$_{2}$-invariant to a state $\left\vert \varphi \right\rangle $
that is Z$_{2}$-noninvariant. It follows that the only
\emph{stochastic} Z$_{2}$-frameness monotone is the trivial one --
the number of parity eigenstates receiving nonzero weight. We call
this quantity the~\emph{chiral spectrum cardinality.} It is clearly
analogous to the Schmidt number, which is a stochastic entanglement
monotone.

Note, however, that whereas two entangled states may differ in Schmidt
number, so that the conversion of one to the other can only be achieved in
one direction, all pairs of states with nonzero Z$_{2}$-frameness have
chiral spectrum cardinality of two, and therefore can be stochastically
converted one to the other in either direction. In this sense, the
restriction of the Z$_{2}$-SSR allows more possibilities for resource
interconversion than the restriction of LOCC for pure bipartite states.

\subsection{Asymptotic transformations}

\label{sec:asymptransfZ2}

Consider a state that has a decomposition into even and odd parity states of
the form $|\psi \rangle =\sqrt{p_{0}}|0\rangle +\sqrt{p_{1}}|1\rangle .$

\begin{theorem} \label{thm:asymptoticz2}
The unique (modulo normalization) measure of Z$_{2}$
-frameness for pure states is
\begin{equation}
F^{\infty }(|\psi \rangle )=-\log \left\vert p_{0}-p_{1}\right\vert .
\label{eq:uniqueframenessmonotoneZ2}
\end{equation}
\end{theorem}

\textbf{Proof.} We assume $|\psi \rangle =\sqrt{p_{0}}|0\rangle +\sqrt{p_{1}}
|1\rangle $ with $p_{0}\geq p_{1}$. Note first that
\begin{equation*}
|\psi \rangle ^{\otimes N}=\sum_{m=0}^{N}\sqrt{p_{0}}^{m}\sqrt{p_{1}}
^{N-m}\sum \left\vert 0\right\rangle ^{\otimes m}\left\vert 1\right\rangle
^{\otimes N-m}
\end{equation*}
where the final sum is over all the ways of having $m$ systems in
state $ \left\vert 0\right\rangle $ and $N-m$ in state $\left\vert
1\right\rangle .$ \emph{\ }\ It is useful to decompose this into
unnormalized states of even and odd parity,
\begin{equation*}
|\psi \rangle ^{\otimes N}=\left\vert \tilde{\chi}_{\text{even}
}\right\rangle +\left\vert \tilde{\chi}_{\text{odd}}\right\rangle ,
\end{equation*}
where $\left\vert \tilde{\chi}_{\text{even}}\right\rangle $ contains the
terms where $N-m$ is even, and $\left\vert \tilde{\chi}_{\text{odd}
}\right\rangle $ contains the terms where $N-m$ is odd. Noting that
\begin{eqnarray*}
\left\vert \left\vert \tilde{\chi}_{\text{even}}\right\rangle \right\vert
&=&\sum_{m|\text{ }N-m\text{ even}}\binom{N}{m}p_{0}^{m}p_{1}^{N-m} \\
&=&\frac{1}{2}\left( (p_{0}+p_{1})^{N}+(p_{0}-p_{1})^{N}\right) \\
&=&\frac{1}{2}\left( 1+(p_{0}-p_{1})^{N}\right) ,
\end{eqnarray*}
and bearing in mind that the normalized states $\left\vert \tilde{\chi}_{
\text{even}}\right\rangle /\left\vert \left\vert \tilde{\chi}_{\text{even}
}\right\rangle \right\vert $ and $\left\vert \tilde{\chi}_{\text{odd}
}\right\rangle /\left\vert \left\vert \tilde{\chi}_{\text{odd}}\right\rangle
\right\vert $ can be transformed by a Z$_{2}$-invariant unitary into $
\left\vert 0\right\rangle $ and $\left\vert 1\right\rangle ,$ we infer that
the standard form of $|\psi \rangle ^{\otimes N}$ is
\begin{equation}
|\psi \rangle ^{\otimes N}=\sqrt{r_{0}}\left\vert 0\right\rangle +\sqrt{r_{1}
}\left\vert 1\right\rangle .  \label{eq:psiN}
\end{equation}
where
\begin{eqnarray}
r_{0} &\equiv &\frac{1}{2}+\frac{1}{2}\left( p_{0}{}-p_{1}{}\right) ^{N},
\label{eq:r0} \\
r_{1} &\equiv &\frac{1}{2}-\frac{1}{2}\left( p_{0}{}-p_{1}{}\right) ^{N}.
\label{eq:r1}
\end{eqnarray}
Note that because we have assumed $p_{0}\geq p_{1},$ it follows that $
r_{0}\geq r_{1}.$

Suppose that the target state has the standard form $|\phi \rangle =\sqrt{
q_{0}}|0\rangle +\sqrt{q_{1}}|1\rangle $ where $q_{0}\geq q_{1}$ We can find
an analogous expression to Eq. (\ref{eq:psiN}) for $\left\vert \phi
\right\rangle ^{\otimes M}.$ \ The condition for the existence of a
reversible transformation between $\left\vert \psi \right\rangle ^{\otimes
N} $ and $\left\vert \phi \right\rangle ^{\otimes M}$ is simply the
condition that their standard forms be equal, namely,
\begin{equation} \label{eq:condition}
\left( p_{0}{}-p_{1}{}\right) ^{N}=\left( q_{0}{}-q_{1}{}\right) ^{M}.
\end{equation}

Suppose that neither $\left\vert \psi \right\rangle $ nor $\left\vert \phi
\right\rangle $ is the degenerate state $\left\vert +\right\rangle $, that
is, assume that $p_{0}\neq p_{1}$ and $q_{0}\neq q_{1}.$ Taking the absolute
value and logarithm on both sides of the condition, we obtain
\begin{equation*}
\lim_{N\rightarrow \infty }\frac{M}{N}=\frac{\log \left(
p_{0}{}-p_{1}{}\right) }{\log \left( q_{0}{}-q_{1}{}\right) }.
\end{equation*}
It follows from Eq. (\ref{eq:uniquemeasure}) that the measure that
determines the rate of asymptotic reversible interconversion is
\begin{equation}
F^{\infty }(|\psi \rangle )=\mathcal{N}\log (p_{0}-p_{1})\;
\end{equation}
for some normalization factor $\mathcal{N}$.

For a pure state $\left\vert \psi \right\rangle $ that is not in the
standard form, so that $p_{0}<p_{1},$ the same reasoning implies that $
F^{\infty }(|\psi \rangle )=\mathcal{N}\log (p_{1}-p_{0}).$ \ It follows
that $F^{\infty }(|\psi \rangle )=\mathcal{N}\log |p_{0}-p_{1}|$ provides a
measure for an arbitrary pure state. \ The normalization $\mathcal{N}$ is a
conventional choice which we take to be $-1.$ \ The base of the logarithm is
also a conventional choice which we take to be $2.$ \ These choices are
discussed below.\strut

Finally, we need to consider the degenerate cases. If $q_{0}\neq
q_{1}$ but $ p_{0}=p_{1},$ then the condition for $\left\vert \psi
\right\rangle ^{\otimes N}\leftrightarrow \left\vert \phi
\right\rangle ^{\otimes M}$, Eq.~\ref{eq:condition}, becomes
\begin{equation*}
\left( q_{0}{}-q_{1}{}\right) ^{M}=0.
\end{equation*}
Consequently, for any finite value of $N,$ the value of $M$ must become
arbitrarily large to satisfy the condition, in other words, the rate becomes
infinite, $M/N\rightarrow \infty$. This means that the degenerate state $
\left\vert +\right\rangle $ can be transformed reversibly into an
arbitrarily large number of copies of any nondegenerate state.

If $p_{0}\neq p_{1}$ but $q_{0}=q_{1},$ then Eq.~\ref{eq:condition}
becomes
\begin{equation*}
\left( p_{0}{}-p_{1}{}\right) ^{N}=0.
\end{equation*}
Consequently, for any value of $M,$ we have $M/N=0$. This means that the
degenerate state $\left\vert +\right\rangle $ cannot be obtained reversibly
from any number of copies of a nondegenerate state.

Both of these facts are captured by defining $F^{\infty }(\left\vert
+\right\rangle )=\infty $ for the degenerate state $\left\vert
+\right\rangle \equiv 2^{-1/2}(|0\rangle +\left\vert 1\right\rangle ).$
The expression for $F^{\infty }$ in Eq.~(\ref{eq:uniqueframenessmonotoneZ2})
can therefore be taken to apply even to the degenerate state. QED.

It is worth noting that $F^{\infty }(|+\rangle )=\infty $ implies that $N$
copies of $|+\rangle $ can be transformed to \emph{one} copy of $|+\rangle $
and vice-versa. Therefore, the state $|+\rangle $ alleviates completely the Z
$_{2}$-SSR. Furthermore, as described in the proof, the asymptotic rate with
which one can produce $\left\vert +\right\rangle $ given any state that is
not $\left\vert +\right\rangle $ is zero. The state $\left\vert
+\right\rangle $ is therefore a very special sort of resource -- it is
sufficient to completely lift the SSR and no amount of any lesser resource
can substitute for it. Although it is the only distinguished state in the
Z$_{2}$ resource theory that has nonzero frameness, the $\left\vert
+\right\rangle $ state cannot play a role analogous to the one played by the
singlet state in entanglement distillation, because we cannot distill any
amount of $\left\vert +\right\rangle $ from any other state.

If not $\left\vert +\right\rangle ,$ then what \emph{does} make a
good choice of standard resource against which to judge the strength
of any given state? Any nondegenerate state with nonzero frameness
will do. We adopt a convention that makes the expression for
$F^{\infty }$ particularly simple. First, we assume the logarithm to
be base $2.$ Second, we take the normalization factor $ \mathcal{N}$
introduced in the proof to be $\mathcal{N}=-1,$ so that $ F^{\infty
}$ has the form presented in
Eq.~(\ref{eq:uniqueframenessmonotoneZ2}). The negative sign is
chosen to ensure that the measure is positive. The unit magnitude of
the normalization implies that the state having unit frameness,
$F^{\infty }=1,$ is the one for which $ p_{0}=3/4,$ that is,
$\sqrt{3}/2\left\vert 0\right\rangle +1/\sqrt{2} \left\vert
1\right\rangle $. The measure $F^{\infty }(\left\vert \psi
\right\rangle )$ then quantifies the number of states of this form
that one can distill from $\left\vert \psi \right\rangle $
asymptotically.

\begin{figure}[tbp]
\includegraphics[scale=.8]{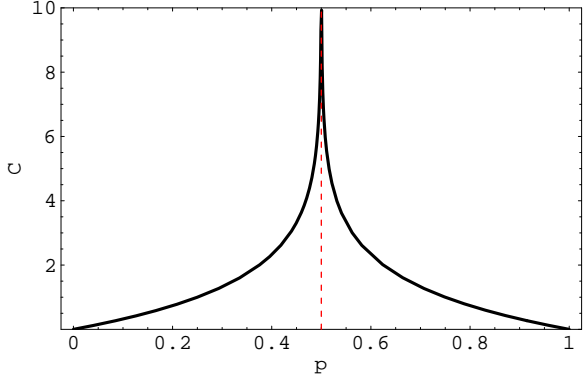}
\caption{A plot of the asymptotic measure $F^{\infty }$ of Z$_2$-frameness
as a function of the probability $p_0$ of the state having even parity.}
\label{fig:f}
\end{figure}

We may, of course, express the measure of Z$_{2}$-frameness entirely in
terms of $p_{0}$, $F^{\infty }(|\psi \rangle )\equiv -\log \left\vert
2p_{0}-1\right\vert .$  A plot of $F^{\infty}$ as a function of $p_{0}$
is provided in Fig.~\ref{fig:f}.

Finally, we note that although $F^{\infty}$ is a deterministic
monotone (by virtue of quantifying the asymptotic rate of conversion
and the fact that any such measure is a deterministic monotone, as
shown in Sec.~\ref {sec:asympframemanip}), it is \emph{not }an
\emph{ensemble} monotone.  $F^{\infty }$ is related to $\mathcal{C}$
by
\begin{equation*}
F^{\infty }(|\psi \rangle )=-\log (1-\mathcal{C}(\left\vert \psi
\right\rangle )),
\end{equation*}
where we drop the absolute value because $\mathcal{C}(\left\vert \psi
\right\rangle )\leq 1.$ Given that the second derivative with respect to $x$
of $-\log (1-x)$ is always positive, $F^{\infty }$ is a convex function of $
\mathcal{C}$ and consequently is not an ensemble monotone.

There is another, more intuitive, way to see
that $F^{\infty }$ cannot be an ensemble monotone.
Theorem~\ref{prop:maxprobZ2} implies that a state $\left\vert \psi \right\rangle
=\sqrt{p_{0}}\left\vert 0\right\rangle +\sqrt{p_{1}}\left\vert 1\right\rangle $
with nonzero frameness can always be converted to the state $\left\vert
+\right\rangle $ with nonzero probability (indeed, the probability is simply
$\mathcal{C}(\left\vert \psi \right\rangle )$). It follows that $
\left\vert \psi \right\rangle $ can be converted to an ensemble of states
that assign nonzero weight to $\left\vert +\right\rangle $. However, given
that $F^{\infty }(\left\vert +\right\rangle )$ is unbounded$,$ the average
fidelity for this ensemble will also be unbounded. It follows that the
average value of $F^{\infty }$ can be increased using Z$_{2}$-invariant
operations and so it fails to be an ensemble monotone.

We see that the property of ensemble monotonicity need not hold for
the unique asymptotic measure of a resource. This result calls into
question the widespread tendency to require any measure of a
resource (such as entanglement or frameness) to be an ensemble
monotone. As discussed in Sec. \ref{sec:framenessmonotones}, to be
operationally well-motivated, it may suffice for a measure of
frameness to be a \emph{deterministic} monotone rather than an
\emph{ensemble} monotone, which is precisely what occurs in the case
of $F^{\infty}$.

We end this section by showing that the measure $F^\infty(\psi )$ has a nice additivity
property.
\begin{proposition} $F^\infty(\psi )$ is strongly additive.
\end{proposition}
\textbf{Proof:} \ Let $|\psi \rangle =\sqrt{p_{0}}|0\rangle +\sqrt{p_{1}}
|1\rangle $ with $p_{0}\geq p_{1}$ and $|\phi \rangle =\sqrt{q_{0}}|0\rangle
+\sqrt{q_{1}}|1\rangle $ with $q_{0}\geq q_{1}.$ The tensor product of these
two states is:
\begin{equation*}
|\psi \rangle |\phi \rangle =\sqrt{r_{0}}|\chi _{0}\rangle +\sqrt{r_{1}}
|\chi _{1}\rangle \;,
\end{equation*}
where $r_{0}\equiv p_{0}q_{0}+p_{1}q_{1},$ $r_{1}\equiv $ $
p_{0}q_{1}+p_{1}q_{0}$, $|\chi _{0}\rangle \equiv (\sqrt{p_{0}q_{0}}
|00\rangle +\sqrt{p_{1}q_{1}}|11\rangle /\sqrt{r_{0}}$ and $|\chi
_{1}\rangle \equiv (\sqrt{p_{0}q_{1}}|01\rangle +\sqrt{p_{1}q_{0}}|10\rangle
/\sqrt{r_{1}}$. \ Noting that $r_{0}\geq r_{1}$ (because $
r_{0}-r_{1}=(p_{0}-p_{1})(q_{0}-q_{1})\geq 0$), and noting that we can
transform $\left\vert \chi _{0}\right\rangle \rightarrow \left\vert
0\right\rangle $ and $\left\vert \chi _{1}\right\rangle \rightarrow
\left\vert 1\right\rangle $ by a Z$_{2}$-invariant unitary, we see that $
\sqrt{r_{0}}\left\vert 0\right\rangle +\sqrt{r_{1}}\left\vert 1\right\rangle
$ is the standard form of the tensor product state. Therefore,
\begin{eqnarray*}
F^\infty(|\psi \rangle |\phi \rangle ) &=&-\log \left\vert
r_{0}-r_{1}\right\vert \\
&=&-\log \left\vert (p_{0}-p_{1})(q_{0}-q_{1})\right\vert \\
&=&-\log \left\vert p_{0}-p_{1}|-\log |q_{0}-q_{1}\right\vert \\
&=&F^\infty(|\psi \rangle )+F^\infty(|\phi \rangle )\;.
\end{eqnarray*}
QED.

\section{Resource theory of the SU(2)-SSR}

\subsection{Frames for orientation}

A reference frame for orientation, commonly called a Cartesian frame, is
associated with SO(3), the rotation group. An element of SO(3) can be given,
for instance, by specifying three Euler angles. We will represent it
instead by a vector $\vec{\theta},$ representing a rotation by $\theta $
about the axis $\hat{\theta}=\vec{\theta}/\theta .$ We will extend the group
of rotations SO(3) to the group SU(2) to allow for spinor representations.
The representation $T$ of $SU(2)$ on a Hilbert space $\mathcal{H}$
determines how the quantum system transforms under rotations,
\begin{equation*}
T(\vec{\theta})=\exp (i\vec{\theta}\mathbf{\cdot \hat{J}}),
\end{equation*}
where $\mathbf{\hat{J}}=(\hat{J}_{x},\hat{J}_{y},\hat{J}_{z})$ is the
angular momentum operator.

The states that are SU(2)-invariant, which we may also describe as
rotationally invariant, are those satisfying
\begin{equation*}
T(\vec{\theta})\rho T^{\dag }(\vec{\theta})=\rho \text{ \ \ \ \ }\forall
\vec{\theta}\in \mathrm{SU(2)}.
\end{equation*}
This is equivalent to the condition
\begin{equation*}
\lbrack \rho ,\hat{J}_{x}]=[\rho ,\hat{J}_{y}]=[\rho ,\hat{J}_{z}]=0.
\end{equation*}

It is useful to decompose the Hilbert space as
\begin{equation}
\mathcal{H}=\bigoplus_{j}\mathcal{M}_{j}\otimes \mathcal{N}_{j},
\label{eq:SU(2)decompofH}
\end{equation}
where $j\in \{0,1/2,1,3/2,...\}$ is the angular momentum quantum
number, the $\mathcal{M}_{j}$ carry irreducible representations of
SU(2), and the $ \mathcal{N}_{j}$ are the multiplicity spaces,
carrying the trivial representation of SU(2).

Relative to this decomposition, the SU(2)-invariant states have the
form \cite{BRS07}
\begin{equation*}
\rho =\sum_{j}p_{j}\frac{I_{\mathcal{M}_{j}}}{\dim (\mathcal{M}_{j})}\otimes
\rho _{\mathcal{N}_{j}},
\end{equation*}
where $I_{\mathcal{M}_{j}}$ is the identity operator on $\mathcal{M}_{j},$ $
\dim (I_{\mathcal{M}_{j}})$ is the dimension of $\mathcal{M}_{j},$ $p_{j}$
is a probability distribution over $j,$ and $\rho _{\mathcal{N}_{j}}$ is an
arbitrary density operator on $\mathcal{N}_{j}.$

We again focus our attention on pure states. Relative to the decomposition
of Eq.~(\ref{eq:SU(2)decompofH}), a general pure state can be written as
\begin{equation*}
\left\vert \psi \right\rangle =\sum_{j,m,\beta }c_{jm\beta }\left\vert
j,m\right\rangle \otimes \left\vert j,\beta \right\rangle ,
\end{equation*}
where the $\left\vert j,m\right\rangle $ form a basis of $\mathcal{M}_{j}$
and the $\left\vert j,\beta \right\rangle $ form a basis of $\mathcal{N}
_{j}. $ \ It transforms under SU(2) as
\begin{equation*}
T(\vec{\theta})\left\vert \psi \right\rangle =\sum_{j,m,\beta }\left( T_{j}(
\vec{\theta})\left\vert j,m\right\rangle \right) \otimes \left\vert j,\beta
\right\rangle ,
\end{equation*}
where $T_{j}(\vec{\theta})$ is the $j$th irreducible unitary representation
of SU(2).

Again, any operation on the $\mathcal{N}_{j}$ can be achieved under the
SU(2)-SSR, therefore an arbitrary pure state can always be transformed into
the standard form
\begin{equation*}
\left\vert \psi \right\rangle =\sum_{j,m}c_{jm}\left\vert j,m\right\rangle ,
\end{equation*}
and we presume this form henceforth. \ Effectively, we are confining
ourselves to the Hilbert space $\mathcal{H}^{\prime }=\bigoplus_{j}\mathcal{M
}_{j}=\mathrm{span}\{\left\vert j,m\right\rangle \}_{j,m}\subseteq \mathcal{H
}.$

We can now write SU(2)-invariant states simply as
\begin{equation}
\rho =\sum_{j}p_{j}\Pi _{j}  \label{eq:SU(2)invariantstate}
\end{equation}
where $\Pi _{j}=\sum_{m}\left\vert j,m\right\rangle \left\langle
j,m\right\vert .$

It is interesting to note that the only pure state that is
SU(2)-invariant is $\left\vert j=0,m=0\right\rangle ,$ because it is
the only one for which the density operator has the form of
Eq.~(\ref{eq:SU(2)invariantstate}). Consequently, it is the only
pure state that is free under the SU(2)-SSR -- every other pure
state is a resource. To see how these resources can be manipulated,
we must derive the form of SU(2)-invariant operations.

\subsection{SU(2)-invariant operations}

We now apply Lemma~\ref{lemma:Ginvariantoperations} to the problem
of characterizing the SU(2)-invariant operations. Recall that the
irreducible representations of SU(2) are labeled by the set of
nonnegative integers and half-integers $J\in \{0,1/2,1,3/2,\dots \}$
and are each of dimension $2J+1.$ The $J$th irreducible unitary
representation of SU(2), $u^{(J)}:\mathrm{ SU(2)}\rightarrow
\mathbb{C}^{2J+1},$ has matrix elements
\begin{equation}
u_{MM^{\prime }}^{(J)}(\vec{\theta})\equiv \left\langle J,M\right\vert e^{-i
\vec{\theta}\mathbf{\cdot \hat{J}}}\left\vert J,M^{\prime }\right\rangle ,
\end{equation}
where $\left\vert J,M\right\rangle $ is the joint eigenstate of $\hat{J}^{2}$
and $J_{z}$ with eigenvalues $J(J+1)$ and $\hbar M.$

It then follows from Lemma~\ref{lemma:Ginvariantoperations} that an
SU(2)-invariant operation has Kraus operators $K_{J,M,\alpha },$
labeled by an irrep $J,$ a basis element $M,$ and a multiplicity
index $\alpha $, satisfying
\begin{equation}
e^{i\vec{\theta}\mathbf{\cdot \hat{J}}}K_{J,M,\alpha }e^{-i\vec{\theta}
\mathbf{\cdot \hat{J}}}=\sum_{M^{\prime }}u_{MM^{\prime }}^{(J)}(\vec{\theta}
)K_{J,M^{\prime },\alpha },\quad \forall \vec{\theta}\in \mathrm{SU(2).}
\label{eq:SU(2)invariantKraus}
\end{equation}
The set of operators $\{K_{J,M,\alpha }|M\}$ for fixed $J$ and
$\alpha $ is sometimes called a \emph{spherical tensor} of rank $J$
(see e.g. p.~569 in~\cite{Messiah}).

The simplest case to consider is the $J=0$ irrep, which is the trivial
representation
\begin{equation*}
u^{(0)}(\vec{\theta})=1.
\end{equation*}
A Kraus operator associated with this irrep, denoted $K_{0,0,\alpha },$
satisfies
\begin{equation*}
e^{i\vec{\theta}\mathbf{\cdot \hat{J}}}K_{0,0,\alpha }e^{-i\vec{\theta}
\mathbf{\cdot \hat{J}}}=K_{0,0,\alpha }\text{ },\quad \forall \vec{\theta}
\in \mathrm{SU(2).}
\end{equation*}
It follows that $K_{0,0,\alpha }$ is an SU(2)-invariant operator and
therefore has the form
\begin{equation}
K_{0,0,\alpha }=\sum_{j}c_{j}^{(\alpha )}\Pi _{j}\,,  \label{Kr}
\end{equation}
where $\Pi _{j}=\sum_{m=-j}^{j}|j,m\rangle \langle j,m|$ and the $
c_{j}^{(\alpha )}$ are complex numbers.

By taking derivatives of Eq.~(\ref{eq:SU(2)invariantKraus}) relative
to the different components of $\vec{\theta}$ and then setting $\vec{\theta}=0$,
one finds that a Kraus
decomposition $\{K_{J,M,\alpha }\}$ can always be found that satisfies
\begin{align}
\lbrack \hat{J}_{z},K_{J,M,\alpha }]& =\hbar M\text{ }K_{J,M,\alpha }
\label{eq:JzactiononM_JM} \\
\lbrack \hat{J}_{\pm },K_{J,M,\alpha }]& =\hbar \sqrt{J(J+1)-M(M\pm 1)}
K_{J,M\pm 1,\alpha }  \label{eq:J+-actiononM_JM}
\end{align}
where $\hat{J}_{\pm }\equiv \frac{1}{\sqrt{2}}(\hat{J}_{x}\pm \hat{J}_{y})$
are the angular momentum raising and lowering operators.

The Wigner-Eckart theorem (see e.g. p.~239 in~\cite{Sakurai})
famously specifies the form of the spherical
tensor operators of rank $J$: a set of operators $\{K_{J,M,\alpha }\}$
satisfy Eqs.~(\ref{eq:JzactiononM_JM}) and (\ref{eq:J+-actiononM_JM}) if and
only if
\begin{eqnarray}
&&\langle j^{\prime },m^{\prime }|K_{J,M,\alpha }|j,m\rangle  \\
&=&(-1)^{j^{\prime }-m^{\prime }}
\begin{pmatrix}
j^{\prime } & J & j \\
-m^{\prime } & M & m
\end{pmatrix}
\langle j^{\prime }\Vert K_{J,\alpha }\Vert j\rangle \;,
\end{eqnarray}
where $f_{J,\alpha }(j^{\prime },j)\equiv \langle j^{\prime }\Vert
K_{J,\alpha }\Vert j\rangle $ does not depend on $m$, $m^{\prime }$ or $M$.
Note that\ for \emph{any} choice of $f_{J,\alpha }(j^{\prime },j)$, the
matrices $K_{J,M,\alpha }$ as defined above satisfy the commutation
relations in Eqs.~(\ref{eq:JzactiononM_JM}) and (\ref{eq:J+-actiononM_JM}).

We therefore conclude that an SU$(2)$-invariant operation admits a Kraus
decomposition $\{K_{J,M,\alpha }\}$ where
\begin{eqnarray}
&&\langle j^{\prime },m^{\prime }|K_{J,M,\alpha }|j,m\rangle
\label{matrixelements} \\
&=&(-1)^{j^{\prime }-m^{\prime }}
\begin{pmatrix}
j^{\prime } & J & j \\
-m^{\prime } & M & m
\end{pmatrix}
f_{J,\alpha }(j^{\prime },j)\;,
\end{eqnarray}
for some choice of $f_{J,\alpha }(j^{\prime },j)$. We require that
$\sum_{J,\alpha}\left|f_{J,\alpha}(j',j)\right|^2\leq 2j+1$ for all
$j, j'\;.$

Recalling that Wigner's $3j$ symbols are defined in terms of
Clebsch-Gordan coefficents by
\begin{equation}
\begin{pmatrix}
j_{1} & j_{2} & j_{3} \\
m_{1} & m_{2} & m_{3}
\end{pmatrix}
\equiv \frac{(-1)^{j_{1}-j_{2}-m_{3}}}{\sqrt{2j_{3}+1}}\left(
j_{1},m_{1},j_{2},m_{2}|j_{3},-m_{3}\right) ,  \label{eq:3jtoCG}
\end{equation}
and that
\begin{equation*}
\left( j_{1},m_{1},j_{2},m_{2}|j,m\right) =\delta _{m,m_{1}+m_{2}}\left(
j_{1},m_{1},j_{2},m_{2}|j,m_{1}+m_{2}\right) ,
\end{equation*}
we conclude that $\langle j^{\prime },m^{\prime }|K_{J,M,\alpha
}|j,m\rangle $ is only nonzero if $m=m^{\prime }-M.$ \  We summarize
the result by the following lemma.

\begin{lemma} \label{lemma:U1invariantopbasic} An arbitrary
SU(2)-invariant operation on $\mathcal{B}(\mathcal{H}^{\prime })$
admits a Kraus decomposition $\{K_{J,M,\alpha }\},$ where $J\in
\{0,1/2,1,3/2,...\},$ $M\in \{-J,...,J\}$ and $\alpha $ is an
integer, such that
\begin{eqnarray}
K_{J,M,\alpha } &=&\sum_{j^{\prime }=0,1/2,1...}\sum_{m=-j^{\prime
}}^{j^{\prime }}\sum_{\;j=|J-j^{\prime }|}^{J+j^{\prime
}}(-1)^{j^{\prime
}-m}  \notag \\
&&\times
\begin{pmatrix}
j^{\prime } & J & j \\
-m & M & m-M%
\end{pmatrix}
\notag \\
&&\times
%c^{(J,\alpha)}_{j^{\prime },j}\;
f_{J,\alpha}(j^{\prime },j)\; |j^{\prime },m\rangle \langle
j,m-M|\;. \label{eq:M_JM}
\end{eqnarray}
where the $2\times 3$ matrix is a Wigner $3j$ symbol and the
function $f_{J,\alpha}(j^{\prime },j)$ does not depend on $m$ or
$M$, and satisfies $\sum_{J,j',\alpha}\left|f_{J,\alpha}(j^{\prime
},j)\right|^2\leq 2j+1$ for all $j\;,$ with equality if the
operation is trace-preserving.
\end{lemma}

%It follows that
%\begin{eqnarray}
%K_{J,M,\alpha } &=&\sum_{j^{\prime }=0,1/2,1...}\sum_{m^{\prime }=-j^{\prime
%}}^{j^{\prime }}\sum_{\;j=|J-j^{\prime }|}^{J+j^{\prime }}(-1)^{j^{\prime
%}-m^{\prime }}  \notag \\
%&&\times
%\begin{pmatrix}
%j^{\prime } & J & j \\
%-m^{\prime } & M & m^{\prime }-M
%\end{pmatrix}
%\notag \\
%&&\times f_{J,\alpha }(j^{\prime },j)\;|j^{\prime },m^{\prime }\rangle
%\langle j,m^{\prime }-M|\;.  \label{eq:M_JM}
%\end{eqnarray}

We recover Eq.~(\ref{Kr}) by noting that for $J=M=0,$ we have
\begin{equation*}
\begin{pmatrix}
j^{\prime } & 0 & j \\
-m & 0 & m
\end{pmatrix}
=\frac{(-1)^{j^{\prime }-m}}{\sqrt{2j+1}}\delta _{j,j^{\prime }},
\end{equation*}
and consequently
\begin{equation*}
K_{0,0,\alpha
}=\sum_{j=0,1/2,1...}\frac{f_{\alpha}(j)}{\sqrt{2j+1}}\Pi _{j},
\end{equation*}
for some amplitudes $f_{\alpha}(j)$.

 A different characterization of SU(2)-invariant
operations is provided in Boileau~\emph{et al.}~\cite{BSLB07}, but
we shall not make use of it here. Determining the connection between
it and the characterization provided above is a subject for future
research.

\subsubsection{SU(2)-invariant unitaries}

In addition to the unitaries defined on the multiplicity spaces
$\mathcal{N} _{j},$ there are SU(2)-invariant unitaries on the
subspace $\mathcal{H} ^{\prime }=\mathrm{span}\{\left\vert
j,m\right\rangle \}_{j,m}=\bigoplus_{j} \mathcal{M}_{j}$. \ Unitary
operations have only a single Kraus operator, so there are unitaries
among the $J=0$ irreducible SU(2)-invariant operations. The single
(unitary) Kraus operator for such an operation has the form
$K_{0,0}= \sum_{j}c_{j}\Pi _{j}$, where $|c_{j}^{(\alpha )}|=1$ for
all $j$. These operations on $\mathcal{H} ^{\prime }$ simply change
the relative phases between the $\mathcal{M}_{j}$ subspaces.

\subsection{A restricted set of SU(2) frame states}

From the very outset, it is clear that there will be several
distinct sorts of resources under the SU(2)-SSR. To see this, note
that there is a distinction between a quantum Cartesian frame (a
state that picks out a triad of orthogonal spatial directions) and a
symmetric quantum direction indicator (a state that only picks out a
single direction in space and is symmetric under rotations about
that direction). Symmetric quantum direction indicators for distinct
directions are clearly inequivalent resources because to transform
one to the other would require the ability to rotate about some
third axis, and the latter operation is forbidden under the
SU(2)-SSR. It then follows that a symmetric quantum direction
indicator is not equivalent to a quantum Cartesian frame because the
latter can only be built out of a \emph{pair} of the former for
distinct directions. So we can already see that two resources under
the SU(2)-SSR need not be interconvertible.

The general problem of the transformation of pure resource states under the
SU(2)-SSR appears to be very difficult and we do not attempt to solve it
completely here. Rather, we restrict our attention to a subset of states.
To define this set, recall the definition of an SU(2)-coherent state. It
is a highest weight state $\left\vert j,m=j\right\rangle _{\hat{n}}$ where $
\hat{n}$ denotes the quantization axis. \ Now\ define $\mathcal{H}_{\hat{n}
}\equiv \mathrm{span}\{\left\vert j,j\right\rangle _{\hat{n}
}|j=0,1/2,1,\dots \}$, the subspace of $\mathcal{H}^{\prime }$ consisting of
all linear combinations of SU(2)-coherent states associated with the same
quantization axis. \ Note that every state in $\mathcal{H}_{\hat{n}}$ except
one is a resource under the SU(2)-SSR. \ The exception is the singlet $
\left\vert 0,0\right\rangle $.\ Consequently, we can define a set of
resource states by removing the singlet, $\mathcal{C}_{\hat{n}}\equiv
\mathcal{H}_{\hat{n}}-\mathrm{span}\{\left\vert 0,0\right\rangle \}.$ \ Note
that only the states of the form $\left\vert j,j\right\rangle _{\hat{n}}$
are symmetric direction indicators for $\hat{n}$ because any linear
combination of these fails to be invariant under rotations about $\hat{n}$
(the physical interpretation of states in $\mathcal{C}_{\hat{n}}$ is unclear
at present). \ We are finally in a position to define the full set of
quantum reference states with which we will concern ourselves here. It is
\begin{equation*}
\mathcal{C}\equiv \bigcup\limits_{\hat{n}\in \mathrm{S}_{2}}\mathcal{C}_{
\hat{n}},
\end{equation*}
the union of the $\mathcal{C}_{\hat{n}}$ for all directions $\hat{n}$ on the
unit sphere $\mathrm{S}_{2},$ that is, all choices of quantization axis.
Note that the set $\mathcal{C}$ is not exhaustive. \ States in $\mathcal{H}
^{\prime }$ assigning nonzero amplitude to any $\left\vert j,m\right\rangle $
with $|m|<j$ are excluded.

We will show that an element of $\mathcal{C}_{\hat{n}}$ and an element of $
\mathcal{C}_{\hat{m}}$ where $\hat{n}\neq \hat{m}$ \ are inequivalent
resources in the sense that one cannot convert one to the other, not even
with probability less than unity.

It follows that there is a continuous infinity of different types of
resources under the SU(2)-SSR. This is similar to what occurs for pure
state entanglement for four qubits. In this sense, the
resource theory for SU(2) frames is degenerate. Nonetheless, one can still
ask what transformations are possible \emph{within} a class
$\mathcal{C}_{\hat{n}}$ and indeed, a nontrivial structure is found to which we turn in
subsequent sections. Here, we prove the existence of the distinct
classes.

\subsubsection{Proof of the existence of inequivalent classes within the
restricted set}

An arbitrary state in $\mathcal{C}_{\hat{n}}$ can be written as $\left\vert
\psi \right\rangle =\sum_{j}c_{j}\left\vert j,j\right\rangle $. \ However,
it can always be transformed, by SU(2)-invariant unitaries, into the
standard form
\begin{equation*}
\left\vert \psi \right\rangle =\sum_{j}\sqrt{p_{j}}\left\vert
j,j\right\rangle ,
\end{equation*}
where $0\leq p_{j}\leq 1$ and $\sum_{j}p_{j}=1.$ Note further that $
|j,j\rangle \otimes |j^{\prime },j^{\prime }\rangle =|j+j^{\prime
}.j+j^{\prime }\rangle $ \ (as a straightforward calculation of
Clebsch-Gordan coefficients confirms) so that multiple systems with states
drawn from $\mathcal{C}_{\hat{n}}$ can also be represented in the standard
form. \ We assume this form in the following.

We wish to determine which, if any, SU(2)-invariant operations
$\mathcal{E}$ can take a state $\left\vert \psi \right\rangle \in
\mathcal{H} _{\hat{n}}$ to a pure state, $\mathcal{E}(\left\vert
\psi \right\rangle \left\langle \psi \right\vert )=\lambda
\left\vert \phi \right\rangle \left\langle \phi \right\vert .$  We
do not assume that $\left\vert \phi \right\rangle \in
\mathcal{H}_{\hat{n}}$ (although it will be shown that this is the
case for SU(2)-invariant operations that take pure states to pure
states). \

Every SU(2)-invariant operation can be written as a convex sum of irreducible
\emph{\ }SU(2)-invariant operations, $\mathcal{E}=\sum_{J,\alpha }$\strut $
w_{J,\alpha }\mathcal{E}_{J,\alpha },$ where the irreducible operations are
labelled by the irreps $J$ and a multiplicity index $\alpha $. \ It suffices
therefore to identify, for a given $J,$ which irreducible SU(2)-invariant
operations $\mathcal{E}_{J}$ can take a pure state $\left\vert \psi
\right\rangle \in \mathcal{H}_{\hat{n}}$ to another pure state. Recalling
Eq. (\ref{Kr}), an irreducible SU(2)-invariant operation $\mathcal{E}_{0}$
associated with $J=0$ has a single Kraus operator $K_{0,0}=\sum_{j}c_{j}\Pi
_{j}$ which clearly takes $\left\vert \psi \right\rangle $ to a pure state
within $\mathcal{H}_{\hat{n}}.$

The interesting case is $J>0.$ \ A given $\mathcal{E}_{J}$ has Kraus
operators $\{K_{J,M}|M\in \{-J,...,J\}\}$ where the $K_{J,M}$ satisfy
Eq.~(\ref{eq:M_JM}). \ The only freedom we have is in the variation of the
function $f_{J}(j,j^{\prime }).$ \ Our question, therefore, is: for
what choice of $f_{J}(j,j^{\prime })$ can one achieve
\begin{equation*}
\sum_{M}K_{J,M}\left\vert \psi \right\rangle \left\langle \psi \right\vert
K_{J,M}^{\dag }=\lambda \left\vert \phi \right\rangle \left\langle \phi
\right\vert ,
\end{equation*}
or equivalently,
\begin{equation}
K_{J,M}\left\vert \psi \right\rangle =h(J,M)\left\vert \phi \right\rangle
\text{ for all }M\in \{-J,...,J\},  \label{eq:Mpsiphi}
\end{equation}
where the functions $h(J,M)$ satisfy
\begin{equation*}
\sum_{M=-J}^{J}\left\vert h(J,M)\right\vert ^{2}=\lambda .
\end{equation*}

\begin{theorem} \label{thm:SU2classes} If the irreducible SU(2)-invariant
map $\mathcal{E}_{J}$ takes a pure state $|\psi \rangle \in \mathcal{H}_{\hat{n}}$ to
another pure state, $|\phi\rangle$,
then the restriction of $\mathcal{E}_{J}$ to $\mathcal{H}_{\hat{
n}}$ must have Kraus operators of the form
\begin{eqnarray}
K_{J,M} &=&0\text{ for }M\in \{-J+1,\dots ,J\}  \notag \\
K_{J,-J} &=&\sum_{j\geq J}c_{j}^{(J)}|j-J,j-J\rangle \langle j,j|\;,
\label{eq:pure2puremjm}
\end{eqnarray}
where the $c_{j}^{(J)}$ are complex coefficients satisfying
$|c_j^{(J)}|^2 \leq 1$ for all $j$, with equality if the operation
is trace-preserving.
\end{theorem}

Note that due to the form of $\mathcal{E}_{J}$, the output state $|\phi
\rangle $ is always in $\mathcal{H}_{\hat{n}}$. Consequently,
SU(2)-invariant maps cannot transform a pure state inside $\mathcal{H}_{\hat{
n}}$ to one outside $\mathcal{H}_{\hat{n}}$ with any probability.

\textbf{Proof.} Suppose that $\left\vert \psi \right\rangle
=\sum_{j}\sqrt{ p_{j}}\left\vert j,j\right\rangle .$ \ From
Eq.~(\ref{eq:M_JM}), we infer that

\begin{equation}
K_{J,M}\left\vert \psi \right\rangle =\sum_{j^{\prime }}\sum_{j=|J-j^{\prime
}|}^{J+j^{\prime }}a_{j,j^{\prime }}^{(J,M)}\left\vert j^{\prime
},j+M\right\rangle   \label{eq:mjm}
\end{equation}
where the amplitudes are
\begin{align}
a_{jj^{\prime }}^{(J,M)}& \equiv (-1)^{j^{\prime }-(j+M)}
\begin{pmatrix}
j^{\prime } & J & j \\
-(j+M) & M & j
\end{pmatrix}
\notag \\
& \times f_{J}(j^{\prime },j)\sqrt{p_{j}}.  \label{eq:ajjM}
\end{align}

Recalling the definition of the Wigner $3j$ symbol in terms of
Clebsch-Gordan coefficients (Eq.~(\ref{eq:3jtoCG})), we note that necessary
conditions for a Wigner $3j$ symbol to be nonzero,
\begin{equation*}
\begin{pmatrix}
j_{1} & j_{2} & j_{3} \\
m_{1} & m_{2} & m_{3}
\end{pmatrix}
\neq 0,
\end{equation*}
are
\begin{align}
\left\vert j_{1}-j_{2}\right\vert & \leq j_{3}\leq j_{1}+j_{2},
\label{eq:in1} \\
\left\vert m_{1}\right\vert & \leq j_{1},  \label{eq:in2} \\
\left\vert m_{2}\right\vert & \leq j_{2}.  \label{eq:in3}
\end{align}
For the particular $3j$ symbol appearing in Eq.~(\ref{eq:ajjM})`to
be nonzero, the necessary conditions are
\begin{align}
\left\vert J-j^{\prime }\right\vert & \leq j\leq j^{\prime }+J
\label{eq:constj} \\
\left\vert -(j+M)\right\vert & \leq j^{\prime }  \notag \\
\left\vert M\right\vert & \leq J\;.  \notag
\end{align}
We point out that these conditions are also \emph{sufficient} for
the $3j$ symbol that appears in Eq.~(\ref{eq:ajjM}) to be non-zero
(see the formula C.~24 in p.~1059 of ref.~\cite{Messiah}). The
latter two constraints can be written as bounds on $M$,
\begin{align}
-(j^{\prime }+j)& \leq M\leq j^{\prime }-j,  \label{eq:constM1} \\
-J& \leq M\leq J.  \label{eq:constM2}
\end{align}
From the first inequality in Eq. (\ref{eq:constj}), we can deduce that $
-(j+j^{\prime })\leq -J$ which implies that the larger of the two lower
bounds on $M$ is the second. \ Similarly, we can deduce from this inequality
that $j^{\prime }-j\leq J$ which implies that the smaller of the two upper
bounds on $M$ is the first. \ All told, we have
\begin{equation}
-J\leq M\leq j^{\prime }-j.  \label{eq:Mbounds}
\end{equation}

Consider first the case wherein the $3j$ symbol is nonzero for only a \emph{
single} value of $M.$ To ensure that this is the case, we must ensure that
the lower and upper bounds on $M$ coincide, that is, we must take
\begin{equation*}
M=j^{\prime }-j=-J.
\end{equation*}
This constraint can be enforced by choosing
\begin{equation*}
f_{J}(j^{\prime },j)=\delta _{j^{\prime },j-J}f_{J}(j).
\end{equation*}
By substituting this choice of $f_{J}(j^{\prime },j)$ into
Eq.~(\ref{eq:mjm}), we obtain Eq.~(\ref{eq:pure2puremjm}), the
allowed form of $\mathcal{E} _{J}$ given in the theorem. To prove
the theorem, we must show that this is the \emph{only} possible form
that $f_{J}(j^{\prime },j)$ can take.

We do so by assuming the contrary and deriving a contradiction.
Suppose that $f_{J}(j^{\prime },j)\neq 0$ for some triple of values
$J$, $j^{\prime }$, $j $ satisfying $j^{\prime }-j>-J$ (so that the
upper and lower bounds in Eq.~(\ref{eq:Mbounds}) do \emph{not}
coincide) and $j\geq |J-j^{\prime }|$ (so that the 3j-symbols are
nonzero). Note that for fixed $J$ and $j^{\prime },$ the $j$ value
for such a triple is in the range
\begin{equation*}
\left\vert J-j^{\prime }\right\vert \leq j<J+j^{\prime }.
\end{equation*}
For such distinguished triples, $a_{jj^{\prime }}^{(J,M)}\neq 0$ and
consequently $K_{J,M}|\psi \rangle \neq 0$, for all $M$ in the range $-J\leq
M\leq j^{\prime }-j$, which includes more than one value.

Suppose $M_{1},$ $M_{2}$ are two distinct values in the given range.
Equation (\ref{eq:Mpsiphi}) implies that $K_{J,M_{1}}|\psi \rangle $
and $ K_{J,M_{2}}|\psi \rangle $ must be proportional to each other.
It follows from Eq. (\ref{eq:mjm}) that for $J,$ $j^{\prime },j$ a
distinguished triple, the state $K_{J,M_{1}}|\psi \rangle $ assigns
nonzero amplitude to the term $\left\vert j^{\prime },m_{1}^{\prime
}\right\rangle $ with $ m_{1}^{\prime }\equiv j+M_{1}$. Similarly,
$K_{J,M_{2}}|\psi \rangle $ assigns nonzero amplitude to the term
$\left\vert j^{\prime },m_{2}^{\prime }\right\rangle $ with
$m_{2}^{\prime}\equiv j+M_{2}.$ However, given the assumed
proportionality of $K_{J,M_{1}}|\psi \rangle $ and $K_{J,M_{2}}|\psi
\rangle $, we must also have the former term in $K_{J,M_{2}}|\psi
\rangle $ and the latter term in $K_{J,M_{1}}|\psi \rangle .$ It
follows that there must be another value of $j$, call it
$j_{\text{max}},$ such that $ K_{J,M_{1}}\left\vert j_{\max
},j_{\max }\right\rangle $ assigns nonzero amplitude to $\left\vert
j^{\prime },m_{2}^{\prime}\right\rangle $ (hence $ f_{J}(j^{\prime
},j_{\text{max}})\neq 0$) and a third value of $j,$ call it $
j_{\text{min}}$, such that $K_{J,M_{2}}|\psi \rangle \left\vert
j_{\min },j_{\min }\right\rangle $ assigns nonzero amplitude to
$\left\vert j^{\prime },m_{1}^{\prime}\right\rangle $ (hence
$f_{J}(j^{\prime },j_{\text{min} })\neq 0$). \ From
Eq.~(\ref{eq:mjm}), we see that we require $j_{\text{max}
}=j+M_{2}-M_{1}$ and $j_{\text{min}}=j+M_{1}-M_{2}.$

Taking $M_{1}=-J$ and $M_{2}=-J+\eta ,$ where $0<\eta \leq j^{\prime
}-j+J,$ \ we have $j_{\text{min}}=j-\eta ,$ and
$j_{\text{max}}=j+\eta .$ \ For the maximum value of $\eta ,$ we
find $j_{\text{min}}=2j-(J+j^{\prime })<j$ and $
j_{\text{max}}=J+j^{\prime }.$ \ We conclude that $f_{J}(j^{\prime
},x)\neq 0 $ for $j_{\text{min}}\leq x\leq j_{\text{max}}.$  In
particular, we see that if we have a triple $J$, $j^{\prime },$ $j$
satisfying $\left\vert J-j^{\prime }\right\vert \leq j<J+j^{\prime
}$ and $f_{J}(j^{\prime },j)\neq 0,$ then the triple $J,j^{\prime
},j_{\text{min}}$\emph{\ }satisfies $ \left\vert J-j^{\prime
}\right\vert \leq j_{\text{min}}<J+j^{\prime }$ and $
f_{J}(j^{\prime },j_{\text{min}})\neq 0.$

Applying this rule recursively, we generate more triples of the form $
J,j^{\prime },j_{n}$ where $j_{n}\equiv 2j_{n-1}-(J+j^{\prime })<j_{n-1},$
and $j_{0}=j.$ \ Because $j_{n}$ decreases at every iteration, it eventually
takes the minimum possible value of $\left\vert J-j^{\prime }\right\vert .$\
Consequently, we can conclude that $f_{J}(j^{\prime },x)\neq 0$ for
\begin{equation*}
\left\vert J-j^{\prime }\right\vert \leq x\leq J+j^{\prime }.
\end{equation*}

If follows from Eq.~(\ref{eq:mjm}) that for $J,$ $j^{\prime }$ that
are part of a distinguished triple, the state $K_{J,M}|\psi \rangle
$ assigns nonzero amplitude to terms $\left\vert j^{\prime
},m^{\prime }\right\rangle $ with $ m^{\prime }=x+M$ where $x$ is in
the given range. \ In particular, the smallest value of $m^{\prime
}$ such that $\left\vert j^{\prime },m^{\prime }\right\rangle $
receives nonzero amplitude is $m_{\text{min}}^{\prime }=\left\vert
J-j^{\prime }\right\vert +M.$ \ However, this depends explicitly on
$M,$ contradicting the assumption, articulated in
Eq.~(\ref{eq:Mpsiphi}), that the normalized form of $
K_{J,M}\left\vert \psi \right\rangle $ is independent of $M.$ QED.

\subsection{Restricting to a fixed quantization axis}

\label{sec:KrausforSU(2)}

We restrict ourselves to the Hilbert space $\mathcal{H}_{\hat{n}}$
containing all linear combinations of SU(2)-coherent states
associated with the quantization axis $\hat{n}$. Given
Thm.~\ref{thm:SU2classes}, an\ irreducible SU(2)-invariant operation
on $\mathcal{H}_{\hat{n}}$ must have a single Kraus operator $K_{J}$
that can be factored as
\begin{equation}
K_{J}=S_{-J}\tilde{K}_{J}  \label{eq:KrausforSU2}
\end{equation}
where
\begin{equation*}
\tilde{K}_{J}=\sum_{j}c_{j}^{(J)}|j,j\rangle \langle j,j|
\end{equation*}
with $|c_{j}^{(J)}|\leq 1$ for all $j$ and
\begin{equation*}
S_{-J}=\sum_{j\geq J}|j-J,j-J\rangle \langle j,j|
\end{equation*}
for some positive integer or half-integer $J.$ \ Clearly, $\tilde{K}_{J}$
changes the relative amplitudes (weights and phases) of the $\left\vert
j\right\rangle ,$ possibly eliminating the amplitude for some, while $S_{-J}$
shifts $\left\vert j,j\right\rangle $ down to $\left\vert
j-J,j-J\right\rangle $.

We have now specified the form, within a given space
$\mathcal{H}_{\hat{n}},$ of irreducible SU(2)-invariant operations
that implement pure-to-pure transformations.\ The most general
SU(2)-invariant operation that implements pure-to-pure
transformations is a sum of these. Denoting the multiplicity index
by $\alpha ,$ a general SU(2)-invariant operation can be written as
$\mathcal{E}=\sum_{J,\alpha } \mathcal{E}_{J,\alpha }$ where
$\mathcal{E}_{J,\alpha }$ is an irreducible SU(2)-invariant
operation associated with the $J$th irrep. Incorporating the
constraint that the trace be nonincreasing, we summarize our result
with the following lemma.

\begin{lemma}\label{lemma:SU2invariantop} An SU(2)-invariant
operation on $\mathcal{H}_{\hat{n}}$ that takes pure states to pure
states admits a Kraus decomposition $\{K_{J,\alpha}\}$, of the form
\begin{equation*}
K_{J,\alpha }=S_{-J}\tilde{K}_{J,\alpha}
\end{equation*}
where $\tilde{K}_{J,\alpha }=\sum_{j}c_{j}^{(J,\alpha)}|j,j\rangle
\langle j,j|$ changes the relative amplitudes of the $|j,j\rangle$
states, possibly eliminating some, and $S_{-J}=\sum_{j\geq
J}|j-J,j-J\rangle \langle j,j|$ shifts the value of $j$ downward by
$J$. The coefficients satisfy $\sum_{J\leq j}\sum_{\alpha}
|c_j^{(J,\alpha)}|^2 \leq 1$ for all $j$, with equality if the
operation is trace-preserving.
\end{lemma}

Notice that the form of the allowed Kraus operators for an SU(2)-invariant
operation on $\mathcal{H}_{\hat{n}}$ is almost the same as the allowed Kraus
operators for an irreducible U(1)-invariant operation, discussed in
Sec.~\ref{sec:RFsforphase}, with $j$ playing the role of $n.$ The only difference
is that the $j$ value can only be shifted downward, whereas $n$ can be
shifted in either direction. The resource theory of SU(2) frames in $
\mathcal{H}_{\hat{n}}$ is consequently very close to that of U(1) frames,
particularly for single-copy transformations. We therefore lean heavily on
the results and proofs provided in Sec.~\ref{sec:RFsforphase} in
describing and justifying the SU(2) resource theory.

Although the Stinespring dilation theorem guarantees that there is a
way of physically implementing these SU(2)-invariant operations by
introducing ancillae in SU(2)-invariant states, implementing SU(2)-invariant
unitaries, and tracing out systems, it is instructive to see how the shift
operation is achieved in this way. In order to shift the $j$ value down by
$J>0$ (i.e. to implement the operation $S_{-J}(\cdot )S_{-J}^{\dag }$), one
simply adds an ancilla in state $\left\vert 0,0\right\rangle $ (the only
pure SU(2)-invariant state), implements the unitary $\left\vert
j,j\right\rangle \left\vert 0,0\right\rangle \rightarrow \left\vert
j-J,j-J\right\rangle \left\vert J,J\right\rangle $ (which is an
SU(2)-invariant operation), and discards the ancilla.

At first sight, one might hope to shift the $j$ value upward by the reverse
of this process. However, one would be required to begin by adding an
ancilla in the state $\left\vert J,J\right\rangle $ where $J>0$, and this
operation cannot be accomplished under the SU(2)-SSR because $\left\vert
J,J\right\rangle $, unlike $\left\vert 0,0\right\rangle ,$ does not come for
free. The difference between the SU(2) and U(1) cases -- shifts being
permitted in both directions for the former and only downward for the latter
-- is a result of the fact that every number eigenstate $\left\vert
n\right\rangle $ can be prepared under the SSR, whereas among the $
\left\vert j,j\right\rangle $ states only the singlet can be prepared under
the SSR.

\subsection{Deterministic single-copy transformations}

\strut We assume the states to be in the standard forms $|\psi \rangle
=\sum_{j}\sqrt{p_{j}}|j,j\rangle $ and $|\phi \rangle =\sum_{j}\sqrt{q_{j}}
|j,j\rangle .$

\begin{theorem}\label{D3} The necessary and sufficient condition for the
transformation $\left\vert \psi \right\rangle \rightarrow \left\vert \phi
\right\rangle $ to be possible by a deterministic SU(2)-invariant operation
is that
\begin{equation}
p_{j}=\sum_{J}w_{J}q_{j+J},
\end{equation}
where the sum varies over all positive integers and half-integers and the
$w_{J}$ form a probability distribution.
\end{theorem}

The proof is simply the one presented in Sec.~\ref{sec:RFsforphase}
for the associated U(1) result but where $k$ is substituted with $J$.

\subsection{Stochastic single-copy transformations}

Again, the situation is analogous to that of the U(1) case and consequently
it is useful to define the \emph{j-spectrum }of $\left\vert \psi
\right\rangle $ as the set of $j$ values to which $\left\vert \psi
\right\rangle $ assigns nonzero probability. \ If $\left\vert \psi
\right\rangle =\sum_{j}\sqrt{p_{j}}\left\vert j,j\right\rangle ,$ then the
set is $\{j|p_{j}\neq 0\}$. \ The cardinality of this set will again be
denoted by $\mathcal{S}(\psi )$, and a list of the elements of the set in
ascending order will be denoted
\begin{equation*}
\text{\textrm{j-}}\mathrm{Spec}(\psi )\equiv (j_{1}(\psi ),j_{2}(\psi
),...,j_{\mathcal{S}(\psi )}(\psi )).
\end{equation*}
The conditions under which a stochastic single-copy transformation is
possible are as follows.

\begin{theorem} \label{prop:criterionstochasticSU(2)}The
transformation $|\psi \rangle \rightarrow |\phi \rangle $ is possible using
stochastic SU(2)-invariant operations if and only if
\begin{equation}
\exists J\in \{0,1/2,1,\dots \}:\mathrm{j}\text{\textrm{-}}\mathrm{Spec}
(\phi )\subset \mathrm{j}\text{\textrm{-}}\mathrm{Spec}(\psi )-J.
\end{equation}
\end{theorem}

Again, the proof follows the one described in the U(1) case, but where the
shifts in $j,$ unlike the shifts in $n,$ can only be made in the downward
direction.

Finally, we also have a result concerning the maximum probability of
transformation which parallels Thm.~\ref{thm:maxprobU1}.

\begin{theorem}\label{S3} If there is only a single value of $J$ such that the
condition $\mathrm{j}$\textrm{-}$\mathrm{Spec}(\phi )\subset \mathrm{j}$
\textrm{-}$\mathrm{Spec}(\psi )-J$ holds, then the maximum probability of
achieving the transformation $\left\vert \psi \right\rangle \rightarrow
\left\vert \phi \right\rangle $ using SU(2)-invariant operations is
\begin{equation*}
P(\left\vert \psi \right\rangle \rightarrow \left\vert \phi \right\rangle
)=\min_{j}\left( \frac{p_{j}}{q_{j-J}}\right) .
\end{equation*}
\end{theorem}

\subsection{Stochastic SU(2)-frameness monotones}

Every stochastic U(1)-frameness monotone that was identified in
Sec.~\ref{sec:stochasticU1monotones}, has an analogue in the case of
the SU(2)-SSR for states restricted to $\mathcal{H}_{\hat{n}}.$ \ We
need only identify $ \left\vert j,j\right\rangle $ with $\left\vert
n\right\rangle $ to define them. However, in addition to these,
there are novel stochastic frameness monotones stemming from the
fact that the $j$-spectrum can only be shifted downward. For
instance, the highest $j$ value in the spectrum, $j_{
\mathcal{S}(\psi )}(\psi)$ clearly cannot be increased and
consequently is a stochastic SU(2)-frameness monotone.

\subsection{Asymptotic transformations}

We now discuss the asymptotic limit, where we are interested in
transformations of the form
\begin{equation}
|\psi \rangle ^{\otimes n}\;\rightarrow \;|\varphi \rangle ^{\otimes
m} \label{abc}
\end{equation}
in the limit where $n$ and $m$ go to infinity. (We switch from the
uppercase $N$ and $M$ of Thms.~\ref{thm:asymptoticmeasureU1}
and~\ref{thm:asymptoticz2} to lowercase $n$ and $m$ in order to
avoid confusion with the azimuthal angular momentum quantum number
$M$.)

Similarly to the U(1) case, we will see that if $\left\vert \psi
\right\rangle $ has a gapless $j$-spectrum, then $|\psi \rangle
^{\otimes n}$ has weights on $j$ that are Gaussian in the limit
$n\rightarrow \infty ,$ and given that the mean and variance of
Gaussian states are additive under tensor product, it follows that
the only features of $\left\vert \psi \right\rangle $ and
$\left\vert \varphi \right\rangle $ that will be significant are the
mean and variance of the distribution over $j$ that they define. We
therefore begin by providing precise definitions of these quantities
and demonstrating that they are ensemble SU(2)-frameness monotones.

Let $|\psi \rangle $ be a state in $\mathcal{H}_{\hat{n}}$. By analogy to
the number operator $\hat{N}$ in the U(1) setting, we define an operator $
\mathcal{J}$ on $\mathcal{H}_{\hat{n}}$ as
\begin{equation*}
\mathcal{J}\equiv \sum_{j=0,\frac{1}{2},1,...}j|j,j\rangle \langle j,j|\;.
\end{equation*}
\textbf{Definition:} The mean of $\mathcal{J}$ for the state $|\psi \rangle $
is:
\begin{equation*}
\mathcal{M}(|\psi \rangle )\equiv 2\langle \psi |\mathcal{J}|\psi \rangle \;.
\end{equation*}
\textbf{Definition:} The variance in $\mathcal{J}$ for the state $|\psi
\rangle $ is:
\begin{equation*}
V(|\psi \rangle )\equiv 4\left[ \langle \psi |\mathcal{J}^{2}|\psi \rangle
-\langle \psi |\mathcal{J}|\psi \rangle ^{2}\right] \;.
\end{equation*}

The factors in the definitions of $\mathcal{M}$ and $V$ have been chosen
such that $\mathcal{M}(\left\vert +\right\rangle )=1$ and $V(\left\vert
+\right\rangle )=1,$ where $\left\vert +\right\rangle \equiv $ $(|0,0\rangle
+|1,1\rangle )/\sqrt{2}$.

Lemma~\ref{lemma:varianceismonotone} of Sec.~\ref{sec:asymptransfU1} proves
that the variance in $\hat{N}$ is an ensemble monotone under the U(1)-SSR. \
A comparison of Eqs. (\ref{eq:krausforU1}) and (\ref{eq:KrausforSU2}) shows
that by identifying $\left\vert j,j\right\rangle $ with $\left\vert
n\right\rangle ,$ the SU(2)-invariant pure-to-pure transformations on the
space $\mathcal{H}_{\hat{n}}$ are mathematically a proper subset of the
U(1)-invariant pure-to-pure transformations, given that they do not allow
upward shifts of $j$. It follows that the variance in $\mathcal{J}$ is an
ensemble frameness monotone over the pure states of $\mathcal{H}_{\hat{n}}$
under the SU(2)-SSR.

Such a result also holds for the mean of $\mathcal{J}$.

\begin{lemma} $\mathcal{M}(|\psi \rangle )$ is an ensemble frameness
monotone on $\mathcal{H}_{\hat{n}}$.
\end{lemma}

\textbf{Proof.} An SU(2)-invariant measurement transforms a state $
\left\vert \psi \right\rangle $ into an ensemble of states. \ For the most
general such measurement, each outcome may be associated with an
SU(2)-invariant operation that has multiple Kraus operators. \ However, as
argued in the proof of lemma \ref{lemma:varianceismonotone}, it suffices to
consider the measurements for which each outcome is associated with a single
Kraus operator.

Suppose the outcome $\mu $ occurs with probability $w_{\mu }$ and is
associated with a Kraus operator $K_{\mu }$ which, by lemma \ref
{lemma:SU2invariantop}, has the form
\begin{equation*}
K_{\mu }=\sum_{j}c_{j}^{(\mu )}|j-J_{\mu },j-J_{\mu }\rangle \langle j,j|\;,
\end{equation*}
where $c_{j}^{(\mu )}$ are complex coefficients and $J_{\mu }$ is a
non-negative integer or half-integer. Note that
\begin{equation*}
\lbrack \mathcal{J},K_{\mu }]=-J_{\mu }K_{\mu }\;.
\end{equation*}
After an outcome $\mu $ has occurred, the state of the system is $|\phi
_{\mu }\rangle =\frac{1}{\sqrt{w_{\mu }}}K_{\mu }|\psi \rangle ,$ where $
w_{\mu }=\Vert K_{\mu }|\psi \rangle \Vert $. It follows that the average of
the mean of $\mathcal{J}$ is
\begin{align}
& \sum_{\mu }w_{\mu }\mathcal{M}(|\phi _{\mu }\rangle )=\sum_{\mu }\langle
\psi |K_{\mu }^{\dag }\mathcal{J}K_{\mu }|\psi \rangle \nonumber\\
& =\sum_{\mu }\langle \psi |K_{\mu }^{\dag }K_{\mu }\left( \mathcal{J}
-J_{\mu }\right) |\psi \rangle \nonumber\\
& \leq \langle \psi |\left( \sum_{\mu }K_{\mu }^{\dag }K_{\mu }\right)
\mathcal{J}|\psi \rangle \leq \mathcal{M}(|\psi \rangle )\;,
\label{eq:monotonicity4mean}
\end{align}
where we have used the fact that $\sum_{\mu }K_{\mu }^{\dag }K_{\mu }\leq I$
. Therefore, the mean of $\mathcal{J}$ is non-increasing on average under
SU(2)-invariant operations. QED.

Another important property of the mean and the variance of $\mathcal{J}$ is
that they are both strongly additive. That is, one can easily check that for
any two states $|\psi \rangle ,|\varphi \rangle \in \mathcal{H}_{\hat{n}}$
we have
\begin{eqnarray*}
\mathcal{M}(|\psi \rangle \otimes |\varphi \rangle ) &=&\mathcal{M}(|\psi
\rangle )+\mathcal{M}(|\varphi \rangle ) \\
V(|\psi \rangle \otimes |\varphi \rangle ) &=&V(|\psi \rangle )+V(|\varphi
\rangle )\;.
\end{eqnarray*}
This property plays an important role in the following theorem.

\begin{theorem}\label{thm:asymptoticsu2}
For states $|\psi \rangle $ and $\left\vert \varphi \right\rangle $
with gapless $j$-spectra, the transformation $|\psi \rangle
^{\otimes n}\;\rightarrow \;|\varphi \rangle ^{\otimes m}$ is
achievable by SU(2)-invariant operations in the limit $n\rightarrow
\infty $ with an optimal rate of
\begin{equation}
\lim_{n\rightarrow \infty }\frac{m}{n}=\min
\Big\{\frac{\mathcal{M}(|\psi \rangle )}{\mathcal{M}(|\varphi
\rangle )},\frac{V(|\psi \rangle )}{ V(|\varphi \rangle )}\Big\}\;.
\label{mncon}
\end{equation}
\end{theorem}

The proof is provided at the end of the section.

In some resource theories, the rate of interconversion between any two
states is provided by a single function over those states. For instance,
in pure state entanglement theory, the entropy of entanglement is such a
function, while in the resource theory of pure states under the Z$_{2}$-SSR,
$F^\infty$ of Eq.~(\ref{eq:uniqueframenessmonotoneZ2}) is such a function,
and in the resource theory of pure states with gapless number spectra under the U(1)-SSR,
the variance is such a function. What the theorem above shows is that in the resource
theory of pure states with gapless $j$-spectra under the SU(2)-SSR, no such
function can be found. Rather, one requires a \emph{pair} of distinct
functions over the pure states, namely, the mean of $\mathcal{J}$ and the
variance of $\mathcal{J},$ in order to deduce the rate of interconversion
between any two states.

Note that if $\mathcal{M}(|\psi \rangle )/\mathcal{M}(|\varphi
\rangle )\neq V(|\psi \rangle )/V(|\varphi \rangle ),$ then the
asymptotic rate of interconversion from $\left\vert \psi
\right\rangle $ to $\left\vert \varphi \right\rangle $ will not be
the inverse of the rate from $\left\vert \varphi \right\rangle $ to
$\left\vert \psi \right\rangle .$ \ To see this, simply note that
the rate in one direction is\emph{\ }$R(\psi \rightarrow \varphi
)=\min \left( \frac{\mathcal{M}(|\psi \rangle )}{\mathcal{M}
(|\varphi \rangle )},\frac{V(|\psi \rangle )}{V(|\varphi \rangle
)}\right) $ while in the other direction it is $R(\varphi
\rightarrow \psi )=\min \left( \frac{\mathcal{M}(|\varphi \rangle
)}{\mathcal{M}(|\psi \rangle )},\frac{ V(|\varphi \rangle )}{V(|\psi
\rangle )}\right) =\left[ \max \left( \frac{ \mathcal{M}(|\psi
\rangle )}{\mathcal{M}(|\varphi \rangle )},\frac{V(|\psi \rangle
)}{V(|\varphi \rangle )}\right) \right] ^{-1}$. But if the
asymptotic rates are not inverses of one another, then the
transformation is not reversible.

Conversely, if the ratio of means is equal to the ratio of
variances, then the rates are indeed inverses of one another. This
result is summarized by the following corollary.

\begin{corollary} \label{corollary}
%\textbf{Corollary:}
For states $|\psi \rangle $ and $\left\vert \varphi
\right\rangle $ with gapless $j$-spectra, the transformation $|\psi \rangle
^{\otimes n}\;\rightarrow \;|\varphi \rangle ^{\otimes m}$ can be achieved
\emph{reversibly} if and only if
\begin{equation*}
\frac{\mathcal{M}(|\psi \rangle )}{V(|\psi \rangle )}=\frac{\mathcal{M}
(|\varphi \rangle )}{V(|\varphi \rangle )}.
\end{equation*}
In that case $\lim_{n\rightarrow \infty }\frac{m}{n}\equiv
\mathcal{M}(|\psi \rangle )/\mathcal{M}(|\varphi \rangle )=V(|\psi
\rangle )/V(|\varphi \rangle )$ is the asymptotic rate of
interconversion.
\end{corollary}

We can therefore separate the set of all pure states with gapless
$j$-spectra into equivalence classes where the equivalence relation
is equality of the ratio of $\mathcal{M}$ to $V.$ \ Within each
equivalence class, reversible asymptotic interconversions are
possible and either $\mathcal{M}$ or $V$ can serve as the unique
measure of frameness (from which the asymptotic rate of
interconversion between any two states can be inferred). Asymptotic
interconversion of states in distinct equivalence classes can only
be achieved irreversibly.

The question arises of whether one can find, in each equivalence
class, a natural convention for a \textquotedblleft gold
standard\textquotedblright\ against which states can be compared. \ One
possibility is the state
\begin{equation*}
\left\vert +_{p}^{(j)}\right\rangle \equiv \sqrt{p}\left\vert
0,0\right\rangle +\sqrt{1-p}\left\vert j,j\right\rangle ,
\end{equation*}
for which $\mathcal{M(}|+_{p}^{(j)}\rangle )=2(1-p)j$ and $V\mathcal{(}
|+_{p}^{(j)}\rangle )=4p(1-p)j^{2},$ so that the ratio
\begin{equation*}
\mathcal{M}(|+_{p}^{(j)}\rangle )/V(|+_{p}^{(j)}\rangle )=\frac{1}{2pj}.
\end{equation*}
Consequently, if we choose $j=\left\lceil 1/r\right\rceil ,$ where $
\left\lceil x\right\rceil $ denotes the smallest integer larger than $x,$
and if we choose $p$ such that $p=1/2rj,$ so that $p\leq 1/2$ by definition,
then the state $\left\vert +_{p}^{(j)}\right\rangle $ can be taken as the
gold standard for the equivalence class with ratio $r.$

Note that for $r\geq 1,$ we have $j=1$, the standard state is of the form $
(1/\sqrt{2r})\left\vert 0,0\right\rangle +\sqrt{(2r-1)/2r}\left\vert
1,1\right\rangle $ and the rate at which a state $\left\vert \psi
\right\rangle $ can be converted to this standard is simply $\mathcal{M}
(\left\vert \psi \right\rangle )(r/(2r-1))=V(\left\vert \psi \right\rangle
)(r^{2}/(2r-1)).$ \ This convention is particularly nice at $r=1,$ where the
standard state is $(|0,0\rangle +|1,1\rangle )/\sqrt{2}$ and the rate is
simply $\mathcal{M}(\left\vert \psi \right\rangle )=V(\left\vert \psi
\right\rangle )$.

\textbf{Proof of theorem~\ref{thm:asymptoticsu2}.}
Because both $\mathcal{M}$ and $V$ are ensemble
frameness monotones, if the transformation is achievable then we must have $
\mathcal{M}(|\psi \rangle ^{\otimes n})\geq \mathcal{M}(|\phi \rangle
^{\otimes m})$ and $V(|\psi \rangle ^{\otimes n})\geq V(|\phi \rangle
^{\otimes m})$ which is equivalent to condition~(\ref{mncon}). This
establishes the necessity of the condition. We now demonstrate its
sufficiency.

Following reasoning parallel to that presented in the proof of
Thm.~\ref{thm:asymptoticmeasureU1} (but refraining from shifting the
$j$ value just yet), we can take $|\psi \rangle ^{\otimes n}$ to the
standard form
\begin{equation}
|\psi \rangle ^{\otimes n}=\sum_{j=nj_{\text{low}}}^{nj_{\text{high}}}\sqrt{
r_{j}}|j,j\rangle \;,  \label{eq:SU(2)psiN}
\end{equation}
where $j_{\text{low}}\equiv j_{1}(\psi )$ and $j_{\text{high}}\equiv
j_{ \mathcal{S}(\psi )}(\psi )$ and
\begin{equation}
r_{j}\equiv \sum \frac{j!}{n_{_{j_{\text{low}}}}!n_{_{j_{\text{low}
}}+1}!\cdots n_{j_{\text{high}}}!}p_{j_{\text{low}}}^{n_{j_{\text{low}
}}}p_{j_{\text{low}}+1}^{n_{j_{\text{low}}+1}}\cdots p_{j_{\text{high}
}}^{n_{j_{\text{high}}}}\;,  \label{eq:SU(2)rj}
\end{equation}
where the sum is taken over all sets of nonnegative integers $n_{_{j_{\text{
low}}}},n_{_{j_{\text{low}}}+1},\cdots ,n_{j_{\text{high}}}$ for which $
\sum_{j^{\prime }=j_{\text{low}}}^{j_{\text{high}}}n_{j^{\prime }}=n$ and $
\sum_{j^{\prime }=j_{\text{low}}}^{j_{\text{high}}}j^{\prime }n_{j^{\prime
}}=j$. In the limit $n\rightarrow \infty $, the $r_{j}$ approach a Gaussian
distribution as long as for all $j\in \{j_{\text{low}},...,j_{\text{high}
}\}, $ $p_{j}>0$ \cite{SVC04b}. The proof is blocked if $p_{j}=0$ for some
$j$ in this range and it is for this reason that our theorem is restricted
to pure states with gapless $j$-spectra.

First, suppose $V(|\varphi \rangle )/V(|\psi \rangle )\geq
\mathcal{M} (|\varphi \rangle )/\mathcal{M}(|\psi \rangle ),$ so
that $n/m=V(\left\vert \varphi \right\rangle )/V(\left\vert \psi
\right\rangle )$. In this case, the $m$-fold product of $\left\vert
\varphi \right\rangle $ has the same variance as the $n$-fold
product of $\left\vert \psi \right\rangle ,$ but a smaller mean
value of $j$. However, by lemma \ref{lemma:SU2invariantop} one can
always reduce the mean value of $\mathcal{J}$ by any integer or
half-integer amount, and this operation does not affect the
variance. Implementing such a shift leaves one with a state that is
arbitrarily close to $m$ copies of $\left\vert \varphi
\right\rangle$ in the limit of $n\rightarrow \infty$. To see that
this is the case, define
\begin{equation*}
J_{0}\equiv \left\lfloor n\mathcal{M}(|\psi \rangle )-m\mathcal{M}(|\varphi
\rangle )\right\rfloor
\end{equation*}
where $\left\lfloor x\right\rfloor $ denotes the largest integer or
half-integer less than $x,$ and define
\begin{equation*}
\left\vert \gamma _{n}\right\rangle \equiv S_{-J_{0}}\left( \left\vert \psi
\right\rangle ^{\otimes n}\right) .
\end{equation*}
Clearly,
\begin{eqnarray*}
\mathcal{M}(\left\vert \gamma _{n}\right\rangle ) &=&n\mathcal{M}(|\psi
\rangle )-J_{0} \\
&\rightarrow &m\mathcal{M}(\left\vert \varphi \right\rangle )
\end{eqnarray*}
in the limit of $n\rightarrow \infty$.

The alternative is that $V(|\varphi \rangle )/V(|\psi \rangle )\leq \mathcal{
M}(|\varphi \rangle )/\mathcal{M}(|\psi \rangle ),$ so that $n/m=\mathcal{M}
(\left\vert \varphi \right\rangle )/\mathcal{M}(\left\vert \psi
\right\rangle )$. In this case, the $m$-fold product of $\left\vert \varphi
\right\rangle $ has the same mean value of $\mathcal{J}$ as the $n$-fold
product of $\left\vert \psi \right\rangle $ but a smaller variance. \ All
that remains to show therefore is that, using SU(2)-invariant operations,
one can reduce the variance by an arbitrary amount while preserving the mean
value of $\mathcal{J}.$ \ (Note that such an operation does not lead to an
increase of either $V$ or $\mathcal{M}$ and so is consistent with the latter
being ensemble frameness monotones.)

The requisite operation involves implementing a measurement on each
copy of $ \left\vert \psi \right\rangle .$ Suppose that the outcomes
of the measurement are labeled by $\mu ,$ the probability of outcome
$\mu $ is denoted by $w_{\mu }$ and the normalized final state
associated with outcome $\mu $ is denoted by $\left\vert \psi _{\mu
}\right\rangle .$ We begin by showing that there exists a
measurement on $\left\vert \psi \right\rangle $ such that the
ensemble of final states has, on average, mean of $\mathcal{J}$
equal to that of $\left\vert \psi \right\rangle ,$
\begin{equation}
\sum_{\mu }w_{\mu }\mathcal{M}(\left\vert \psi _{\mu }\right\rangle )=
\mathcal{M}(\left\vert \psi \right\rangle ),  \label{eq:Ms}
\end{equation}
and a variance satisfying
\begin{equation}
\sum_{\mu }w_{\mu }V(|\psi _{\mu }\rangle )=\frac{m}{n}V(|\varphi \rangle ).
\label{eq:Vs}
\end{equation}

We assume that each outcome $\mu $ is associated with an operation $\mathcal{
E}_{\mu }$ defined by a single Kraus operator of the form $K_{\mu
}=\sum_{j}c_{j}^{(\mu )}\left\vert j\right\rangle \left\langle j\right\vert $
. This is an SU(2)-invariant measurement, by virtue of the Kraus
operators being of the form outlined in lemma~\ref{lemma:SU2invariantop}.
However, for no outcome $\mu $ does the measurement incorporate a nontrivial
shift operation $S_{-J}$.  The constraint that the overall operation be
trace-preserving implies that $\sum_{\mu }K_{\mu }^{\dag }K_{\mu }=I,$ or
equivalently, that $\sum_{\mu }|c_{j}^{(\mu )}|^{2}=1$ for all $j.$ \ Note
that this constraint is satisfied if one takes $c_{j}^{(\mu )}=u_{j\mu }$
where $u$ is a unitary matrix.
Such a measurement does not change the mean of $\mathcal{J}$ on average
because it saturates the inequality in Eq.~(\ref{eq:monotonicity4mean}).

Recalling that the maximum $j$ value to which $\left\vert \psi \right\rangle
$ assigns nonzero probability is $j_{\max },$ it suffices to consider the
operation $\mathcal{E}_{\mu }$ on $\mathrm{span}(\left\vert j\right\rangle
\left\langle j\right\vert :j\leq j_{\max })$. If the $c_{j}^{(\mu )}
$ are to be the components of a unitary matrix, then the range of $\mu $
must also be $0$ to $j_{\max }.$ Now consider two extreme cases.

(i) $\mathcal{E}_{\mu }$ is defined by the unitary matrix $u_{j\mu }=\delta
_{j,\mu }.$ \ In this case, $\left\vert \psi _{\mu }\right\rangle
=\left\vert j=\mu \right\rangle \left\langle j=\mu \right\vert $ so that $
V(\left\vert \psi _{\mu }\right\rangle )=0$ for all $\mu ,$ and consequently
$\sum_{\mu }w_{\mu }V(\left\vert \psi _{\mu }\right\rangle )=0.$
Meanwhile, the mean of $\mathcal{J}$ is the same on average, $\sum_{\mu
}w_{\mu }\mathcal{M}(\left\vert \psi _{\mu }\right\rangle )=\sum_{\mu
}w_{\mu }\mu =\sum_{\mu }|\left\langle \mu |\psi \right\rangle |^{2}\mu =
\mathcal{M}(\left\vert \psi \right\rangle ).$

(ii) $\mathcal{E}_{\mu }$ is defined by the unitary Fourier matrix $u_{j\mu
}=j_{\max }^{-1/2}\exp [i2\pi \mu j/j_{\max }]$. In this case, $\left\vert
\psi _{\mu }\right\rangle $ differs from $\left\vert \psi \right\rangle $
only by the phases of the $\left\vert j\right\rangle $ terms, so that $
V(\left\vert \psi _{\mu }\right\rangle )=V(\left\vert \psi \right\rangle )$
for all $\mu ,$ and consequently $\sum_{\mu }w_{\mu }V(\left\vert \psi _{\mu
}\right\rangle )=V(\left\vert \psi \right\rangle )$.

Because there exists a continuous path between any two
unitaries~\footnote{We thank Larry Bates, Peter Lancaster and
Peter Zvengrowski for bringing this to our attention;
in particular we thank Larry Bates and Peter Lancaster for showing us
(explicitly) several different continuous paths that connect the Fourier matrix
$u_{j\mu}=j_{\max }^{-1/2}\exp [i2\pi \mu j/j_{\max }]$ with the identity.}, and
because the average variance is a continuous function of the unitary matrix $
u$, we conclude that for every variance in the range $0$ to $V(\left\vert
\psi \right\rangle )$, there exists a unitary matrix $u$ on the path
connecting $\delta _{j,\mu }$ and $j_{\max }^{-1/2}\exp [i2\pi \mu j/j_{\max
}]$ that achieves this variance on average. In particular, we can find a
measurement that yields $\sum_{\mu }w_{\mu }V(\left\vert \psi _{\mu
}\right\rangle )=\frac{m}{n}V(|\varphi \rangle ).$

After performing this measurement on each of the $n$ copies of $|\psi
\rangle $, one obtains the final state
\begin{equation*}
|\chi _{n}\rangle =|\psi _{0}\rangle ^{\otimes n_{0}}\otimes |\psi
_{1}\rangle ^{\otimes n_{1}}\otimes \cdots |\psi _{N-1}\rangle ^{\otimes
n_{j_{\max }}}\;,
\end{equation*}
where in the limit $n\rightarrow \infty $ we have $n_{\mu }\rightarrow
w_{\mu }n$ for $\mu =0,1,...,j_{\max }$. Hence, in the limit $n\rightarrow
\infty $ we have
\begin{eqnarray*}
V(|\chi _{n}\rangle ) &=&\sum_{\mu =0}^{j_{\max }}n_{\mu }V(|\psi _{\mu
}\rangle ) \\
&\rightarrow &n\sum_{\mu =0}^{j_{\max }}w_{\mu }V(|\psi _{\mu }\rangle ) \\
&=&mV(|\varphi \rangle ).
\end{eqnarray*}
In addition,
\begin{eqnarray*}
\mathcal{M}(|\chi _{n}\rangle ) &=&\sum_{\mu =0}^{j_{\max }}n_{\mu }\mathcal{
M}(|\psi _{\mu }\rangle ) \\
&\rightarrow &n\sum_{\mu =0}^{j_{\max }}w_{\mu }\mathcal{M}(|\psi _{\mu
}\rangle ) \\
&=&n\mathcal{M}(|\psi \rangle ) \\
&=&\mathcal{M}(|\varphi \rangle ^{\otimes m})\;,
\end{eqnarray*}
where we have used the fact that the mean of $\mathcal{J}$ is unchanged as $
\left\vert \psi \right\rangle \rightarrow \left\vert \psi _{\mu
}\right\rangle $. Hence, in the limit $n\rightarrow \infty $, $|\chi
_{n}\rangle $ and $|\varphi \rangle ^{\otimes m}$ have the same mean value
of $\mathcal{J}$ and the same variance.

Therefore, if it can be shown that both $\left\vert \chi _{n}\right\rangle $
and $|\varphi \rangle ^{\otimes m}$ approach Gaussian states in the limit $
n\rightarrow \infty ,$ then it follows that these approach the \emph{same}
state in this limit. \ Clearly, \ $|\varphi \rangle ^{\otimes m}$ approaches
a Gaussian by the same argument establishing that $|\psi \rangle ^{\otimes n}
$ does. \ Similarly, each factor state of $\left\vert \chi _{n}\right\rangle
$ of the form $|\psi _{\mu }\rangle ^{\otimes n_{\mu }}$ approaches a
Gaussian because for each $\mu ,$ $n_{\mu }\rightarrow \infty $ as $
n\rightarrow \infty .$ \ It remains only to show that a tensor product of
Gaussians is also Gaussian.

Consider the tensor product $|\psi _{1}\rangle ^{\otimes n_{1}}\otimes |\psi
_{2}\rangle ^{\otimes n_{2}}$, where $|\psi _{\mu }\rangle ^{\otimes n_{\mu
}}=\sum_{j}r_{j}^{(\mu )}\left\vert j,j\right\rangle \left\langle
j,j\right\vert $ and the $r_{j}^{(\mu )}$ are Gaussian distributions over $
j. $ \ Clearly,
\begin{equation*}
|\psi _{1}\rangle ^{\otimes n_{1}}\otimes |\psi _{2}\rangle ^{\otimes
n_{2}}=\sum_{j,j^{\prime }}r_{j}^{(1)}r_{j^{\prime }}^{(2)}\left\vert
j+j^{\prime },j+j^{\prime }\right\rangle \left\langle j+j^{\prime
},j+j^{\prime }\right\vert .
\end{equation*}
Defining $j^{\prime \prime }\equiv j+j^{\prime }$ and $x\equiv j-j^{\prime
}, $ we have
\begin{equation*}
|\psi _{1}\rangle ^{\otimes n_{1}}\otimes |\psi _{2}\rangle ^{\otimes
n_{2}}=\sum_{j^{\prime \prime }}\tilde{r}_{j^{\prime \prime }}\left\vert
j^{\prime \prime },j^{\prime \prime }\right\rangle \left\langle j^{\prime
\prime },j^{\prime \prime }\right\vert ,
\end{equation*}
where
\begin{equation*}
\tilde{r}_{j^{\prime \prime }}=\sum_{x}r_{(j^{\prime \prime
}+x)/2}^{(1)}r_{(j^{\prime \prime }-x)/2}^{(2)}.
\end{equation*}
In the limit of $n\rightarrow \infty ,$ this is a convolution of two
Gaussians, which is also a Gaussian.  Note that the variance
(respectively mean) of the convolution is equal to the sum of the
variances (respectively means) of the components, as is required for
consistency with the additivity of $V$ and $\mathcal{M}$ under
tensor product. The argument clearly generalizes to the tensor
product of an arbitrary number of Gaussians, implying that
$\left\vert \chi _{n}\right\rangle $ approaches a Gaussian. \ QED.

\section{Conclusions}

A superselection rule is a restriction on operations. \ It may arise from
the practical circumstance of lacking a reference frame for some degree of
freedom. \ The nature of this degree of freedom -- in particular its
associated symmetry group -- determines the set of operations that are
forbidden by the superselection rule. Superselection rules therefore admit
of degree: the more operations they forbid, the stronger they are.

 There is a strict ordering by strength
of the three SSRs we consider in this article. If $\mathfrak{O}[G]$
denotes the set of operations that are forbidden under a G-SSR, then
$\mathfrak{O}[\textrm{Z}_2] \subset \mathfrak{O}[\textrm{U(1)}]
\subset \mathfrak{O}[\textrm{SU(2)}]$.  At an abstract level, this
clearly follows from the fact that Z$_2$ is a subgroup of U(1) which
is a subgroup of SU(2).  A physical explanation of the ordering,
however, requires us to go beyond the particular restrictions --
lack of reference frames for chirality, optical phase, and
orientation -- that we have chosen to emphasize in this article as
illustrations of each type of SSR. (For instance, the operations
that are forbidden by lacking a frame for chirality are not a subset
of those forbidden by lacking a frame for orientation because
without a shared reference frame for chirality, Bob cannot tell
whether a glove he receives from Alice would be described as left or
right by her, whereas if he lacks a shared reference frame for
orientation, he can still do so.)

Fortunately, a triple of restrictions that \emph{do} provide a
physical explanation of the ordering can easily be provided. As
noted in the article, in addition to its significance in optics, the
U(1)-SSR also describes the restriction that Alice and Bob face when
they share a single direction in space. Taking $\hat{z}$ to be their
shared axis, what they lack is knowledge of the angle between their
$\hat{x}$ axes. Operations that are forbidden when sharing a single
axis are a strict subset of those that are forbidden when sharing no
axis (the restriction leading to an SU(2)-SSR). Similarly, in
addition to characterizing the lack of a chiral reference frame, a
Z$_2$-SSR characterizes the restriction that arises if Alice and Bob
share a $\hat{z}$-axis and know the angle modulo $\pi$ between their
respective $\hat{x}$-axes.  In this case, they certainly know more
than if they knew nothing of the angle between their $\hat{x}$-axes
and consequently the operations that are forbidden are a strict
subset of those that arise in the latter case.  In summary, an
SU(2)-SSR) is a stronger restriction than that of a U(1)-SSR which
in turn is stronger than that of a Z$_{2}$-SSR.

 We have shown that the extent of
manipulations that one can make upon the resources defined by each
restriction (quantum states that stand in for the missing reference
frames) depends on the strength of the restriction. Given a single
copy of any pure state that acts as a Z$_2$-resource, there is a
nonzero probability of transforming it into a single copy of any
other such state. By contrast, for pure states that act as
U(1)-resources, there are many pairs for which such a transformation
is not possible (in either one or both directions). The impossible
cases are even more numerous for pure states that act as
SU(2)-resources. Similarly, arbitrarily many copies of any pure
state that acts as a Z$_@$-resource can be transformed
\emph{reversibly} into any other such state with some nonzero rate,
whereas only for certain classes of pure U(1)-resource states is
such asymptotic reversible interconversion possible, and for pure
SU(2)-resource states, the classes are smaller still.

The resource theory for quantum reference frames therefore provides
another example, in addition to that of the resource theory of multipartite
entanglement, of the generic phenomenon that the ease of resource
manipulations decreases with the strength of the restriction.

There are a great many open questions that remain concerning the
manipulation of quantum reference frames. \ In the context of phase
references, the problem of finding the maximum probability with which one
can transform a single copy of some state into a single copy of another has
only been solved under restrictive conditions. Furthermore, the problem of
characterizing when asymptotic transformations can be achieved with nonzero
rate and when they can be achieved reversibly has only been solved
completely for states with gapless number spectra. \ Similar comments apply
for the subset $\mathcal{C}_{\hat{n}}$ of Cartesian frame states.

Extending our results to arbitrary states in the SU(2)-resource theory is
likely to be a very difficult task.\ Note, however, that a feature of the
subspace $\mathcal{H}_{\hat{n}}$ that simplifies the resource analysis is
that it is closed under tensor products. Another subspace that is closed in
this fashion is $\mathcal{H}_{\hat{n},0}\equiv $ \textrm{span}$\{\left\vert
j,m=0\right\rangle :j=0,1,2,...\}$ because $\left\vert
j_{1},m_{1}=0\right\rangle $ and $\left\vert j_{2},m_{2}=0\right\rangle $
only couple to states $\left\vert j,m\right\rangle $ where $m=m_{1}+m_{2}=0.$
\ Furthermore, subspaces of the form $\mathcal{H}_{\hat{n},m}\equiv $
\textrm{span}$\{\left\vert j,m\right\rangle :j\geq m\}$ (where $j$ values
are integer or half-integer according to $m$), which are simply the various
eigenspaces of $J_{\hat{n}},$ although not closed under tensor product
nonetheless have the nice feature that the tensor product of a state from $
\mathcal{H}_{\hat{n},m}$ and a state from $\mathcal{H}_{\hat{n},m^{\prime }}$
is confined to $\mathcal{H}_{\hat{n},m+m^{\prime }}.$ It follows that the
theories of frame manipulations on these subspaces are likely to be more
tractable than the completely general theory and consequently a promising
avenue for future research.

Another direction in which this work may be extended is towards resource
theories for reference frames associated with other groups. \ Any sort of
reference frame can be considered, but particularly interesting
possibilities include: reference orderings (associated with the permutation
group)~\cite{BarWis03,KorKem04,JWBVP06}, inertial frames (associated with
the Lorentz group) \cite{BarTer05}, frames for global positioning
(associated with the Heisenberg-Weyl group)~\cite{AhaKau84}, or even exotic
possibilities such as frames for the color degree of freedom in quantum
chromodynamics (associated with SU(3)).

There are also many aspects of resource theories that we have not addressed
here. For instance, this article has only been concerned with single-copy
and asymptotic transformations. \ Transformations between multiple but
finite numbers of copies have not been considered. \ More importantly, we
have restricted our attention to pure states. \ In practice, resources are
always mixed to some extent and one of the some significant problems is to
determine the extent to which one can purify a resource. Furthermore, if
the experience from entanglement theory is any guide, many interesting and
surprising phenomena are likely to arise in the context of mixed
states. \ One can already see, however, that the parallels to entanglement
theory will be limited. Specifically, because there are no SU(2)-invariant
pure states with $j>0$, a mixed SU(2)-invariant state in a subspace $
\mathcal{H}_{j}$ with $j>0$ does not admit any convex decomposition into
SU(2)-invariant pure states. \ It follows that although the SU(2)-invariant
mixed state is not a resource, the elements of every convex decomposition of
this state into pure states \emph{are} resources. It is therefore a bad
idea, for instance, to attempt to define a frameness monotone for such mixed
states by the convex roof extension -- a frameness of
formation must be defined differently from the
entanglement of formation.

A strong motivation for the present work is that every novel
resource theory provides an interesting new perspective on its
brethren. Besides the case of quantum reference frames, resources
that have seen some attention of late include: purity as a resource
for doing mechanical work~\cite{Opp02,Hor03}, nonGaussian states as
resources for overcoming a restriction to Gaussian
operations~\cite{EisPle03}, and nonlocal boxes or super-quantum
correlations as resources in the context of quantum
theory~\cite{BLMPPR05,BarPir05,JonMas05}. Even if one's interest is
confined to a particular resource theory, such as the theory of
entanglement, studying alternative resource theories as foils to the
one of interest may well provide a faster route to progress.

That being said, it is also hoped that the present work and its like
will provide a bit of an antidote to the pernicious notion that the
theory of entanglement somehow provides the deepest insights into
the foundations of quantum theory. Not so; the restriction with
respect to which the resource of entanglement is defined -- local
operations and classical communication -- is a \emph{practical
}rather than a foundational restriction. The universe doesn't care
especially for classical channels. \emph{We} care because it is at
present much more difficult to equip distant parties with a quantum
channel than it is to equip them with one that is classical, and
consequently anything that can substitute for the former given the
latter is of great practical value to us. \ The restriction of LOCC
is no different in kind from that of failing to have a sample of
some particular reference frame. \ Nor is entanglement theory
particularly distinguished: it is just one of \emph{many} resource
theories and many of its features are quite generic. It is hoped
that the detailed examples provided in this article will drive this
point home and prompt the quantum information community to spend
less time on the increasingly esoteric details of entanglement
theory and more time on exploring basic questions about other
physical resources. They are likely to be rewarded with unexplored
country.

\emph{Acknowledgments:---} We acknowledge valuable discussions with
Matthias Christandl, Renato Renner, Jonathan Oppenheim, Stephen
Bartlett, Terry Rudolph, Barry Sanders, David Meyer and Nolan
Wallach. We would like also to thank Howard Wiseman for early
discussions concerning reference frames as a unipartite resource,
Peter Turner for clarifying the content of the Wigner-Eckart
theorem, Dave Kielpinski for suggesting a simplification of one of
our proofs, and Ali Rezakhani for help with the figures. G.G.
acknowledges support from NSERC. R.W.S. acknowledges support from
the Royal Society.

\strut

\end{document}